\documentclass[final,5p,times,twocolumn]{elsarticle}
\usepackage{amsfonts,amssymb}
\usepackage{amsmath,amscd,amsthm}
\usepackage{xcolor}
\usepackage{bbm}
\usepackage{algorithm}
\usepackage{algpseudocode}
\usepackage{tikz}
\usepackage{dsfont}
\usepackage[breaklinks,pagebackref,unicode]{hyperref}

\newcommand{\Int }    {\displaystyle \int}
\newcommand{\Si}{\boldsymbol{\sigma}}
\newcommand{\AL}{\boldsymbol{\alpha}}
\newcommand{\C}{\boldsymbol{\chi}}
\newcommand{\vect}{\mathbf{b}}
\newcommand{\flow}{\mathbf{X}}
\newcommand{\pos}{\mathbf{x}}
\newcommand{\SP}{{\cal S}p{\cal S}t}
\newcommand{\be}{\begin{equation}}
\newcommand{\de}{\end{equation}}

\makeatletter
\def\ps@pprintTitle{%
  \let\@oddhead\@empty
  \let\@evenhead\@empty
  \def\@oddfoot{}%
  \let\@evenfoot\@oddfoot}
\makeatother

\newtheorem{propositionav}{Proposition 5.5}
\newtheorem{theoremav}{Theorem 1.1}
\newtheorem{theorem}{Theorem}
\newtheorem{definition}{Definition}
\newtheorem{property}{Property}
\newtheorem{lemma}{Lemma}
\newtheorem{corollary}{Corollary}
\newtheorem*{conjecture}{Conjecture}

\begin{document}

\title{Spontaneous stochasticity and the Armstrong-Vicol passive scalar}
\author[1]{Wandrille Ruffenach}
\address[1]{ENS de Lyon, CNRS, LPENSL, UMR5672, 69342, Lyon cedex 07, France.}
\author[2]{Eric Simonnet \corref{cor1}} \ead{Eric.Simonnet@univ-cotedazur.fr}
\cortext[cor1]{Corresponding author}
\address[2]{Institut de Physique de Nice and INRIA Sophia-Antipolis, CNRS, Univ. C\^ote d'Azur}
\author[2]{Nicolas Valade}


\date{\today}
\begin{abstract}
Spontaneous stochasticity refers to the emergence of intrinsic randomness in deterministic systems under singular limits, a phenomenon conjectured to be fundamental in turbulence. Armstrong and Vicol \citep{AV23,AV24} recently constructed a deterministic, divergence-free multiscale
vector field arbitrarily close to a weak Euler solution, proving that a passive scalar transported by this field exhibits anomalous dissipation and lacks a selection principle in the vanishing diffusivity limit.

{\it This work aims to explain why this passive scalar exhibits both Lagrangian and Eulerian spontaneous stochasticity.}

Part I provides a historical overview of spontaneous stochasticity, details the Armstrong-Vicol passive scalar model, and presents numerical evidence of anomalous diffusion, along with a refined description of the Lagrangian flow map.

In Part II, we develop a theoretical framework for Eulerian spontaneous stochasticity. We define it mathematically, linking it to ill-posedness and finite-time trajectory splitting, and explore its measure-theoretic properties and the connection to RG formalism. This leads us to a well-defined {\it measure selection principle} in the inviscid limit.
Within this limit, we identify a structured family of probability measures, with building blocks the Dirac measures responsible for the well-posed inviscid limits. This approach allows us to rigorously classify universality classes based on the ergodic properties of regularisations. To complement our analysis, we provide simple yet insightful numerical examples.

Finally, we show that the absence of a selection principle in the Armstrong-Vicol model corresponds to Eulerian spontaneous stochasticity of the passive scalar. We also numerically compute the probability density of the effective renormalised diffusivity in the inviscid limit. This Eulerian behaviour contrasts sharply with the Kraichnan model, where only Lagrangian spontaneous stochasticity is possible. We argue that the lack of a selection principle should be understood as a measure selection principle over weak solutions of the inviscid system.

\end{abstract}
\maketitle

\addtocontents{toc}{\protect\setcounter{tocdepth}{1}}  
\section{Introduction}\noindent
\addtocontents{toc}{\protect\setcounter{tocdepth}{3}} 

This project can be understood as a very modest attempt to reconcile two perspectives -- those of mathematicians and physicists -- on the nature of spontaneous stochasticity.

This work is motivated in large part by a recent mathematical proof establishing the existence of a dissipative anomaly in a passive-scalar transport system \cite{AV23,AV24}, 
as well as the absence of a selection principle. Similar results have been observed in recent studies of advection-diffusion partial differential equations (PDEs) \cite{Drivas_Elgindi,colombo,Titi2023,burczak2023anomalous}.

The main message of this paper is twofold. On the one hand, the absence of a selection principle, as seen by mathematicians, is often perceived negatively as a mathematical pathology of the model in the inviscid limit.
The consequence is that such a pathology further distances the scientific community from a deeper understanding of turbulence, adding yet another layer of mystery to the connection between the Navier-Stokes equations and the Euler equations.

On the other hand, the notion of spontaneous stochasticity is  not so well defined among physicists: it is often identified with Lagrangian spontaneous stochasticity as introduced by \cite{Chaves2003}.
Alternatively, it might also be confused sometimes with a notion of classical chaos or with a problem of representing complexity -- for instance, the need to represent complexity in a probabilistic manner.
As a result, spontaneous stochasticity may be wrongly perceived as merely a methodological or representational issue rather than as a genuine physical phenomenon in its own right.

The goal here is to reconcile and clarify these perspectives.
For mathematicians, the absence of a selection principle should not lead to the rejection of the model and is certainly not a mathematical pathology. Its very existence is of a physical nature. Whether it is more or less realistic is a question of a different kind.
In the system of  \cite{AV23,AV24}, as well as by \cite{burczak2023anomalous}, 
it appears in a much more realistic manner than ever seen before.
Remarkably, it can be analysed mathematically, largely thanks to the development of convex integration over the past decade
\cite{hprinciple,Isett,Admi_Onsager,BuckVic}.

For physicists, the fact that the absence of a selection principle appears in linear PDEs of passive scalar and is, at its core, what they refer to as "Eulerian spontaneous stochasticity" should be recognised as a significant milestone.
In fact, the role that these recent models of passive scalar transport play in understanding Eulerian spontaneous stochasticity is reminiscent of the role that the Kraichnan model \cite{Kraichnan} has played in the turbulence community in understanding the dissipative anomaly and Lagrangian spontaneous stochasticity \cite{Chaves2003,Drivas17}.
\\

The paper is structured in two mostly independent parts. First, we provide a rather detailed historical retrospective of the concept of spontaneous stochasticity. We then present a description of the Armstrong-Vicol model \cite{AV23,AV24}. The main result concerning the dissipative anomaly -- equivalently Lagrangian spontaneous stochasticity -- is then studied numerically.

The second part is more mathematically oriented and addresses the framework of {\it Eulerian spontaneous stochasticity}. We first provide a mathematical definition of spontaneous stochasticity in a very general setting, i.e., from the perspective of dynamical systems (Sections \ref{secDef}--\ref{which}). This notion is intentionally broader than possibly more restrictive interpretations of spontaneous stochasticity.

We then discuss the various implications of this definition in finite dimensions, particularly a measure selection principle (Section \ref{secmea}) and the crucial role played by singularities (Section \ref{which}). Next, we present nontrivial examples in finite dimensions (Subsections \ref{Ex1}--\ref{Ex4}). Finally, we examine the Armstrong-Vicol system by discussing the recent result obtained by \cite{AV24} on the absence of a selection principle (Section \ref{Theirproof}). We also provide an alternative mathematical proof based on our own definition of Eulerian spontaneous stochasticity (Section \ref{secMAIN}).
Additionally, we highlight numerical signatures of Eulerian spontaneous stochasticity, from the associated renormalisation sequence (Section \ref{num}).

\tableofcontents
\part{}

\section{Historical review}\noindent

The modern notion of {\it spontaneous stochasticity} has been quite elusive since the first time it
was coined as such by \cite{Chaves2003}. It emerged from a long and intricate history mostly
related to turbulence and the Navier-Stokes equations. 
Spontaneous stochasticity is the property that a deterministic system, in some singular limit, can exhibit an intrinsic source of randomness.
It has often been wrongly identified
with the phenomenon of chaos in nonlinear systems: it is not, not even close... 
The confusion indeed arises simply because in turbulence: chaos and spontaneous stochasticity conspire together.

There are in fact two different parallel stories: a Lagrangian tale
and an Eulerian one. It seems we are still far from reconciliating the two.
\subsection{Lagrangian spontaneous stochasticity}\label{subsec_lss}
 One can probably root the concept 
with the work of Richardson on turbulence diffusion \cite{Richardson1926}.
In essence, the separation ${\bf R}_{12}(t) = {\bf R}_1(t)-{\bf R}_2(t)$ between two close fluid particles in a given flow would satisfy $\frac{d}{dt} {\bf R}_{12} = {\bf v}(t,{\bf R}_1)
-{\bf v}(t,{\bf R}_2)$ or taking the norm $\rho := ||{\bf R}_{12}||, \frac{d}{dt} \rho^2 =
2 \rho \cdot \Delta {\bf v}$. If the fluid velocity is smooth 
$||{\bf v}(t,{\bf R}_1)
-{\bf v}(t,{\bf R}_2)|| \propto ||{\bf R_1-R_2}||$ (Lipschitz), then one would obtain
$\frac{d}{dt} \rho^2 \propto \rho^2$. This yields an exponential growth.
In a turbulent fluid however, velocities are not smooth. It is known from Kolmogorov
\cite{Kolmogorov1941}
that, neglecting intermittent effects, a reasonable approximation is the K41 velocity
having $\frac13$ H\"older regularity: $||{\bf v}(t,{\bf R}_1)
-{\bf v}(t,{\bf R}_2)|| \propto ||{\bf R_1-R_2}||^{\frac{1}{3}}$. It implies that 
one has $\frac{d}{dt} \rho^2 \propto (\rho^2)^\frac23$. This yields a totally different behaviour
where now, $\rho$ grows as a power of time, even more important this is true no matter how close are the two particles initially. This is the famous Richardson's law:
$\langle \rho^2 \rangle \propto t^3$ observed numerically in a consistent way
, see e.g. \cite{Falkovich2001,salazar2009,BEC} and references therein.
It is much less clear at the experimental level  \cite{Jullien2000}. 

Meanwhile, one of the fundamental aspect of turbulence, already suggested
by Taylor \cite{Taylor}, is the  fact that kinetic energy can be dissipated 
for fluid of "infinitesimal viscosity". It was indeed phrased as a conjecture by
Onsager \cite{Onsager49} for 3D Euler, that we would like to quote: 
"{\it It is of some interest to note that in principle, turbulent dissipation as
described could take place just as readily without the final assistance by
viscosity. In the absence of viscosity, the standard proof of the conservation
of energy does not apply, because the velocity field does not remain differentiable! In fact it is possible to show that the velocity field in such "ideal"
turbulence cannot obey any LIPSCHITZ condition of the form $|{\bf v(r'+r)-v(r')}| < {\rm Const.}r^n$,
for any order $n$ greater than 1/3; otherwise the energy is conserved.}"

This remarkable phenomenon of "dissipative anomaly" (sometimes referred to as "the zeroth law of turbulence") also appears in simpler models and in particular
in passive scalar models: linear advection-diffusion equations for a scalar field
$\theta=\theta(t,{\bf x})$ taking the simple form $\partial_t \theta + {\bf v} \cdot \nabla \theta = 
\kappa \Delta \theta$, and where the velocity ${\bf v}(t,{\bf x})$ and initial condition $\theta(0,{\bf x})$ are prescribed. The dissipative anomaly then is expressed mathematically simply as
$\limsup_{\kappa \to 0} \kappa ||\nabla \theta||_{L^2_{x,t}}^2 > 0$.
A synthetic model of turbulence which has played a key role in turbulence studies is
the so-called Kraichnan model \cite{Kraichnan}. It is again an advection-diffusion system where the
velocity mimics some key aspects of velocity fields solving 3-D Navier-Stokes at high Reynolds number. It is a random white-in-time, correlated in space, Gaussian field having two free parameters
controlling its spatial regularity over a given range of scales.
In a remarkable series of works \cite{Gawedzki98,Chaves2003,Cardy2008}, Gaw\c{e}dzky and colleagues were able to establish
using the Kraichnan ensemble that indeed the classical determinism of Lagrangian trajectories breaks down
in the limit of small diffusion. In particular, they showed that dissipative anomaly was
the sole consequence of the phenomenon of spontaneous stochasticity: the Lagrangian particles have nonunique trajectories even when starting from the same initial condition and fixed
realisation of the velocity, and this was essentially a consequence of the spatial
roughness of the velocity. This work was a breakthrough in the understanding on two seemingly different phenomena: spontaneous stochasticity and dissipative anomaly. Some decades later, this equivalence was also established in more general cases by the work of \cite{Drivas17} as a Lagrangian fluctuation-dissipation theorem. In simple words:
spontaneous stochasticity and anomalous diffusion are equivalent, at least for passive scalars.

Of importance is to recall that either in the work of Richardson or later, 
the notion of spontaneous stochasticity must be called in fact {\it Lagrangian spontaneous stochasticity}. "Lagrangian" was later on becoming quite implicit and dropped.
The story of course does not end there. Even more fascinating aspects emerged starting from
the year 1969.
\subsection{Eulerian spontaneous stochasticity}\label{subsec_ess}
In the year 1969, E.N.Lorenz came with a surprising statement that we quote:
"{\it It is proposed that certain formally deterministic fluid systems which possess many
scales of motion are observationally indistinguishable from indeterministic systems;
specifically, that two states of the system differing initially by a small “observational
error” will evolve into two states differing as greatly as randomly chosen states of the
system within a finite time interval, which cannot be lengthened by reducing the
amplitude of the initial error}" \cite{Lorenz69}. Reading twice, this has exactly the same flavor that
Richardson particle pair separation, but it is now the full (Navier-Stokes) Eulerian field which is involved! The existence of such a {\it finite-time} predictability barrier is of course deeply connected to 3-D turbulence itself, in particular it requires the typical Kolmogorov $k^{-5/3}$ spectrum.
This work has been barely noticed for decades, probably shadowed by the famous 1963 paper \cite{Lorenz63} on unpredictability (sensitive dependence to initial conditions) of low-order chaos.
As explained in \cite{Rotunno_Snyder08,Palmer14}, in the 70s, people also believed at that time that the atmospheric spectrum was essentially 2-D with a $k^{-3}$ spectrum and a predictability barrier more consistent with classical chaos. Nowadays, the situation has changed dramatically: computers routinely
solve atmospheric scales much below 100km where a clear $k^{-5/3}$ spectrum emerges. The intuition
of Lorenz becomes highly relevant. 
A name has been given to this sensitivity in 3-D multiscale flows: Palmer \cite{Palmer14} called it the real butterfly effect\footnote{Which, incidentally, was originally a seagull.}. We refer to it as Eulerian spontaneous stochasticity.
Interestingly, the term Eulerian spontaneous stochasticity was not coined until much later -- first introduced by G. Eyink at the APS 2016 conference and only formally discussed in \cite{Eyink_Bandak20}.

When this phenomenon occurs, some Lyapunov exponents -- possibly finite-time ones -- are expected to diverge as the Reynolds number increases. This behaviour contrasts with classical chaos, where Lyapunov exponents may be positive but remain bounded.

For all these reasons, people started to take a closer look at Lorenz's idea.
The first immediate systems were the 2-D surface quasi-geostrophic equations 
which possesses some direct $k^{-5/3}$ cascade \cite{Rotunno_Snyder08},\cite{Palmer14}
and more recently \cite{Nicolasetal24}, as well as the Kelvin-Helmholtz instability in 2-D Navier-Stokes
\cite{Simon_Jeremie_AM20} and Rayleigh-Taylor 3-D turbulence \cite{biferale2018rayleigh} (see also \cite{crisanti1993intermittency} for shell models). Indeed, these results confirm the intuition of Lorenz to various extents. A more detailed review of known results can be found in \cite{Eyink_Bandak20}.

Until now, we do not have clearly stated what this concept is. In general, ideal fluids or
wave dynamics are ill-posed. Typical examples are the inviscid Burgers equations and the incompressible Euler equations. It is of common practice to regularise such systems by adding other terms like (hyper) viscosity and/or small noise terms, e.g.  Navier-Stokes
or viscous Burgers equations. The main question is then: how such systems behave in the limit of vanishing regularisation? This is commonly referred to as the {\it inviscid limit}. Whenever a system becomes random in the inviscid limit, we say it is spontaneously stochastic. A difficult and important question is then to understand how the statistics, in the inviscid limit, depend on the way the system has been regularised.
One of the hope in the case of Euler/Navier-Stokes is that for a wide class of regularisations, those statistics are independent of the set of regularisations considered, therefore exhibiting a form of {\it universality}. 
These questions have been tackled thoroughly by A.Mailybaev and colleagues for the last 10 years 
using the concept of {\it renormalisation group} (RG) \cite{Maily2012,Maily2016,AM_Ra23,AM_Rb23,AM_24} (see also \cite{Eyink_Bandak20}). The RG approach is the one akin to the dynamical system theory approach for the universality of the Feigenbaum period-doubling cascade \cite{Feigenbaum1976}.
These studies mainly focus on simpler shell models of turbulence, like Sabra or dyadic systems which possess many scales (of motion). 
Physicists and mathematicians widely recognise these shell models for sharing many desirable properties with the Euler, Navier-Stokes, or Burgers equations such as power laws, multiscaling \cite{Gilson1998,Mailybaev2012}, making them much easier alternatives given the considerable difficulty of directly tackling the formidable 3-D incompressible Euler equations. The reason to consider a RG approach is that ideal fluids have many symmetries, associated with spatio-temporal self-similar (fractal) structures. Therefore, it is possible to define a particular self-consistent notion of (canonical) regularisations \cite{AM_24}. The RG is then interpreted as a genuine dynamical system in the space of flow maps, where the successive group iterates reflect the level of regularisation of the inviscid system. The lesson  is that, provided there exists some RG attractor (it can be a fixed point), one can infer some universal statistical properties
of the inviscid limit provided a given set of regularisations belong to the basin of attraction. We would like to insist on another misunderstanding that one must not look at
a fixed regularisation, say Navier-Stokes at fixed Reynolds number $Re$ but rather at the inviscid limit $Re \to \infty$. However, signatures of spontaneous stochasticity are visible well before
the limit is actually taken as pointed out in \cite{Eyink_Bandak20}.

In order to exhibit spontaneous stochasticity, the inviscid system must necessarily be ill-posed. Well-posedness is
existence, unicity and continuous dependence on the initial conditions.
In this case, this is rather the nonuniqueness of the inviscid system which is responsible
for the ill-posedness and as a consequence loss of continuity w.r.t. initial conditions
\footnote{It is easy to find infinite-dimensional systems that satisfy existence and uniqueness but are not well-posed.}. To fix the idea, let us write an inviscid system, say Euler equations as 
$\dot x = f(x), x(0)=x_0$. For smooth initial data, it is possible
to establish some (local) well-posedness until some time $T^\star > 0$. One of the most important
open question in mathematical fluid mechanics is whether $T^\star < +\infty$ or not:
{\it does three-dimensional incompressible Euler flow with smooth initial conditions develop a singularity with infinite vorticity after a finite time?} \cite{FrischBec1}. When
such thing happens, the solution blows up in finite-time: nonuniqueness emerges as a pure manifestation
of the solution hitting a (non-Lipschitz) singularity in finite-time. It turns out
that this is precisely what is observed in simpler models of Euler equations, e.g. in log-lattice Sabra
\cite{Ciro} or for Rayleigh-Taylor instability \cite{mailybaev2017toward}. The deep connection between spontaneous stochasticity and the presence of singularities in the inviscid system has been rigorously investigated by \cite{Drivas21} and
\cite{Drivas24} in simpler finite-dimensional dynamical systems. This is also the approach adopted
in the first part of this work although in a more general framework, see Section \ref{sec_ESS}. 

Assuming the solution reaches some singularity in finite-time,
its post blow-up life must face nonuniqueness:  which solutions
of the inviscid system among the infinity of them is chosen in the inviscid limit?
This selection must a priori depend on the way the system is regularised.
This {\it selection principle} is one of the fundamental questions underlying the concept of spontaneous stochasticity. In this work, we convey the idea that one must not look for a "good/physical" regularisation but rather ask for the universality and/or robustness of families of regularisations.
 This is a difficult question, but again it is best addressed using  RG-like approaches.

\subsection{The convex integration revolution}\label{subsec_convex}
Until now, we have only described the physicist viewpoint. What do mathematicians say about this subject? 
Strictly speaking the notion of spontaneous stochasticity has not been addressed
in those terms. Rather, there has been a strong focus on the Millennium price regarding global well-posedness of
the 3D incompressible Navier-Stokes. In particular, central is the understanding of weak solutions
not only for Navier-Stokes, (non)uniqueness of  Leray weak solutions \cite{Leray}, but for the Euler 
weak solutions as well. For the last case, a great achievement of the recent years is the mathematical proof of the Onsager conjecture (see Onsager quote in Section \ref{subsec_lss}). 

The story begins with the Nash-Kuiper $C^1$ isometric embedding theorems for the torus \cite{Nash,Kuiper}. A detailed review can be found in \cite{hprinciple}. Roughly speaking, isometric embedding of the torus involves taking a flat piece of paper and bending it into the shape of a doughnut. However, anyone attempting this would fight against Gauss's Theorema Egregium, which states that Gaussian curvature must be preserved unless the paper is stretched. In simpler terms, this "rigidity" makes it impossible to transform a flat sheet into a sphere or a doughnut without bending --something most people have experienced once.

Crucially, such transformations typically require $C^2$ regularity. This led to an intriguing question posed to J. Nash: could it still be possible to achieve this using only $C^1$ transformations? Surprisingly, the answer was a resounding yes. Striking images can be found nowadays showing how the doughnut looks like \cite{Borelli}: it involves successive embeddings
yielding $C^1$ fractals. It was indeed the birth of convex integration.

What is the link with Euler equations? It took some time before people started to realise there was a strong analogy with Onsager's conjecture. This analogy is a suitable variant of the $h$-principle -- $h$ for homotopy -- found by Gromov \cite{Gromov}, and developed by DeLellis and Sz\'ekelyhidi 
\cite{hprinciple}. The Egregium Gauss theorem is now played by the conservation of energy for solutions
having enough regularity above the critical 1/3 H\"older exponent. This rigid part
(the analogy is the $C^2$ folding of a piece of paper) is the easiest to prove \cite{CET94}. 
Proving the nonuniqueness of dissipative weak solutions in 3-D Euler equations turns out to be much more 
difficult: this is the flexible part of the Onsager's conjecture. It took almost a decade to climb the ladder of weak dissipative solutions regularity from $C^0$ up to $C^{1/3}$ by improving the convex integration schemes until the
conjecture was finally proven by Isett \cite{Isett} and Buckmaster et al. \cite{Admi_Onsager} for the admissible case.

Convex integration schemes have recently attained some high level of sophistication (intermittent schemes) so that it is now possible to handle weak solutions of the 3-D Navier-Stokes as well in the groundbreaking work \cite{BuckVic}. In particular, the viscous Laplacian term is adding much difficulties for controlling the various error estimates necessary for scheme convergence.
While the Armstrong-Vicol transport system \cite{AV23} -- on which our work is based -- also shares common ground with convex integration schemes, it innovates by employing sophisticated techniques of fractal homogenisation.

Convex integration schemes can be interpreted as an  inviscid limit.
Rather than looking at the inviscid limit $Re \to \infty$ in Navier-Stokes where
the Laplacian acts as a regularisation mechanism, one is
considering a family $({\bf u}_k, p_k,\mathring R_k)$ of smooth solutions of the so-called Euler-Reynolds equations:
$\partial_t {\bf u}_k + {\bf u}_k \cdot \nabla {\bf u}_k + \nabla p_k = {\rm div}~
\mathring R_k$, where $\mathring R_k$ is a symmetric trace-free tensor. The regularisation mechanism is now "homogenisation" conveyed by the
Reynolds-stress term (in fact, an effective averaging of the small scales is induced by the integral operator ${\rm div}^{-1}$).
By properly renormalising the solutions, it is possible to construct a 
convergent sequence in some H\"older space $C^\alpha$, $\alpha < \frac13$:
\\
\resizebox{\columnwidth}{!}{
\begin{tikzpicture}
  \node (A0) at (0,0) {$({\bf v}_0,p_0,\mathring R_0)$};
  \node (A1) at (3,0) {$({\bf v}_1,p_1,\mathring R_1)$};
  \node (Ak) at (4.5,0) {$\cdots$};
 \node (Ak1) at (6,0) {$({\bf v}_k,p_k,\mathring R_k)$};
  \node (Ak2) at (9.5,0) {$({\bf v}_{k+1},p_{k+1},\mathring R_{k+1})$};
  \node (A) at (12.5,0) {~~$ \cdots ~~({\bf v},p,\mathring R)$};
  \draw[->, bend left] (A0) to node[above] {${\cal R}_0$} (A1);
  \draw[->, bend left] (Ak1) to node[above] {${\cal R}_{k}$} (Ak2);
\end{tikzpicture},}
where ${\rm div}~\mathring R$ is zero weakly.
The renormalisation operators ${\cal R}_k$ take a complicate form, where both time and space are rescaled. They also involve Nash decompositions and the use of inverse flow maps, see e.g. \cite{ReviewConvex}. Their expressions are typically of the form (\ref{RenormR}) in
Section \ref{sechom}. Obtaining convergence of the Euler-Reynolds sequence $({\bf v}_k, p_k, \mathring{R}_k)$ in a H\"older space $C^\alpha$ is a difficult task.
The closer $\alpha$ is to 1/3 the harder it is to control the various error terms.

The sequence of Euler-Reynolds equations above 
can be interpreted as some controlled regularisations of the inviscid Euler equations. To this respect, both RG approach and convex integration schemes share a common conceptual basis. The approach in \cite{AM_24end}, focusing on the Sabra shell model instead of a PDE, yields simpler renormalisation operators in the autonomous form ${\cal R}_k = 
\underbrace{{\cal R} \circ \cdots \circ {\cal R}}_{\tiny k~\mbox{times}}$. There are few drawbacks in convex integration schemes however.
First, they do not handle the Cauchy initial value problem (see however \cite{Mengual}), second the scale separation involved
is always hypergeometric and seems mandatory in order to control the many estimates.

\subsection{Nonuniqueness and the inviscid limit of Navier-Stokes}
As explained before, the necessary ingredient for having Eulerian spontaneous stochasticity is to have nonunique solutions of the initial value Cauchy problem in the inviscid system. It is thus important to mention first very interesting tools and concepts which have been developed by the mathematicians to handle the lack of a known global uniqueness result for the 3-D Navier-Stokes and/or to formalise the notion of ensemble average in turbulence.
One is the notion of {\it statistical solution} introduced by Foias and Vishik \cite{Foias1972,Foias1976,Vishik1977}, see \cite{Foias2013}.
Other notions of multivalued semigroups have been developed by Sell \cite{Sell1973} and
in particular the notion of {\it generalised semiflows} by Ball \cite{Ball1997}, see also
\cite{Simsen2008,James2003}. All these concepts have a strong Eulerian flavor since they consider a phase space, dynamical system point of view.

More Lagrangian-oriented mathematical results have also been obtained, starting 
from the important DiPerna-Lions theory \cite{diperna1989ordinary},
and the extension by L.Ambrosio \cite{ambrosio2004transport}, as well as the introduction of {\it superposition solutions} by \cite{Flandoli2009}. They give
deeper insights on the well-posedness of (nonautonomous) ODEs and their
natural link with linear transport equations and conservation laws. It is
discussed in Section \ref{subEuler_Lag}. 

With respect to the inviscid limit in Navier-Stokes, the work of DiPerna and Majda \cite{DiPerna1987} explicitly addresses the inviscid limit w.r.t. to viscosity for the deterministic 
3-D Navier-Stokes. They introduce
the notion of {\it measure-valued solutions} based on Young measures.
Their main theorem is a convergence result in a rather weak sense that any weak solution
converges in subsequence to a measure-valued solution. 
Unfortunately, this result does not rule out the possibility that this measure-valued solution collapses into a Dirac mass. Only strong numerical evidence suggests that the emergence of oscillations and concentrations of the solution in the limit would disrupt any Dirac-like measure, leading to a nontrivial one. The approach of 
\cite{Brenier1989} is innovative in that it gives a probabilistic view 
of the selected
solutions in the inviscid 3-D Navier-Stokes limit. The key idea is to exploit a variational formulation of the Euler equations, as a least action principle over volume-preserving flow maps in physical space. Unlike  previous studies, the framework is inherently Lagrangian and yields a natural notion of probability measure in the space of Lagrangian flow maps -- what is referred to as 
{\it generalised flows}. These ideas have been investigated thoroughly in \cite{Thalabard2020}.

Last, it is important to mention the work of S.Kuksin on the 2-D Navier-Stokes equations \cite{kuksin} forced by additive noise.
The noise must scale like the square root of viscosity in order to obtain physically meaningful solutions in the inviscid limit.
Such inviscid limit is called the {\it Eulerian limit}.
The striking result is the emergence of a selection mechanism toward a genuinely random process $U^\omega(\cdot)$ solving the 2-D Euler equations pathwise, with energy and enstrophy
becoming random variables. If $u_\nu^\omega$ is the random Navier-Stokes field, then a subsequential limit is obtained of the form
$\lim_{\nu_j \to 0} \lim_{T \to \infty} u_\nu^\omega(T+\cdot) =
U^\omega(\cdot)$ with the limits taken in that order.
While this resembles spontaneous stochasticity, it is not quite the same. Fixing the initial condition leads to deterministic dynamics. 
Only in the long-time limit does the system lose its memory, and what is left is a stochastic outcome shaped by the Casimir invariants revealing a deep link to the statistical mechanics of the 2-D Euler flow \cite{RSM}.

\section{Anomalous diffusion in the Armstrong-Vicol passive scalar}\label{sec_AV}
We introduce here the passive scalar model studied in \cite{AV23,AV24}. 
All the quantities are precisely defined in Appendix \ref{AppdefI}.
\\\\
We consider the following Cauchy problem for the advection-diffusion 
equation:
\be \label{ad_dif0}
({\cal P}_\kappa): \left\{ 
\begin{array}{l}
\partial_t \theta^\kappa + {\bf b} \cdot \nabla \theta^\kappa = \kappa \Delta
\theta^\kappa ~~{\rm in}~ (0,\infty) \times \mathbb{T}^2 \\\\
\theta^\kappa(0,\cdot)  = \theta_0~~{\rm on}~~\mathbb{T}^2
\end{array}\right..
\de  
The energy equality for a divergence-free vector field ${\bf b}$ writes as
\be \label{energyeq}
||\theta_0||^2_{L^2} - ||\theta^\kappa(1,\cdot)||^2_{L^2} = 2 \kappa ||\nabla \theta^\kappa||_{L^2((0,1)\times \mathbb{T}^2)}^2.
\de 
In particular, if the velocity field ${\bf b}$ has enough spatial regularity, say 
in $L_t^1 C_x^{0,1}$, then $\lim_{\kappa \to 0} \kappa ||\nabla
\theta^\kappa||_{L^2}^2 = 0$, implying that the energy is conserved. 
The main question is to understand the limit $\kappa \to 0$ of the passive
scalar $\theta^\kappa$ when the velocity field lacks regularity, typically being only H\"older continuous.
For the sake of consistency, we explicitly recall the main Theorem 1.1 of \cite{AV24}:
\begin{theoremav}[Anomalous dissipation of scalar variance] Let
$d \geq 2$ and $\alpha \in (0,1/3)$. There exists
\be \label{bdef}
{\bf b} \in C_t^0 C_x^{0,\alpha}([0,1] \times
\mathbb{T}^d) \cap C_t^{0,\alpha} C_x^0([0,1]\times \mathbb{T}^d)
\de  
which satisfies $\nabla \cdot {\bf b}(t,\cdot) = 0$, $\forall t \in (0,\infty)$ such that,
for every mean-zero initial datum $\theta_0 \in H^1(\mathbb{T}^d)$, the family of unique solutions $\{\theta^\kappa\}_{\kappa > 0} \in C([0,1];L^2(\mathbb{T}^d))$ of (\ref{ad_dif0}) satisfy
\be \label{anomal0}
\limsup_{\kappa \to 0} 
\kappa ||\nabla  \theta^\kappa||^2_{L^2((0,1)\times \mathbb{T}^d)}
\geq \varrho^2 ||\theta_0||^2_{L^2(\mathbb{T}^d)},
\de 
for some constant $\varrho = \varrho(d,\theta_0) \in (0,1]$ which depends only on 
$d$ and the ratio $||\theta_0||_{L^2(\mathbb{T}^d)}/||\nabla \theta_0||^2_{L^2(\mathbb{T}^d)}$.
\end{theoremav}
The loss of regularity in the velocity field ${\bf b}$ leads to a breakdown of energy conservation in the limit -- manifested as an effective anomalous dissipation term (\ref{anomal0}) in the transport equation.
Although the proof is quite involved, it yields an explicit and 'controlled' construction of the transport PDE, typical of convex integration approaches although it differs in many aspects (see Subsection \ref{subsec_convex}), making it especially valuable. 
The strength of the derived estimates allows for additional important results, discussed in Part II.
\\\\
The main idea is to show that there exists 
a well-posed advection-diffusion PDE close to (\ref{ad_dif0}) in a well-defined sense.
This {\it effective} system is defined in the following way:
\be \label{tm}
\left\{\begin{array}{l}
\partial_t \theta_m + {\bf b}_m \cdot \nabla \theta_m = \kappa_m \Delta \theta_m~~{\rm in}~~
(0,\infty) \times \mathbb{T}^2, \\\\
\theta_m(0,\cdot) = \theta_0
\end{array}\right.,
\de
where $m \in \{m^\star,\cdots,M\}$ and $\kappa_m$
are \emph{renormalised diffusivities} satisfying
\begin{equation}\label{avseq}
\kappa_{m-1}  =  \kappa_m + c_0 \frac{\epsilon_m^{2\beta}}{\kappa_m},~~
\kappa_M  =  \kappa \in \bigcup_{m \geq 1} \left[\frac12,2\right]\epsilon_m^{\frac{2 \beta}{q+1}},
\end{equation}
and
\be 
\epsilon_m^{-1} = \big \lceil \Lambda^{\frac{q^m}{q-1}} \big \rceil ,~q = \frac{\beta}{4(\beta-1)},~\beta = \alpha + 1.
\de 
System (\ref{tm}) can  be seen as a {\it regularised}  version of (\ref{ad_dif0}) and
describes the effective behaviour of the passive scalar at scales larger than $\epsilon_m$.
As explained in the previous section, the key mechanism is that the scale $\epsilon_m$
can homogenise at scale $\epsilon_{m-1}$ provided well-defined constraints are met.
In order to achieve this, a multiscale fractal vector field ${\bf b}$ must 
be defined in an ad-hoc recursive way 
\footnote{this recursive construction is characteristic
of all convex integration schemes.}, namely
$
{\bf b}_m = \sum_{k = 0}^m {\bf v}_k,
$
with ${\bf b}_0 = 0$ and where ${\bf v}_k \in C^\infty$ consists of many oscillatory components at scale $\epsilon_k$
(see previous section). It is shown in \cite{AV23}
that ${\bf b}:=\lim_{m \to \infty} {\bf b}_m$ is well-defined and  satisfies (\ref{bdef}).
\subsection{Homogenisation of a simple shear flow and constraints on the scales}\label{sechom}
The way the velocity field ${\bf b}$ is constructed is very involved and consists of space-time fractal structures with periodic shear flows alternating in time.
It is a multiscale flow with infinitely many scales separated hyper-geometrically.
Our aim is to explain in simple terms the homogenisation mechanism of the velocity field
constructed by \cite{AV24}. For this, it is useful to adopt a Lagrangian viewpoint
and to express the dynamics of particles transported by a given shear flow. In addition, 
it gives for free the various scaling constraints for the velocity field in \cite{AV24}.
The diffusion process with diffusivity $\kappa$ yields the following SDE:
\begin{align}
&dX_t= 2\pi a_m \epsilon_m \cos\left( \dfrac{2\pi}{\epsilon_m} Y_t\right) +\sqrt{2\kappa} \, d W^1_t,~~
dY_t= \sqrt{2\kappa} \, d W^2_t, \nonumber \\ &(X_0,Y_0) = (0,0).
\end{align}
The amplitude $a_m$ corresponds to the size of the velocity Lipschitz norm. In fact, 
in order for the velocity to be H\"older with exponent $\alpha$, one must take
$a_m = \epsilon_m^{\alpha-1}$. The other quantities are the 
two independent Wiener processes $W_t^{1,2}$, and $\epsilon_m$ is the typical length over which the shear flow varies. Some standard
calculations, further simplified by our choice of initial condition, give
\begin{align} \label{eq:effdiff_shear}
\begin{array}{llll}
\mathbb{E}X_t^2  & =  2\left(\kappa+\dfrac{1}{2}\dfrac{a_m^2 \epsilon_m^4}{\kappa} \right) t + \left(   \dfrac{a_m \epsilon_m^3}{2\pi\kappa} \right)^2 \left( e^{- \omega_m t} -1 \right) \\
& \left. +\dfrac{1}{12}\left(e^{-4 \omega_m t}-1\right) \right),  ~~~ \omega_m := \frac{4\pi^2 \kappa}{\epsilon_m^2},  \\
\mathbb{E}Y_t^2 &=  2\kappa t.
\end{array}
\end{align} 

\begin{figure}[htbp]
\centerline{\includegraphics[width=\columnwidth]{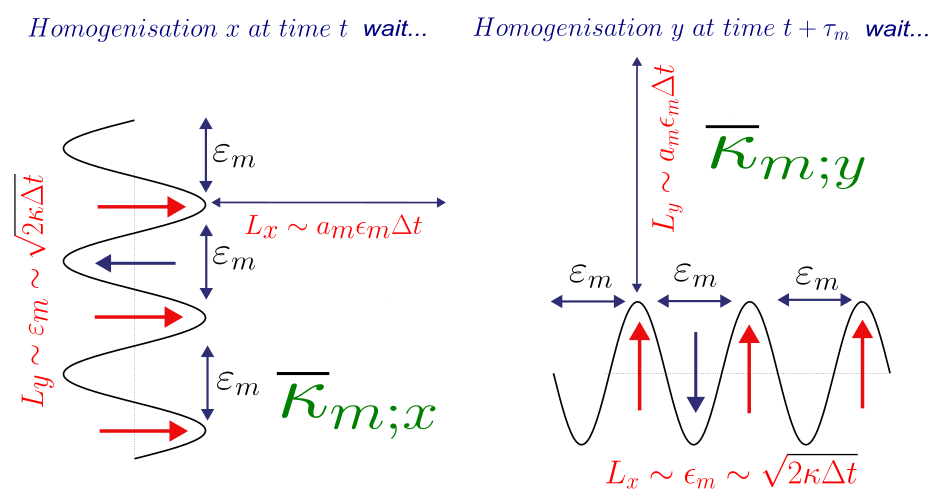}}
\caption{A sketch of homogenisation at scale $\epsilon_m$.
The time $\Delta t$ is the time needed to see the shear at scale $\epsilon_m$, namely 
to have a displacement along $y$ of order $\epsilon_m$
giving $\epsilon_m \sim \sqrt{2 \kappa \Delta t}$. 
It gives the effective diffusivity along $x$: $\overline{\kappa}_{m,x} = \frac{\langle L_x^2 \rangle}{\Delta t}
= \kappa + a_m^2 \epsilon_m^2 \Delta t = \kappa + \frac{a_m^2 \epsilon_m^4}{\kappa}$.
Then one rotates the field after waiting some time $\tau_m \gg \Delta t$,
that is waiting long enough for homogenisation to occur. Then one proceeds again in the
$y$ direction.
}\label{LagHom}
\end{figure}

By direct inspection, the particle displacement in the $x$ direction yields an effective diffusivity
$\kappa^\ast= \kappa +  \dfrac{a_m^2 \epsilon_m^4}{2\kappa}$  (to be compared with (\ref{avseq})).
Let us call $\tau_m$ the typical time for homogenisation to occur. It must be much larger than the relaxation time $\omega_m^{-1}$. The first constraint
is therefore $\tau_m \gg \frac{\epsilon_m^2}{\kappa}$. Let us call $\epsilon_{m-1}$
the typical particle displacement during this homogenisation process: it is
$\epsilon_{m-1}^2 \propto \left( \kappa +  \dfrac{a_m^2 \epsilon_m^4}{2\kappa} \right) \tau_m \gg \left( \kappa +  \dfrac{a_m^2 \epsilon_m^4}{2\kappa} \right) \omega_m^{-1}
= \frac{\epsilon^2_m}{4\pi^2} + \frac12 \left( \frac{a_m \epsilon_m^3}{2\pi \kappa}
\right)^2$ giving the second constraint: $\epsilon_{m-1} \gg \frac{a_m \epsilon_m^3}{ \kappa}$. Those constraints correspond exactly to the constraints used in \cite{AV24}.
This homogenisation process is illustrated in a simplified form in Fig.~\ref{LagHom}.

In summary, the process $(X_t,Y_t)$ can be interpreted, at large times, as an effective diffusive process with diffusivity $(\kappa^\ast,\kappa)$. 
From the expression of $\kappa^\ast$, it is evident that even in the limit of vanishing diffusivity, the effective diffusivity $\kappa^\ast$ 
remains strictly positive and can even diverge. However, this does not imply that the passive scalar undergoes anomalous diffusion in the sense of 
Theorem 1.1 in \cite{AV24}. 
Indeed, the effective diffusive description is only valid on time scales much larger
than $\tau_m$ and length scales much larger 
than $\epsilon_{m-1}$, preventing the observation of finite effective diffusivity on time scales of order unity in the vanishing $\kappa$ limit.

A last remark is that, in order for the homogenisation process to cascade to larger scales, much care is required. Take for instance a naive zonal superposition of shear flows: 
${\displaystyle {\bf b}_M = 
\sum_{m=0}^M a_m \epsilon_m \cos \left( \frac{2\pi}{\epsilon_m} y \right)
\left(\begin{array}{l} 1 \\ 0 \end{array} \right)}$. Using the previous calculation, one can show that 
the mean particle displacement is 
\begin{align}
\mathbb{E}X_t^2  = 2 {\cal K} t + \sum_{m =0}^M \sum_{n=0}^M \frac{a_m \epsilon_m^3}{2\pi \kappa} \frac{a_n \epsilon_n^3}{2\pi \kappa}\left( \dfrac{2+\left(\frac{\epsilon_m}{\epsilon_n}\right)^2+\left(\frac{\epsilon_n}{\epsilon_m}\right)^2 }{4\left(1 +\left(\frac{\epsilon_m}{\epsilon_n}\right)^2+\left(\frac{\epsilon_n}{\epsilon_m}\right)^2 \right)}- \dfrac{17}{12} \delta_{n,m} \right) \nonumber \\ + o_{t \rightarrow \infty}\left(e^{-at} \right) \nonumber 
\end{align}
with ${\displaystyle {\cal K} = \kappa + \frac{1}{2\kappa} \sum_{k=0}^M a_m^2 \epsilon_m^4}$ and $a>0$ depends on the choice of sequence $(\epsilon_n)_{0\leq n\leq M}$.

This computation demonstrates that superimposing shear flows at different length scales increases the effective diffusivity ${\cal K}$ beyond that of a single shear flow obtained earlier. However, for the same reasons as before, this superposition alone does not lead to anomalous diffusion. Furthermore, the expression for ${\cal K}$ reveals that the problem at scale $m$ is not simply homogenised into the problem at scale $m-1$; instead, all scales are homogenised simultaneously. As such, this straightforward superposition of shear flows does not enable anomalous diffusion. To address this issue, the authors of~\cite{AV24} introduced a crucial modification: they added the shear flow at scale $m$ in the reference frame of the one at scale $m-1$. This adjustment allows for anomalous diffusion, as demonstrated in \cite{AV24}.

Additionally, in the framework discussed earlier, the effective diffusive regime is achieved only in the $x$-direction. To extend this homogenisation to the $y$-direction, one must introduce an advection field that alternates in time. The time period of alternation between $x$- and $y$-directed shear flows must be sufficiently large to ensure that the effective diffusive regime is fully realised as depicted in Fig.~.\ref{LagHom}.

\subsection{Renormalisation group description} \label{RGd}
Let us call $\Si_m = (\epsilon_m,a_m,\tau_m,\tau_m',\tau_m'')$. The construction
in \cite{AV23} requires three different timescales $\tau_m \ll \tau_m' \ll \tau_m''
\ll a_{m-1}^{-1} \ll \tau_{m-1}$. One can consider three levels of
descriptions for the velocity field ${\bf b}$. The simplest one is to consider only $\tau_m$, namely
the timescale
on which the flow alternates (see previous discussion). The second level of description
is to introduce a second timescale $\tau_m''$ which is another homogenisation timescale
due to the change of the Lagrangian reference frame. The last level of description
introduce another intermediate timescale $\tau_m'$ for regularising the time cutoffs. It is needed to actually
prove the results in \cite{AV23} but is more technical in nature. We thus provide
a renormalisation group description involving $\tau_m$ and $\tau_m''$ only.
One can in principle write some more general formula at the expense of much heavier notations. 
\\\\
We introduce a two-timescale function related to the two timescales $\tau_m,\tau_m''$:
$$
\C(t_1,t_2) =  \left( \mathds{1}_{[k,k+1)}(t_1)  \mathds{1}_{[l,l+1)}(t_2) \right)_{(k,l) \in \mathbb{Z}^2}
$$
as well as the vector field to be rescaled:
\begin{align}
{\bf W}_0({\bf x}) = ({\bf w}_k({\bf x}))_{k \in \mathbb{Z}},~~ & \nonumber \\ {\rm with}~{\bf w}_k({\bf x}) = 
\left( \begin{array}{c}0\\ \sin 2\pi x_1 \end{array} \right) \mathds{1}_{k \in 2\mathbb{Z}}+ 
\left( \begin{array}{c} \sin 2\pi x_2 \\ 0 \end{array} \right) \mathds{1}_{k \in 2\mathbb{Z}+1} \nonumber
\end{align}
It consists of simple alternating shear flows, more complicated choices are possible.
Lastly, we write the Lagrangian flow map $s \mapsto {\bf X_v}(s;{\bf x},t)$:
$$
\frac{d{\bf X}_{\bf v}}{ds} = {\bf v}(s,{\bf X}_{\bf v}),~~{\bf X_v}(t;{\bf x},t) = {\bf x}.
$$
We use the compact vector notation
${\bf X}^{-1}_{\bf v}(t;{\bf x},{\bf s}) = \left({\bf X}^{-1}_{\bf v}(t;{\bf x},s_l)\right)_{l \in \mathbb{Z}}$ 
for some vector ${\bf s} = (s_l)_{l \in \mathbb{Z}}$ and ${\bf n} = (n)_{n \in \mathbb{Z}}$
One can thus express the renormalisation map where
$\sigma = (\epsilon,a,\tau,\tau'')$:
\be  \label{RenormR}
{\cal R}_{\Si}[{\bf v}]: {\bf v} \mapsto
{\bf v} + a \epsilon \C \left(\frac{t}{\tau},\frac{t}{\tau''} \right)
\cdot {\bf W}_0 \left( \frac{1}{\epsilon} {\bf X}^{-1}_{\bf v}
(t;{\bf x},\tau'' {\bf n}) \right).
\de
The vector field ${\bf b}_m$ at scales above $\epsilon_m$ using ${\bf b}_0 = {\bf 0}$ 
is therefore
\be 
{\bf b}_m = \left(
{\cal R}_m \circ {\cal R}_{m-1} \circ \cdots \circ {\cal R}_1 \right)[{\bf 0}].
\de 
One can also express the passive scalar $\theta_m$ through the
standard Feynman-Kac/backward Lagrangian averaging:
\be 
\theta_m^\kappa = {\cal F}_\kappa[{\bf b}_m] \theta_0,
\de 
where 
$({\cal F}_\kappa[{\bf v}] \theta_0)(t,{\bf x}) = 
\mathbb{E} \left[ \theta_0({\bf Y}^\kappa[{\bf v}](0|t,{\bf x})) \right]$ with
${\bf Y}_s \equiv {\bf Y}^\kappa[{\bf v}] (s|t,{\bf x})$ satisfying the SDE
$$
d{\bf Y}_s = {\bf v}(t,{\bf Y}_s) ds + \sqrt{2 \kappa} d{\bf W}_s,~~{\bf Y}_t = {\bf x}.
$$
The main focus is therefore to investigate the double limit
$$
\lim_{\kappa \to0,M \to \infty} {\cal F}_\kappa [{\cal R}_M \circ \cdots \circ {\cal R}_1[{\bf 0}]] \theta_0.
$$
The index $M \to \infty$ is a measure of the system regularisation (e.g., similar to a Galerkin truncation), and $\kappa \to 0$ is the singular limit. This
is a double limit. However, it is possible to show (see \cite{AV23,AV24}) that it is equivalent to consider a single limit $\kappa \to 0$, where now
$M=M(\kappa)$ depends on $\kappa$. The parameter $M$ typically must scale like some $\log \log$ function of $\kappa$ (see Eq. (\ref{avseq})).

\section{Anomalous diffusion: from theory to numerics}
We present here the main numerical result on the anomalous diffusion.
We recall from \cite{Drivas17} that it is equivalent to have Lagrangian spontaneous stochasticity. To be more precise, one considers the backward It\^o SDE associated
with the passive scalar transport equation:
\be \label{eq:BackLag}
d{\bf X}_s^\kappa = {\bf b}({\bf X}_s^\kappa,s) ds + \sqrt{2 \kappa} d{\bf W}_s, {\bf X}_t^\kappa= {\bf x}, s \leq t.
\de
Here ${\bf X}_s^\kappa := {\bf X}^\kappa(s;{\bf x},t), s \leq t$ corresponds to the particle position ${\bf X}_s^\kappa$ at a previous time $s \leq t$ such that it goes through position ${\bf x}$ at time $t$. Let us denote $\Theta_s^\kappa := \theta({\bf X}_s^\kappa,s)$ where $\theta$ satisfies the
transport equation (\ref{ad_dif0}). It satisfies a martingale property: 
$\theta^\kappa({\bf x},t) = \Theta_t^\kappa = \mathbb{E}[\Theta_s^\kappa] = \mathbb{E}[\Theta_0^\kappa] = \mathbb{E}[\theta_0({\bf X}_0^\kappa)]$.
It turns out that the fluctuations obey
$$
\Theta_t^\kappa- \Theta_0^\kappa  = \sqrt{2\kappa} \int_0^t
d {\bf W}_s \cdot \nabla \Theta_s^\kappa.
$$
It gives after squaring:
$$
{\rm Var} \left[\theta_0({\bf X}_0^\kappa)\right] = 2\kappa \int_0^t \mathbb{E}\left[~|\nabla \Theta_s^\kappa|^2 \right].
$$
The fluctuation theorem of \cite{Drivas17} is the integrated version 
$\langle \cdot \rangle = \int_{\mathbb{T}^2} d{\bf x}$ over the physical domain (where one uses the divergence-free property of the velocity field):
\be
\frac12 \langle {\rm Var} \left[\theta_0({\bf X}_0^\kappa) \right]  \rangle = \kappa \int_0^t 
\langle |\nabla \theta^\kappa|^2 \rangle~ds = \kappa ||\nabla \theta^\kappa||^2_{L^2((0,t) \times \mathbb{T}^2}.
\de 
Breaking of the Lagrangian flow map determinism, i.e. Lagrangian spontaneous stochasticity is therefore perfectly equivalent to having anomalous diffusion!

\subsection{Numerical implementation of the Armstrong-Vicol model}
Numerically simulating the passive scalar presents significant challenges, primarily due to extreme scale separation in the velocity field $\vect_m= \nabla \times \phi_m$. A key bottleneck arises from the hypergeometric nature of the problem. For instance, with the parameters used in the proof of anomalous diffusion in~\cite{AV24}, where $\Lambda \geq 2^7$, numerical simulations become infeasible. Relaxing the condition on $\Lambda$ enables computation but still restricts the number of accurately represented scales to at most 3--4 with our choices of parameters detailed in Appendix~\ref{ALGO}.

\begin{figure*}[htbp]
\centerline{\includegraphics[width=2\columnwidth]{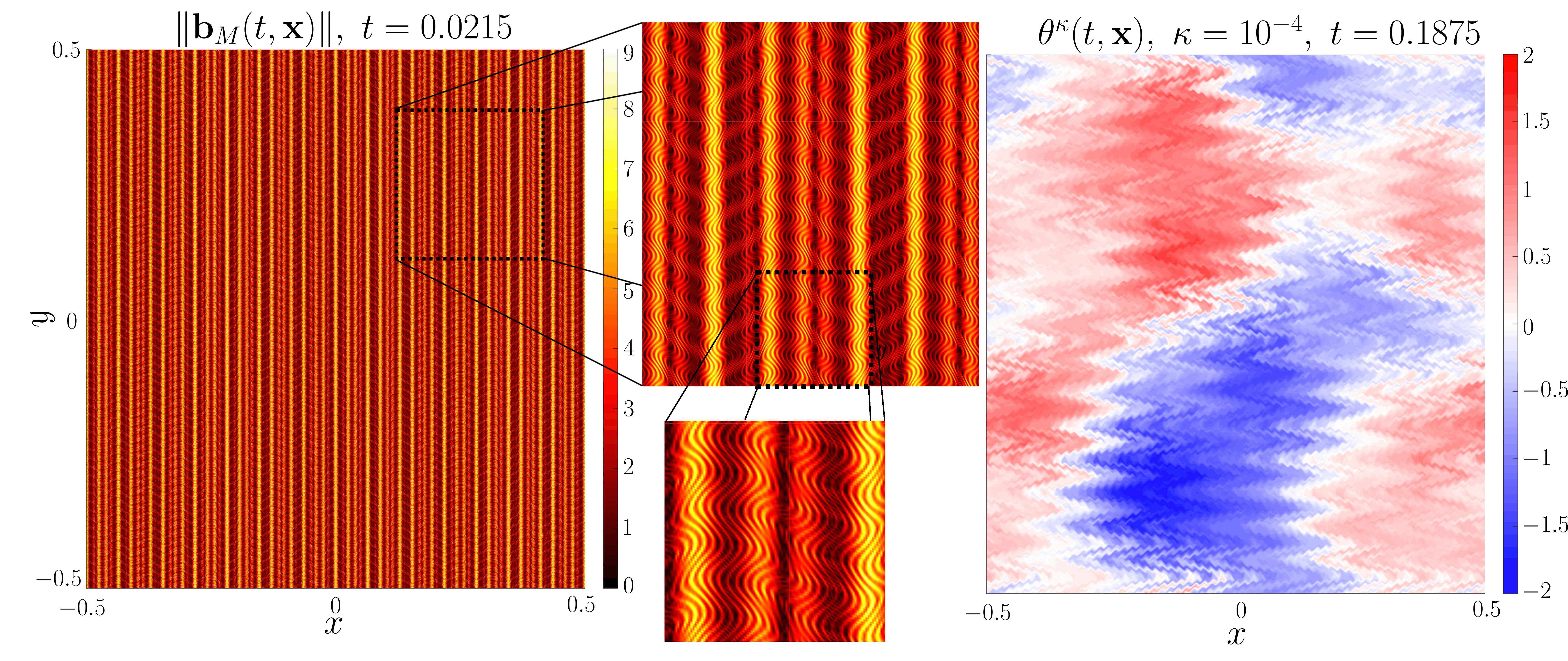}}
\caption{Left: Norm of the vector field at a given time and zoom on a given region to highlight the multiscale nature of the flow. Parameters used in the simulations are given in Appendix~\ref{ALGO}. Right: Solution of the passive scalar equation with advection $\vect_M$ {and initial condition a smooth random gaussian field detailed in Appendix~\ref{ALGO}}.} 
\label{fig:avfield}
\end{figure*} 

Another major difficulty arises from the nonlocal temporal dependencies in both the past and future, preventing a straightforward forward-in-time propagation as in traditional Cauchy initial value problems. Instead, the causality structure necessitates the use of a memory-based strategy to account for these dependencies. {The numerical implementation and parameter choices are given in Appendix.~\ref{ALGO}. In Fig.~\ref{fig:avfield}, we present typical snapshots of the multiscale vector field and passive scalar field obtained numerically. }

Interestingly, all convex integration schemes encounter similar numerical implementation challenges but are subject to additional constraints (e.g. Nash decompositions). We provide an algorithmic outline of our approach, with full self-contained details in Appendix \ref{ALGO}.

\begin{figure}[htbp]
\centerline{\includegraphics[width=\columnwidth]{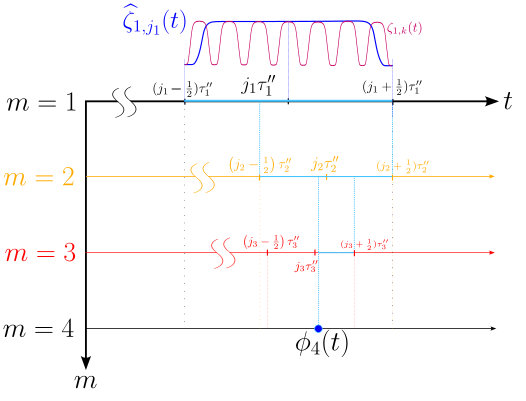}}
\caption{
Schematic representation of the non-local in time dependence of streamfunction $\phi_m$ with respect to $\phi_{p}, \, p< m$. In this example, one wish to compute the value of $\phi_4(t,\pos)$ at a given time. At each sub levels $m$, the light blue line represent the interval of time on which one need to know $\phi_{m<4}$  in order to compute $\phi_4(t)$. It is clear from this picture that the value of $\phi_4(t)$ depends on both posterior and ulterior values of $\phi_{m<4}$, making its computation difficult.}
\label{fig:RecurPhi}
\end{figure} 

\subsection{Lagrangian spontaneous stochasticity: Numerical results}
We aim to illustrate the connection between the anomalous diffusion of the passive scalar and Lagrangian spontaneous stochasticity, as discussed in the previous section and studied in detail in~\cite{Drivas17}. The numerical methods used to compute the Armstrong-Vicol vector field, integrate the passive scalar equation~\eqref{ad_dif0}, and solve the backward Lagrangian dynamics~\eqref{eq:BackLag} are described in Appendix~\ref{ALGO}. This appendix also provides details on the simulation parameters, relevant notations, and a pseudo-code Algorithm~\ref{alg:cap} designed to facilitate the reproduction of our results and support further investigations.

\begin{figure*}[htbp]
\centerline{\includegraphics[width=\columnwidth]{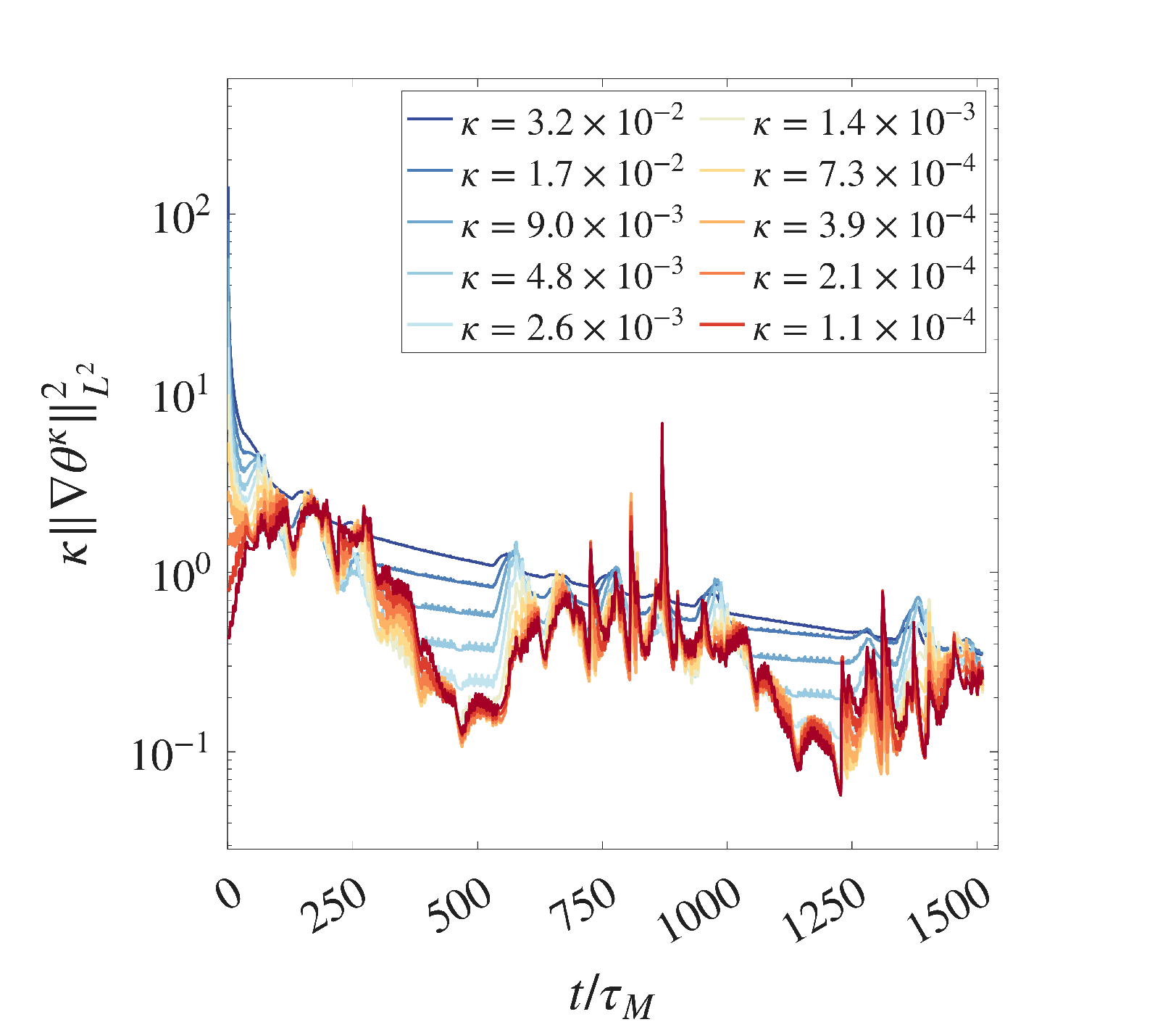} 
\includegraphics[width=\columnwidth]{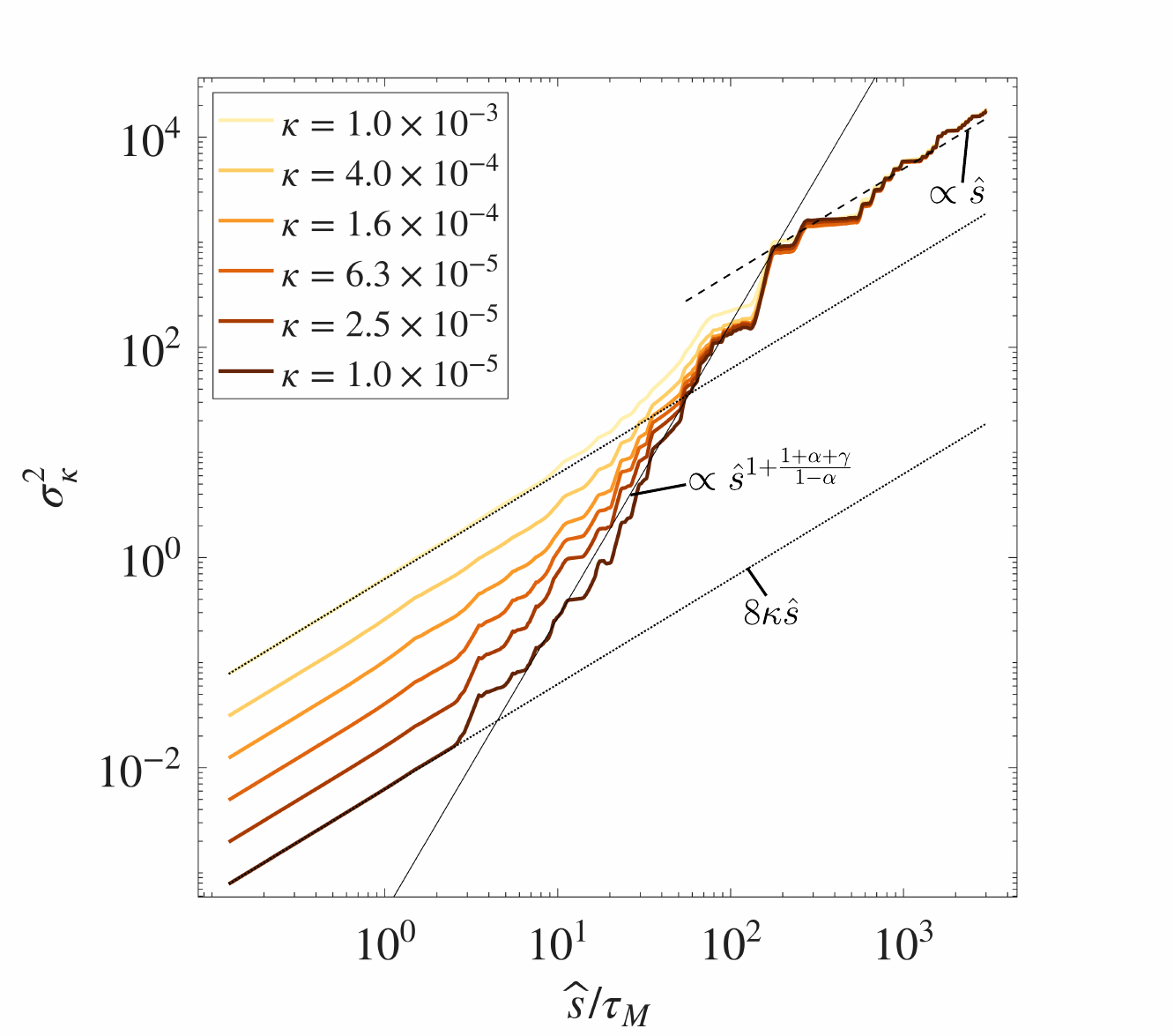}}
\caption{(Left) Diffusion rate of the passive scalar field over time for different values of $\kappa$. Since the initial condition is independent of $\kappa$, the initial dissipation rate is proportional to $\kappa$. However, after a transient regime of order $10^{2}\tau_M\simeq2\tau_1$, the dissipation rate becomes independent of diffusivity and remains strictly positive, it is anomalous diffusion. (Right) Variance of the backward Lagrangian dynamics \eqref{eq:BackLag} as a function of $\hat{s}=t_{f}-s$ for different values of $\kappa$. 
Each colored line represents the variance of ${\bf X}^{ \kappa}_{s}$ as a function 
of $\hat{s}/\tau_M$. The black dotted lines correspond to the expected diffusive scaling 
$  8 \kappa \hat{s} $ {for small $\hat{s}$}, while the solid black line represents the Richardson super-diffusive regime
$\sigma_\kappa^2 \propto \hat{s}^{1 +\frac{1+\alpha +\gamma}{1 -\alpha }}$ predicted in~\cite{AV24}. Finally, the dotted black line in the uppermost part of the figure is a linear scaling $\sigma_\kappa^2 \propto \hat{s}$ corresponding to an effective diffusive regime at large times.}
\label{Anom1}
\end{figure*}

In Fig.~\ref{Anom1}, the left panel displays the diffusion rate $ \kappa \| \nabla \theta^\kappa\|^2_{L^2(\mathbb{T}^2)}$ as a function of time for different values of $\kappa$. Since the initial condition of the passive scalar equation is independent of the diffusivity, the initial diffusion rate is proportional to $\kappa$. However, after a transient regime of order $ \tau''_M$, the dissipation rate reaches a finite value that is independent of diffusivity, namely anomalous diffusion.

On the right panel of Fig.~\ref{Anom1}, we present the behaviour of the variance of {backward} Lagrangian trajectories passing through $(0,0)$ at time $t_f$:
$$ 
\sigma^2_\kappa(s):= \mathbb{E}^{(1,2)} \left[  \left.\left| {\bf X}^{\kappa,(1)}_s  -{\bf X}^{\kappa,(2)}_s    \right|^2 \right|{\bf X}^{\kappa,(1,2)}_{t_{f}}=0   \right], ~{0\leq s \leq t_f}
$$ 
as a function of $\hat{s}/\tau_M:= (t_{f}-s)/\tau_M$ for different values of $\kappa$, where $t_{f}=0.5$ is the final time of the simulation. For small values of $\hat{s}/\tau_M$, the variance follows a purely diffusive
scaling, $\sigma^2_\kappa = 8 \kappa \hat{s}$, and is therefore strongly dependent on 
$\kappa$. However, after a few small-scale turnover times 
$\tau_M$, $\sigma^2_\kappa$ enters into a superdiffusive regime, corresponding to the Richardson regime. 
As noted in \cite{AV24} (though not proven there), the expected Richardson regime depends on the regularity of the advecting vector field. Specifically, it is expected that $\sigma^2_\kappa \propto \hat{s}^{1+ \frac{1+\alpha +\gamma}{1-\alpha}}$, where $\gamma = \frac{(1+\alpha)(1 - 3\alpha)}{5\alpha + 1}$. The exponent of this power law equals $3$ when $\alpha=1/3$, which corresponds to the classical Richardson regime {of} three dimensional turbulence. The solid black line in the right panel of Fig.~\ref{Anom1} represents this superdiffusive scaling (up to an arbitrary multiplicative constant) and show good agreement with our numerical simulations.  Most importantly, during this superdiffusive regime, the variance of Lagrangian trajectories $\sigma^2_\kappa$ becomes independent of $\kappa$. This means that the Lagrangian trajectories are indeed spontaneously stochastic. This superdiffusive regime last until $\hat{s}/\tau_M \simeq 10^2$, which is also the duration of the transient regime for the dissipation rate on the right panel of Fig.~\ref{Anom1}.  
For large values of $\hat{s}/ \tau_M \geq 10^2\simeq 2 \tau_1/ \tau_M$, $\sigma^2_\kappa$ once again exhibits diffusive behaviour. Indeed in Fig.~\ref{Anom1}, for large $\hat{s}$, $\sigma_\kappa^\kappa$ is growing linearly in $\hat{s}$. The black dotted line in Fig.~\ref{Anom1} corresponds to an effective diffusivity on the order of $8 \times 10^{-3}$ and appears to be independent of $\kappa$ within the range shown.
As seen in the figure, the effective diffusivity remains finite in the vanishing diffusivity limit, which is consistent with a finite diffusion rate as $\kappa \to 0$. Exploring this effectively diffusive regime is crucial for comparing it to the renormalised diffusivity sequence of~\cite{AV24}. Unfortunately, due to limited spatial resolution, reliably estimating an effective diffusivity is highly challenging. At this stage, we cannot provide a convincing numerical measurement of these diffusivities.

In summary, our numerical results for the Armstrong-Vicol model are in agreement  with the prediction made in~\cite{AV24} regarding the anomalous diffusion of the passive scalar.
Additionally, by investigating the Lagrangian counterpart of the passive scalar, 
we have provided further insight into the spontaneous stochasticity of the Lagrangian tracers. We also provide an external link to various movies, including the advection of Wandrille's cat in action:  \url{https://www.youtube.com/playlist?list=PLH-V8wbBnQjbTkHdJVxenflTYnN5gDC7B&jct=pcAkkDgSTfYZur1uhx9sCw}.

\clearpage
\part{}

\section{Spontaneous Stochasticity (\texorpdfstring{$\SP$}{Sp})}\label{sec_ESS}
This section aims to provide a more theoretical discussion on the notion of spontaneous stochasticity
with a principal focus on the phase space description that people often referred to as Eulerian
in the context of PDEs.
\subsection{Definitions}\label{secDef}
This section introduces a general mathematical definition of spontaneous stochasticity, applicable to both finite-dimensional (ODEs) and infinite-dimensional dynamical systems.
Given our focus on a phase-space dynamical framework, in the context of PDEs, it must be interpreted as a truly Eulerian notion, in contrast to the Lagrangian interpretation of spontaneous stochasticity.
In finite-dimensional systems, such a distinction, becomes irrelevant, allowing us to simply refer to "spontaneous stochasticity". Throughout the remainder of this work, we will primarily use the acronym $\SP$.\\

Consider some family of regularised systems called $({\cal P}_\kappa), \kappa > 0$, expressed in the following 
abstract autonomous form:
\be \label{generic}
({\cal P}_\kappa): ~\dot x = f(x,\kappa),~x(0) = x_0 \in H,
\de
where $H= \mathbb{R}^d$ is a finite-dimensional Hilbert space with $\langle x,y\rangle$
its scalar product and the corresponding norm $||x||^2 = \langle x,x \rangle$ and we assume $(x,\kappa) \mapsto f(x,\kappa)$ at least continuous.
In the following, we will often use the flow map notation:
$$
\Phi_t[f(\cdot,\kappa)] x_0 ~\mbox{solution of (${\cal P}_\kappa$) at time $t$ with initial condition}~~x_0.$$
Time $t$ is always assumed to be such that Peano existence theorem holds for all times in $[0,t]$.
The "ideal/inviscid" system is denoted $({\cal P}_0)$ and corresponds to $\kappa = 0$
associated with some vector field $f_0(\cdot)$ and $f(\cdot,\kappa)$ is
a regularisation of size $\kappa$, to fix the ideas, we make the following hypothesis
\be 
\makebox[0pt][l]{\tag{H0}}
\label{eq:H0}
\forall \kappa >0~({\cal P}_\kappa) \mbox{~is well-posed and}~ \lim_{\kappa \to 0} ||f(\cdot,\kappa)-f_0(\cdot)||_\infty = 0.
\de 
Here $||\cdot||_\infty$ is the sup norm in $x$.
We do not make any assumptions about the well-posedness of $({\cal P}_0)$ for $\kappa = 0$
nor about $f_0$, except that it must be continuous to ensure existence until time $t$.
Let ${\cal O}: H \to \mathbb{R}$ denotes some observable,
which is assumed continuous.
For some fixed $t$ and initial condition $x_0$, we now introduce the regularisation function:
\be \label{generic_F}
{\cal A}: \mathbb{R}^+ \to \mathbb{R}: \kappa \mapsto {\cal O}( \Phi_t[f(\cdot,\kappa)] x_0).
\de 
 Consequently from Hypothesis (\ref{eq:H0}), ${\cal A}$ is well-defined for $\kappa > 0$. 
In principle, one should write ${\cal A}_{t,x_0,{\cal O}}$. In order to avoid heavy notations we prefer to keep the light notation ${\cal A}$. The regularisation function ${\cal A}$ not only reflects how system (\ref{generic}) is observed through ${\cal O}$, but most importantly describe how the inviscid system $({\cal P}_0)$ is regularised
by the given family $(f(\cdot,\kappa))_{\kappa > 0}$.
The operator $U_t$ defined by $(U_t \mathcal{O})(x) = \mathcal{O}(\Phi_t[f(\cdot,\kappa)] x)$ is known as the Koopman operator and is widely used in ergodic theory \cite{Lasota1994}. However, we will not use this formalism, as our primary interest lies in the dependence on $\kappa$.

\begin{definition}[Spontaneous Stochasticity]\label{SPST}
We assume that hypothesis (\ref{eq:H0}) holds.
The system $({\cal P}_\kappa)$ is said to exhibit spontaneous stochasticity if there exists an observable ${\cal O}:H\to \mathbb{R}$ which is continuous, an initial condition $x_0$, and some time $t < +\infty$
such that the function (\ref{generic_F}) satisfies 
\begin{equation}\label{killeress}
 -\infty <  \liminf_{\kappa \to 0} {\cal A}(\kappa) < \limsup_{\kappa \to 0} {\cal A}(\kappa) < + \infty. \tag{\mbox{$\SP$}}
\end{equation}
\end{definition}\noindent

This definition of {\it spontaneous stochasticity} depends solely on the $\kappa \to 0$ asymptotic behaviour of the regularisation function ${\cal A}$.  
By an abuse of language, we will frequently say that 'this system is $\SP$.' This always implies that there exists a triplet $(t,x_0,{\cal O})$ such that the family of regularisations $({\cal P}_\kappa)$ satisfies Definition \ref{SPST} or that the regularisation function ${\cal A}$ satisfies (\ref{killeress}). The meaning will always be clear from the context.
We will also often write ${\cal A}^\pm$ for the two limsup/liminf (see also notations
(\ref{Anotations}) later on). 

Despite the simplicity of the formulation, several non-trivial consequences can be derived. 
\begin{itemize}
\item  Definition \ref{SPST} is equivalent to have finite-time separation of selected trajectories in phase space. It is also the same that having a lack of selection principle in
the inviscid limit. This is discussed in Subsection \ref{subsec1}
\item  We investigate how measures behave in the inviscid limit using ergodic dynamical system theory. Our approach, though conceptually similar to RG methods, is far more straightforward. Instead of depending on an observable, we extend the regularisation function ${\cal A}$ to $d$-dimensional curves in phase space. We then reformulate the inviscid limit in terms of weak convergence of pushforward probability measures. This perspective not only reveals when and how universality emerges but also confirms that Definition \ref{SPST} is indeed nonempty! A full discussion can be found in Subsection \ref{secmea}.
\item We also derive a necessary condition on $({\cal P}_\kappa)$ for having $\SP$. This allows us to discriminate regions in phase space where one can expect
nonuniqueness in the inviscid limit. This is discussed in Subsection \ref{which}
together with simple albeit nontrivial examples.
\end{itemize}

\subsection{Finite-time separation for arbitrarily close solutions and lack of selection principle}\label{subsec1}
\begin{figure}[htbp]
\centerline{\includegraphics[width=0.9\columnwidth]{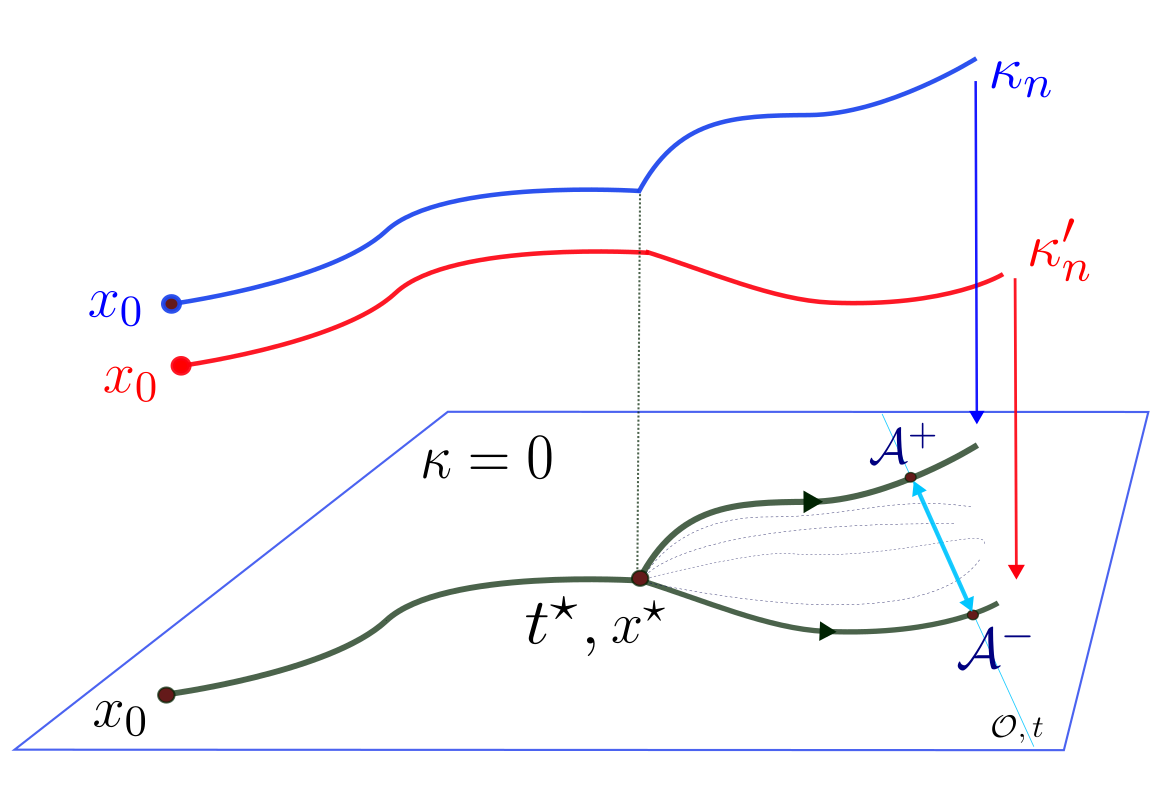}}
\caption{A sketch view of trajectory splitting in the singular limit $\kappa \to 0$.
One can construct a multi-valued flow which gather all possible flows associated with convergent $\kappa$ subsequences (at least two exist). These flows come 
from the well-posed regularised systems (\ref{generic}) in the limit $\kappa \to 0$. Here, there is a particular (finite) time $t^\star$ for which the system meets a non-Lipschitz singularity $x^\star$.
For $t > t^\star$, one has $-\infty < {\cal A}^- < {\cal A}^+ < +\infty$ for the observable ${\cal O}$ at time $t$ shown by a thin blue line.}
\label{split}
\end{figure}

We discuss here how in Definition \ref{SPST}, (\ref{killeress}) relates to two well-known concepts in the inviscid limit $\kappa \to 0$: the 
finite-time separation of trajectories (B) and the lack of selection principle (C).
\begin{property}(Equivalence)\label{equivdef}
Under hypothesis (\ref{eq:H0}), i.e., well-posedness of $({\cal P}_\kappa)_{\kappa > 0}$
and the existence of a triplet $(t,x_0,{\cal O})$ where ${\cal O}$ is continuous, one has 
$$
(\ref{killeress}) \Longleftrightarrow (B) \Longleftrightarrow (C).
$$
\end{property}
\noindent{\bf Proof}: see Appendix \ref{proof2flows} \qed
\\
The inviscid limit for $({\cal P}_\kappa)$ thus selects at least two distinct limsup/liminf trajectories starting from the same $x_0$ among infinitely many trajectories in the inviscid system, which differ at time $t$. Due to that, one has automatically unbounded finite-time Lyapunov exponents, meaning that the
quantity (\ref{fts}) diverges in the inviscid limit:
\begin{align} \label{fts}
\lim_{\kappa \to 0} ~\Biggl| \langle \nabla &{\cal O}(\Phi_t[f(\cdot,\kappa)] x_0),  \Biggr. \nonumber \\ &\left. {\cal T} {\rm exp} \left(
\Int_0^t \frac{\partial f(\cdot,\kappa)}{\partial x}(\Phi_s[f(\cdot,\kappa)] x_0)~ds \right)
v \rangle\right| = + \infty,
\end{align}
where ${\cal T}$ is the time-ordering operator.
We give a simple sketch of this scenario in Fig.~\ref{split}.

\subsection{A measure-theoretic description: selected measures and universality}\label{secmea}
Definition \ref{SPST} introduces the concept of the regularisation function $\kappa \mapsto {\cal A}(\kappa)$. A natural question is to understand how the values of 
${\cal A}(0)$ are distributed: given ${\cal A}$, can we define a well-posed probability 
measure associated with ${\cal A}(0)$? However, this question remains incomplete without specifying 
an ambient measure. We introduce the normalised Lebesgue measures: 
\be \label{defleb}
\lambda_\kappa(B) = \frac{1}{\kappa} {\rm Leb}([0,\kappa] \cap B),~\mbox{for all Borel sets $B \subset \mathbb{R}$}.
\de
To define a probability measure, we push forward $\lambda_\kappa$ by the mapping
$\kappa \mapsto {\cal A}(\kappa)$ and study the limit as $\kappa \to 0$. We denote these measures
by $\mu_\kappa$ in the following discussion.
The new family of probability measures $\{ \mu_\kappa \}$ satisfies, for all test functions $F \in C_b(\mathbb{R})$:
\be \label{mukdef} 
\mu_\kappa = {\cal A}_\star \lambda_\kappa~: 
\langle \mu_\kappa,F \rangle = \int_\mathbb{R} F(s) d\mu_\kappa(s) = \frac{1}{\kappa} \int_0^\kappa F({\cal A}(s))~ds.
\de 
This family of measures is tight, ensured by the boundedness of ${\cal A}$ from Definition \ref{SPST}. Consequently, as $\kappa \to 0$, Prokhorov's Theorem guarantees convergence at least along subsequences. However, uniqueness is not necessarily assured; in other words, the weak limit may not exist in general, except along subsequences.
Given this, one may ask: What happens if one uses a different ambient measure rather than the (normalised) Lebesgue one $\lambda_\kappa$ (\ref{defleb})? Can uniqueness be recovered?

Consider a simple example where the regularisation function exhibits rapid oscillations as $\kappa \to 0$, take for instance ${\cal A}(x) = \Re(e^{i \frac{1}{x}})$. In this case, Definition \ref{SPST} trivially holds  with ${\cal A}^\pm = \pm 1$.
Let us compute the weak limit $\mu_\kappa$ defined w.r.t. to the ambient measures  $(x^\alpha)_\star \lambda_\kappa, \alpha > 0$.
Applying the dominated convergence theorem and  the Riemann-Lebesgue lemma, one obtains
$$
\begin{array}{c}
{\displaystyle \langle {\cal A}_\star (x^\alpha)_\star \lambda_\kappa,F \rangle = \lim_{\kappa \to 0} \frac{1}{\kappa} \int_0^\kappa F(e^{i x^{-\alpha}})~dx =} \\
{\displaystyle \lim_{T \to \infty} \frac{1}{\alpha} \int_1^\infty F(e^{i T \theta}) \theta^{-\frac{1}{\alpha}-1} d\theta  = \frac{1}{2\pi} \int_0^{2\pi} F(e^{i \theta}) d\theta,\quad T = \kappa^{-\alpha}.}
\end{array}
$$
Thus the weak limit is the uniform Haar measure on $\mathbb{S}^1$ (its real part yields the arcsin law).
Notably, this limit is independent of $\alpha$, suggesting that the resulting probability measure may be largely insensitive to the choice of the ambient measure, or alternatively to the specific form of ${\cal A}$ (since $({\cal A} \circ h)_\star \lambda_\kappa = {\cal A}_\star (h_\star \lambda_\kappa)$), even asymptotically. 
It appears that this phenomenon of statistical robustness is far more general than this particular case and is explained in Subsection \ref{Bebuattak}.

Now consider a case with log-periodic behaviour, such as
${\cal A}(x) = e^{i \log x}$. A similar calculation yields using $F(z)=\sum_{k \geq 0} F_k z^k$:
\begin{align}
\frac{1}{\kappa} \int_0^\kappa F(e^{i \log x})~dx 
= \sum_{k \in \mathbb{Z}} \frac{F_k}{1+ik} e^{-i T k},  \nonumber \\  F_k= \frac{1}{2\pi} \int_0^{2\pi} F(e^{i\theta}) e^{-i k \theta}d\theta,  T = -\log \kappa.
\end{align}
Apart from the constant term $F_0= F(0)$, the other terms introduce fast oscillations and the limit $T \to \infty$ is undefined. If one pushes forward the measure $\left(e^{-\frac{1}{x}}\right)_\star \lambda_\kappa$ instead
of $\lambda_\kappa$ 
then one recovers the power-law situation above with the weak limit being
the Haar measure.
What does it means? If one observes oscillations at a rate that is too fast relative to their frequency, a sort of "stroboscopic effect" occurs, resulting in some 
partial information along a given subsequence. 
The whole statistic is somehow dispatched among the various subsequences. It is indeed the situation for the renormalised diffusivity sequence associated with the Armstrong-Vicol passive scalar: due to the hypergeometric separation, the usual $\kappa$ scaling is not the correct one, one numerically observes that the renormalised sequence has the expected $\log |\log \kappa |$ scaling; see Section \ref{num}, Fig.~\ref{fig:betaLBall}.
\\\\
These issues should not come as a surprise: "bad sampling gives bad statistics". Bad sampling is mathematically translated as a bad choice of an ambient measure when pushed forward by the regularisation function.
These observations lead us to question whether modifying how we sample the regularisation function always guarantees the existence of a weak limit. A closely related question is whether a stronger notion of $\SP$ should be adopted to ensure a {\it Measure Selection Principle}, meaning the probability measure supported by ${\cal A}(0)$ is a weak limit. Answering these questions appears highly challenging and possibly out of reach. However, we introduce a powerful formalism that not only addresses them but also establishes a link with recent RG approaches for probability kernels \cite{AM_Rb23,AM_Ra23}.

In particular, we will show that when $\SP$ holds in the sense of Definition \ref{SPST}, a Measure Selection Principle applies to a nearby regularisation, potentially a rescaled version of the original. The upcoming discussion is quite abstract and heavily influenced by the Bebutov flow and ergodic theory concepts. We consider the following result the most significant contribution of this paper, independently of the other parts addressing the Armstrong-Vicol model.
\subsubsection{Generalised formulation}
The aim here is to reformulate Definition \ref{SPST} by working directly
at the level of the flow in the space  $H=\mathbb{R}^d$ rather than looking at it through an observable ${\cal O}$. We still retain the finite-dimensional hypothesis. 
\begin{definition}[Regularisation space and Bebutov flow]\label{RegBebu}
Let $C_b=C_b(\mathbb{R}^+;\mathbb{R}^d)$ be the space of bounded continuous functions.
Let $(t,x_0), x_0 \in H=\mathbb{R}^d$ be fixed with $t < \infty$. The space of all regularisations of the inviscid problem $({\cal P}_0)$ with singular behaviour at $\kappa=0$ mapped to infinity is:
\begin{align} \label{Regdef}
{\cal R}_{\rm eg}:= \left\{ \gamma \in C_b~|~\exists f ~\mbox{satisfying (\ref{eq:H0}) such that}~ \right. \nonumber \\ \left. \gamma(\tau) = \Phi_t[f(\cdot,\frac{1}{\tau})] x_0,~\tau \geq 0
\right\}.
\end{align}
We equip ${\cal R}_{\rm eg}$ with the compact-open topology, i.e. the uniform convergence on compact sets.
We define the so-called Bebutov flow on the space ${\cal R}_{\rm eg}$ as:
\be 
{\cal S}_\tau: {\cal R}_{\rm eg} \to {\cal R}_{\rm eg}, \gamma \mapsto {\cal S}_\tau \gamma,~({\cal S}_\tau \gamma)(t) = \gamma(t+\tau).
\de 
and the associated projection:
\be \label{Projno}
\Psi: {\cal R}_{\rm eg} \to \mathbb{R}^d: \Psi(\gamma) = \gamma(0).
\de 
In particular, one has $\Psi({\cal S}_\tau \gamma) = ({\cal S}_\tau \gamma)(0) = \gamma(\tau)$.
\end{definition}\noindent
Note that $\gamma$ in Eq. (\ref{Regdef}) is extended by continuity to be defined for $\tau=0$, namely $f(\cdot,\infty)$ is implicitly assumed to be a well-defined (not interesting) limit. The following does not depend on this convention since one is looking at the (interesting) singular limit $\tau \to \infty$. Moreover, the mapping itself $\kappa \to 0$ to $\tau \to \infty$ is arbitrary, we just fix one. \\
{\bf Remark}: 
The Bebutov flow is a classical construction that represents any continuous function as the realisation of a dynamical system in an infinite-dimensional (topological) phase space (see Theorem 7.2.1 in \cite{Lasota1994}). It is often said that $\gamma$ is the trace of the (Bebutov) dynamical system. We call this interpretation {\it canonical} because it always holds no matter $\gamma$.
Note that the infinite-dimensional dynamical system ${\cal S}_\tau$ and the mapping $\Psi$ are well-defined and in particular continuous in the compact-open topology. However, ${\cal S}_\tau$ has an unusual feature as phase space and dynamical time $\tau$ are intrinsically linked. A strikingly similar situation arises in self-similar blowups in shell models \cite{Dombre1998, CiromeSimon}, contrasting with the classical PDE framework, where functional space and dynamical time of the propagators remain uncoupled in general.
There are multiple ways to introduce a (semi)group structure including the interesting possibility to define a skew-product flow. Another well-known approach is to build a discrete semigroup based on the symmetries of the inviscid system: this is the renormalisation group (RG).
\\

To emphasise the connection with the previous sections, we denote
${\cal A} = ({\cal A}_1,\cdots,{\cal A}_d)$ with ${\cal A}(\kappa) = \gamma(1/\kappa) 
, \gamma \in {\cal R}_{\rm eg}$.
Assume that one wants to know the probability that ${\cal A}(0)$ is in a particular region of $\mathbb{R}^d$, i.e. $\Pr({\cal A}(x) \in B | x \in [0,\kappa])$, where $B$ is a Borel set of $\mathbb{R}^d$. It is, as before, described by the pushforward of the normalised Lebesgue measure $\lambda_\kappa$, namely the resulting probability measure is $\mu_\kappa = {\cal A}_\star \lambda_\kappa$:
\begin{align}
\langle \mu_\kappa,F \rangle = \frac{1}{\kappa} \int_0^\kappa F({\cal A}(x)) dx = T \int_T^\infty F(\gamma(\tau)) \frac{d\tau}{\tau^2}, \nonumber \\ T:= \frac{1}{\kappa},~\mbox{for all}~F \in C_b(\mathbb{R}^d;\mathbb{R}). \nonumber
\end{align}
A frequently asked question is: what observables ${\cal O}$ in Definition \ref{SPST} are eligible for detecting $\SP$?
This is answered by the following simple result:
\begin{property}\label{ObsSP}
Let $S_0 = \left\{ \Phi_t[f_0(\cdot)] x_0 \right\}$, and 
$({\cal O},t,x_0)$ satisfying (\ref{killeress}) then
${\cal O}|_{S_0} \neq {\rm cst}$ and ${\cal O}(S_0)$ is a closed interval.
\end{property}\noindent
{\bf Proof}: see Appendix \ref{ProofObsSP}.
\\
\subsubsection{All probability measures are attainable by spontaneous stochasticity}

Despite the rather abstract concepts and results which follow, this section is driven by pragmatism: we want to know which statistics can be attained by spontaneous stochasticity.
The following is a measure-theoretic approach, namely instead of focusing on the dynamics of the Bebutov flow, we will investigate the statistical properties of the flow.
We first define the probability measure space with support the available values at time $t$ of the
inviscid system $({\cal P}_0)$:
\be \label{calM0}
{\cal M}_0 = {\cal P}(S_0),~~S_0 = \{ \Phi_t[f_0(.)] x_0\}.
\de
where ${\cal S}_0$ is compact (and also connected by Kneser's theorem).
Before proceeding, we need some useful Lemma:
\begin{lemma}\label{Birkhoff}
Assume that $\lim_{T \to \infty} \frac{1}{T} \int_1^T F(\gamma(\tau)) d\tau = \bar F$, for all $F \in C_b(\mathbb{R}^d;\mathbb{R})$, then
\be \label{chockbardebz}
\lim_{T \to \infty} \int_1^\infty F(\gamma(\tau T)) \frac{d\tau}{\tau^2} = \bar F.
\de 
\end{lemma}\noindent
{\bf Proof}: it is given in Appendix \ref{proofBirkhoff}.
\\

We finally define the family of probability measures:
\be 
\langle \mu_T,F \rangle = \int_1^\infty F \circ \Psi({\cal S}_{\tau T} \gamma) \frac{d\tau}{\tau^2},~~\forall F \in C_b(\mathbb{R}^d;\mathbb{R}).
\de
It is tight due to the boundedness of $\gamma$. By applying Lemma \ref{Birkhoff}, tightness 
together with Prokhorov's Theorem, we can define the probability measure space ${\cal M}(\gamma) \subset {\cal P}(\mathbb{R}^d)$, equipped with the weak topology:
\begin{align} \label{Mg}
{\cal M}(\gamma) = \left\{ \mu~ |~\exists T_n \to \infty, \langle \mu,F \rangle = \lim_{T_n \to \infty} \langle \mu_{T_n},F \right.
\rangle  \nonumber \\ \left. = \lim_{T_n \to \infty} \frac{1}{T_n} \int_0^{T_n} F \circ \Psi({\cal S}_\tau \gamma) d\tau \right\}.
\end{align}
Here the equalities are understood to hold for all $F \in C_b(\mathbb{R}^n;\mathbb{R})$.
Sometimes, we will use the equivalent writing $\mu = \lim_{T_n \to \infty} \frac{1}{T_n}
\int_0^{T_n} \delta_{\gamma(s)} ds$ (limit understood in the weak topology).
We can state the interesting result on the structure of ${\cal M}(\gamma)$:
\begin{property}\label{propMg}
The probability measure space ${\cal M}(\gamma) \subset {\cal M}_0$ is compact and is such that either one of the two possibilities occurs:
\begin{itemize}
  \item ${\cal M}(\gamma)$ reduces to a singleton containing a measure $\mu$ and $\gamma$ is necessarily generic w.r.t. $\mu$. In such a case, 
  if $\mu$ is non-Dirac then $({\cal P}_\kappa)$ is $\SP$ and the system selects $\mu$ in the inviscid limit $\kappa \to 0$.
 \item ${\cal M}(\gamma)$ contains a non-empty (possibly a continuum) set of measures where $\gamma$ is such that it is non-generic for all  measures in 
 ${\cal M}(\gamma)$. Convexity is not guaranteed.
\end{itemize}
\end{property}\noindent
{\bf Proof}:  see Appendix \ref{alternative}.
\\
In the case ${\cal M}(\gamma)$ does not reduce to a singleton, we will say that
 \begin{center}
 {\it $\gamma$ is non-renormalisable}.
 \end{center}
 We will give two nontrivial examples of non-renormalisable $\gamma$,
 giving either a convex or nonconvex set ${\cal M}(\gamma)$ not reducing to a singleton. They  correspond to typical regularisations which oscillate too slowly w.r.t. the ambient Lebesgue measure:
\begin{itemize}
\item $\gamma(s) = e^{i \log s}$. One obtain the set of measures ${\cal M}(\gamma) = \{ \mu_a \}_{a \in [0,2\pi]}$ with $\langle \mu_a,F \rangle = \int_0^1 F(e^{i a} s^i) ds$.
This set is nonconvex (see details in Appendix \ref{alternative}).
\item $\gamma(s) = \tanh( s \sin \log s)$. One obtains ${\cal M}(\gamma) = \{ \mu_a\}_{a \in [\sigma,1-\sigma]}$ with $\sigma = \frac{1}{1+e^\pi}$ and $\mu_a = a \delta_{+1} + (1-a) \delta_{-1}$. This set is convex (see details in Appendix \ref{exconv} ).
\end{itemize}

We now collect all the probability measure spaces ${\cal M}(\gamma)$ into one big set
\be 
{\cal M} = {\bigcup_{\gamma \in {\cal R}_{\rm eg}} {\cal M}(\gamma)},
\de 
still equipped with the weak topology. In the following, we show that ${\cal M}$ is in fact equal to ${\cal M}_0$ and as a consequence compact and convex. It means that all probability measures on the compact set $S_0$, can be attained by ad-hoc regularisations.
We first show that for all $x \in {\cal S}_0$
there is a regularisation $\gamma_x \in {\cal R}_{\rm eg}$ such that 
${\cal M}(\gamma_x) = \{\delta_x \}$. In other words, for this $\gamma_x$ there is a
classical selection principle to a given solution of $({\cal P}_0)$, i.e. no spontaneous stochasticity. The next step is to build regularisations that attain any convex combinations of these Dirac measures. These measures are the {\it extreme} points, or "pure states" of ${\cal M}_0$. 
Since the closed convex hull of $\{ \delta_x \}_{x \in S_0}$ is precisely ${\cal M}_0$, then one concludes that ${\cal M}_0 ={\cal M}$. 
This is the following theorem:
\begin{theorem}\label{M0M}
Let ${\cal M}_0 = {\cal P}(S_0)$, see Eq. (\ref{calM0}), and denote ${\cal E} = \{ \delta_x \}_{x \in S_0}$, one has ${\cal M}_0 = \overline{\rm co}({\cal E})$
\footnote{${\displaystyle {\rm co}(A) = \left\{ \sum_{i=1}^k \theta_i a_i | k \in \mathbb{N}, \theta_i \geq 0, \sum_{i=1}^k \theta_i = 1, ~a_i \in A \right\}}$} and 
\be 
{\cal M} = {\cal M}_0.
\de 
As a consequence ${\cal M}$ is {both} compact and the closed convex hull of its extreme points ${\cal E}$. 
\end{theorem}\noindent
{\bf Proof}: see Appendix \ref{M0Mproof}.
\\\\
An interesting corollary is that one can construct an almost explicit regularised problem, $({\cal P}_\kappa)$, which contains all statistical behaviours and for which a probability measure cannot be defined in the inviscid limit $\kappa \to 0$.
\begin{corollary}[]\label{gun}
There exists a non-renormalisable regularisation $\gamma_{\rm un}$ such that 
$$
{\cal M}(\gamma_{\rm un}) = {\cal M}_0.
$$
\end{corollary}\noindent 
{\bf Proof}: see Appendix \ref{proofgun}.
\subsubsection{The Bebutov attack: from RG to universality classes}\label{Bebuattak}
Let us summarise our work. We defined a space, ${\cal R}_{\rm eg}$, consisting of continuous bounded curves $\gamma:\mathbb{R} \to \mathbb{R}^d$, 
whose asymptotics describe how regularisations behave as they approach the inviscid system. On this space, equipped with the compact-open topology, we introduced a well-defined infinite-dimensional dynamical system known as the Bebutov flow, which shifts these curves while taking a given regularisation as the initial condition.
One can then analyse the Bebutov flow's dynamics in terms of invariant sets, attractors, and basins of attraction. For instance, a fixed attracting point corresponds to a selection principle for all regularisations within its basin toward a specific solution of the inviscid system. Spontaneous stochasticity, on the other hand, corresponds to more complex invariant sets. The connection with RG approaches thus becomes much clearer. However, in our finite-dimensional setting, everything can be rigorously defined.

Rather than studying the Bebutov dynamics directly, we focused on its statistical properties and the measures supported on the invariant sets. A measure selection principle arises whenever a probability distribution can be defined in the inviscid limit. We encoded these statistical properties in a broader space of probability measures, ${\cal M} = \bigcup_{\gamma \in {\cal R}_{\rm eg}} {\cal M}(\gamma)$, where weak subsequential limits exist, see Eq. (\ref{Mg}). Remarkably, this space coincides with the space of all probability measures associated with the inviscid system,
${\cal M}_0:= {\cal P}(S_0), S_0=\{ \Phi_t[f_0(\cdot)] x_0 \}$.
The key consequence is that one can construct a regularisation to target any probability measure in ${\cal P}(S_0)$. As far as we know, this correspondence has not been established before.

\begin{figure}[hptb]
\centering
\centerline{\includegraphics[width=\columnwidth,keepaspectratio]{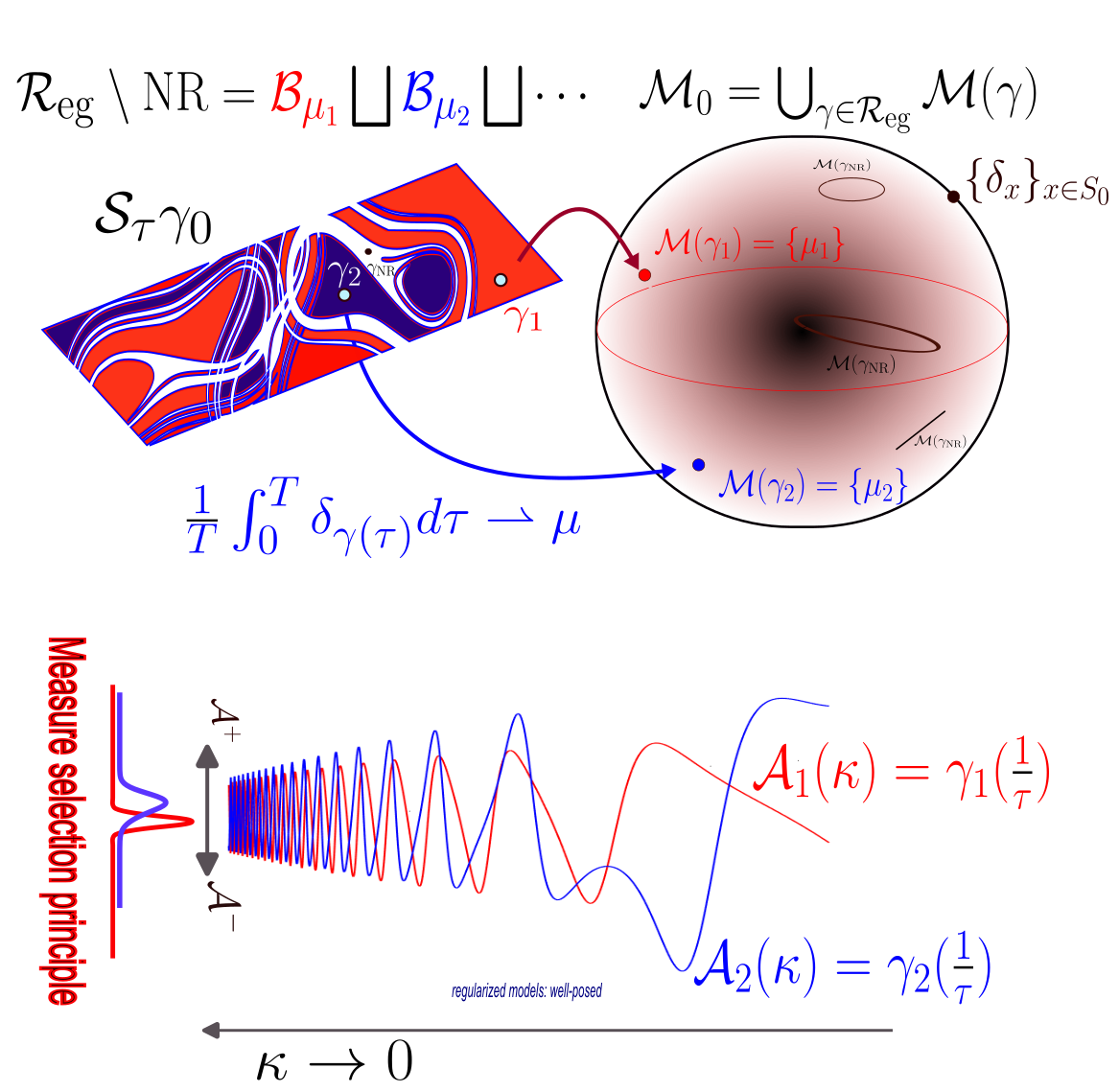}} 
\caption{A tale of spontaneous stochasticity.
We present our results graphically. On the left, the infinite-dimensional space of all regularisations is denoted ${\cal R}_{\rm eg}$, where each point $\gamma$ represents a specific regularisation. These regularisations evolve under the Bebutov flow. For each $\gamma$, we define a set-valued map $\gamma \mapsto {\cal M}(\gamma)$ using ergodic subsequential limits.
If ${\cal M}(\gamma)$ consists of a single element, then we can define a unique probability measure on the space of solutions of the inviscid system. Otherwise, the regularisation is classified as non-renormalisable (NR). The set of probability measures obtained coincides exactly with the space of probability measures for the inviscid system, denoted ${\cal M}_0 = {\cal P}(S_0)$, which is depicted on the right. This space is convex and compact, with its extreme points corresponding to the Dirac measures $\delta_x, x \in S_0$. By definition, these specific regularisations are not $\SP$.
Finally, by excluding non-renormalisable regularisations $\gamma_{\rm NR}$, we obtain a well-defined partition of the space, structured around genericity sets ${\cal B}_\mu$ under the Bebutov flow. 
The diagram highlights two sets in red and blue, each representing a universality class ${\cal B}_\mu$. 
These classes contain all regularisations that converge to the same measure $\mu$ in the inviscid limit. Spontaneous stochasticity occurs as soon as ${\cal M}_0$ undergoes a set bifurcation from the well-posed singleton regime!}
\label{SPRG}
\end{figure}

It is then possible to define universality classes for free.
Consider a given measure $\mu \in {\cal M}_0$, we define its generic set as
\begin{align}\label{B_mu}
{\cal B}_\mu := \left\{  \gamma \in {\cal R}_{\rm eg} | \lim_{T \to \infty} \frac{1}{T}
\int_0^T F(\Psi({\cal S}_\tau \gamma))d\tau = \langle \mu,F \rangle 
\right\},
\end{align} 
for all $F \in C_b(\mathbb{R}^d;\mathbb{R})$.
From the previous results, the filtered partition applies:
\be 
{\cal R}_{\rm eg} \setminus \{\gamma \mbox{ non-renormalisable} \} = \bigsqcup_{\mu \in {\cal M}_0} {\cal B}_\mu,
\de 
where ${\cal B}_\mu$ is never empty, possibly reducing to a singleton modulo the translation by the Bebutov flow. This set is precisely playing the role of basins of attractions in classical RG approaches but at a statistical level. Even more importantly, spontaneous stochasticity can appear provided regularisations have the property to "mix" but in the  weakest possible form: ergodicity. By Definition \ref{RegBebu} and Eq. (\ref{Projno}), this partition remains independent of the chosen dynamical system, as Birkhoff averages do not involve ${\cal S}_\tau$ at the end. However, the dynamics in ${\cal R}_{\rm eg}$ still depend on the specific choice of ${\cal S}_\tau$. Our method is just offering a trick to frame the problem through dynamical systems in a very intuitive way. We give a sketch of the scenario in Fig.~\ref{SPRG}.
\\

In this section, $(t,x_0)$ are kept fixed. If one makes $t$ varies, a typical scenario can occur where the inviscid system is (locally) well-posed until $t \leq t^\star$ (the blow-up time).
In this case, one observes a {\it set bifurcation} where ${\cal M}$ switches from a singleton to a larger set necessarily containing non-Dirac measures. Even more spectacular scenario can occur where a cascade of set bifurcations is triggered as $t$ increases. A simple example is given in Section \ref{Ex4}.
This simple fact leads to a very provocative statement: 
\begin{center}
{\it If ${\cal M}_0$ is not a singleton, then there exist regularisations for which $({\cal P}_\kappa)$ is $\SP$!}
\end{center}
This result is a direct consequence of Theorem \ref{M0M}: one can always construct a renormalisable regularisation by taking two arbitrary points $x \neq y$ in $S_0$ to produce $\theta \delta_x +
(1-\theta) \delta_y$, $\theta \in (0,1)$. As a consequence, one can root $\SP$ in the lack of uniqueness for $({\cal P}_0)$. 
This question is addressed in Section \ref{which}.
\\

Many further developments are possible:
\begin{itemize}

\item 
We note that non-renormalisable $\gamma$ may be interpreted as a form of spontaneous stochasticity but at the measure level. However, care must be taken as it may simply reflect a poor sampling choice--specifically, an inadequate ambient measure used to define probabilities. The question remains open, as it is unclear whether there exist non-renormalisable regularisations that remain irreducible under any reparametrisation.
\begin{conjecture}
For all non-renormalisable $\gamma \in {\cal R}_{\rm eg}$, there exists some continuous measurable $h$
 with $h(\tau) \to \infty, \tau \to \infty$, such that ${\cal M}(\gamma \circ h)$ reduces to a singleton. 
\end{conjecture}\noindent
We do believe this conjecture to be true in which case $\SP$ must be "idempotent". If it is not, then the Bebutov pushforward $({\cal S}_\tau)_\star$ becomes nontrivial so that $\SP$ is observed at the measure level, say "level-1" $\SP$. 

\item  A key question is the size of these universality classes. In particular, the presence of chaos in the inviscid system is likely to play a significant role. 
Consider the following formal scenario: suppose ${\cal S}_\tau$ is a chaotic dynamical system with a compact attractor ${\cal A}$ and an SRB measure $\mu_{\rm SRB}$ supported on it. Then the basin of attraction ${\cal B}(\mu_{\rm SRB})$ is expected to contain a large class -- measure-theoretically -- of regularisations $\gamma_0$ that are attracted to ${\cal A}$ in the compact-open topology.
Non-generic (non-prevalent) regularisations are expected to form much smaller sets or may even fail to be renormalisable.
Even more importantly, the choice of a specific ambient measure does not matter, as long as it is absolutely continuous with respect to Lebesgue. For any $\gamma_0 \in {\cal B}(\mu_{\rm SRB})$, ${\cal M}(\gamma_0) = \{ \mu_{\rm SRB} \}$, even after reparameterising $\gamma_0$ -- provided the reparameterisation preserves absolute continuity. 
This provides evidence that regularisations satisfying stronger mixing properties than ergodicity are statistically more stable.
It is highly reminiscent of the results obtained in \cite{Drivas24}.

\item As already observed, the probability measures in ${\cal M}$ depend on time $t$ and the initial condition $x$ from $({\cal P}_0)$, and should be denoted by $\mu_{t,x}$. These limiting measures are known in the literature as {\it Young measures}. Given the intimate connection between ODEs and transport equations, it is essential to take into account the dependence on $(t,x)$. In particular these measures can be used to build dissipative weak solutions of the transport PDE. We will investigate these aspects in a near future.
\end{itemize}

\subsection{Stochastic versus deterministic regularisations}\label{stochno}
In many cases, regularisations of $({\cal P}_0)$ involve randomisation (see e.g., in the PDE case \cite{Bandaketal_PRL24}, \cite{Simon_Jeremie_AM20} and \cite{Drivas24} in the ODE case). This introduces two small parameters: the noise amplitude $\epsilon$, and a cut-off parameter -- such as viscosity -- represented here by $\kappa$. In our framework, this leads to a broader class of regularised problems, denoted $({\cal P}_{\epsilon,\kappa}^\omega)$, constrained by ${\cal P}_{0,0}^\omega = {\cal P}_0$, making the link with the inviscid deterministic model. The parameter 
$\epsilon$ controls the noise intensity with realisations $\omega$. Deterministic regularisations correspond to taking the limit $\epsilon \to 0$ first, then $\kappa \to 0$. Reversing this order -- taking $\kappa \to 0$ first -- typically yields a single regularisation by noise. The two limits do not commute in general. However, in practice one considers a joint limit $\epsilon,\kappa \to 0$, where one must clearly define how the two parameters are linked -- for instance, by setting $\kappa = \kappa(\epsilon)$ and then analyzing the limit as a single inviscid limit when the noise amplitude $\epsilon$ goes to zero.

Assume for a moment, one would like to rephrase Definition \ref{SPST} for stochastic/randomised regularisations. Assume that $\kappa = \kappa(\epsilon)$, then we define $\mu_{\epsilon}(B)$ as the probability that
${\cal O}(u_{\epsilon,\kappa})$ solution of 
$({\cal P}_{\epsilon,\kappa}^{\omega})$ at time $t$ and initial condition $x_0$ is in some Borel set  $B \subset \mathbb{R}$.

\begin{figure}[htbp]
\centerline{\includegraphics[width=0.9\columnwidth]{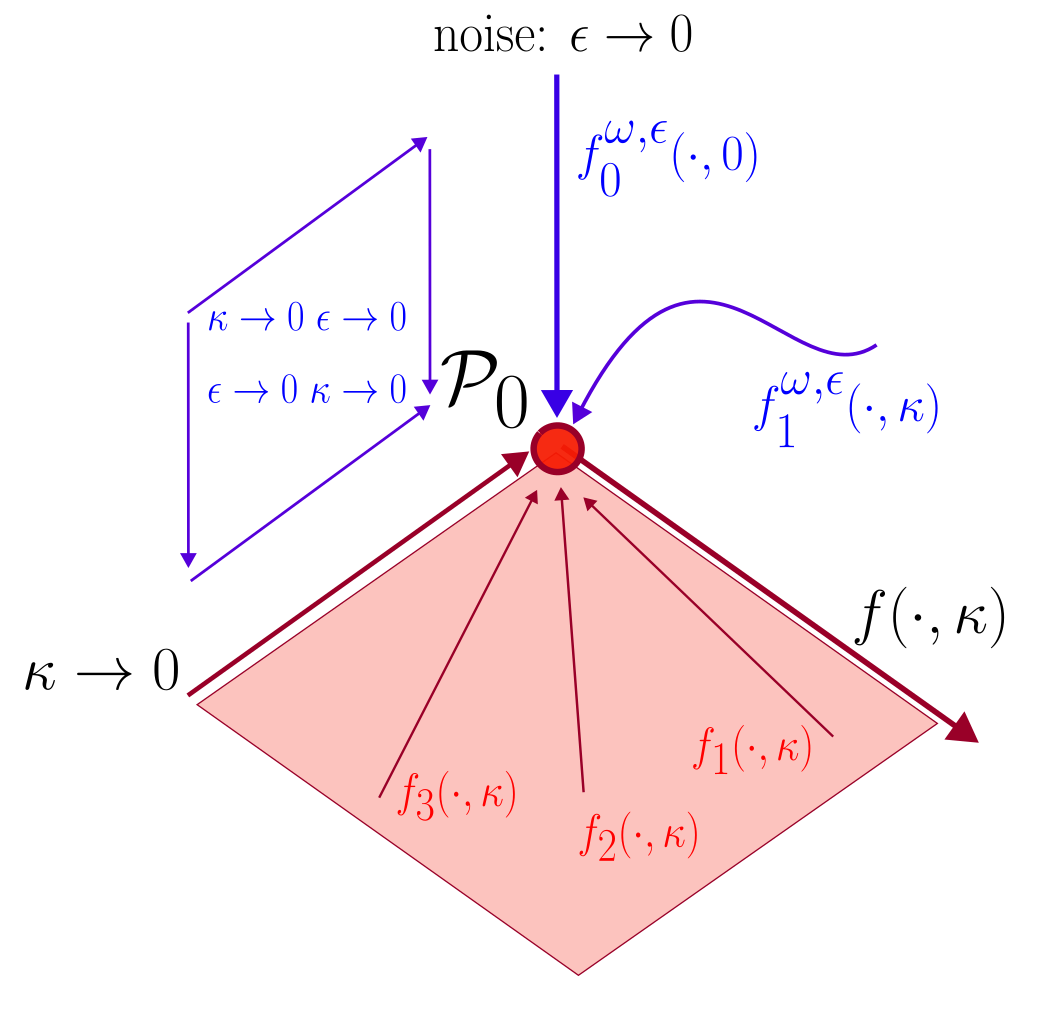}}
\caption{We focus on deterministic regularisations, represented by the red plane, with a single inviscid limit $\kappa \to 0$. More general stochastic regularisations take the form $f^{\omega,\epsilon}(\cdot,\kappa)$, where $\epsilon$ controls the noise amplitude and $f^{\omega,0}(\cdot,0) = f_0(\cdot)$. In such cases, one must specify the relation between $\epsilon$ and $\kappa$ and consider the limit $\epsilon \to 0$ with $\kappa = \kappa(\epsilon)$. Note that each of these limits can yield different selection mechanisms within ${\cal P}_0$.
Most of our results (Subsections \ref{secmea} and \ref{which}) can be extended to this general framework.}
\label{doublelim}
\end{figure}

Denote $\mu$ all subsequential limits (in the weak topology) of  $\mu_{\epsilon}$ then hypothesis (\ref{eq:H0}) would become $\mu \in {\cal P}({\cal O}_0)$, the
set of probability measures with support ${\cal O}_0$
and where ${\cal O}_0 := \{ {\cal O}(\Phi_t[f_0(\cdot)]x_0) \}$ and
Definition \ref{SPST} would take the form
\be \label{SPST_S}
\inf_\mu \operatorname*{inf}_{x \sim \mu} x < 
\sup_\mu \operatorname*{sup}_{x \sim \mu} x.
\de 
Note that the essential inf/sup are not needed since the support of the measures $\mu_\epsilon$ are connected by Kneser theorem and ${\cal O}$ is continuous. 
This definition covers the important case where a weak limit exists: $\mu_\epsilon \rightharpoonup \mu$ with $\mu$ non-Dirac.
However, as in Definition \ref{SPST}, we also allow more general situations where no such limit exists. 
All these probabilistic elements, in fact, arise naturally from our deterministic framework as well. This was the focus of Subsection \ref{secmea}. Notably, the discussion does not depend on whether the finite-$\kappa$ regularisation is random or deterministic. To keep the exposition simple and avoid introducing extra notation, we purposely restrict ourselves to deterministic regularisations and a single-parameter limit. We give a sketch in Fig.~\ref{doublelim}. However, we would like to insist again on a fundamental aspect of spontaneous stochasticity.
\\\\
{\it 
Contrary to common intuition, randomising the regularisation is not required for stochastic effects to appear.}
\\

It is necessary to clarify an important point regarding the notion of randomness. A standard protocol in turbulence consists in sampling a set of initial conditions for different Reynolds numbers -- that is, in our setting, for different values of $1/\kappa$ (and, in our case, for the same initial condition $x_0$). The empirical average then amounts to estimating the pushforward of a background measure, such as a normalised Lebesgue measure, that is, an estimate of the integral like in (\ref{mukdef}).

Indeed, (spontaneous) stochasticity may arise when regularisations approach the inviscid system in a rapidly oscillating and erratic way, even if the vector fields themselves are deterministic and converge in the sense of (\ref{eq:H0}). While such oscillations can generate genuine randomness, this phenomenon can only occur in the singular limit. In order for this randomness to persist at the limit as stochastic behaviour, it is necessary that the inviscid system be ill-posed. This condition is both essential and is addressed in Subsection \ref{which}. If the inviscid system is well-posed, such stochastic effects cannot arise; the mechanism is effectively excluded. Yet ill-posedness, on its own, does not guarantee stochastic behaviour. The absence of a selection mechanism in the inviscid limit -- linked to the regularisation -- is also required. 
The function $x \mapsto \sin \frac{1}{x}$ as $x \to 0$ provides a useful analogy: it exemplifies how regularised systems can fail to converge to a single limit, instead exhibiting different limiting behaviours along different subsequences -- mirroring what may occur in the inviscid limit. Likewise, randomness in the limit enters through the choice of these subsequences. This gives rise to a probability space $(\Omega, \mathcal{F}, \mu)$, where $\Omega$ is the set of converging sequences $x_n \to 0$ for which $\lim \sin \frac{1}{x_n}$ exists. Defining $X(\omega) = \lim_{n \to \infty} \sin \frac{1}{x_n}$ yields a random variable whose distribution is the arc sine law.

Lastly, it is worth noting that the autonomous case does, in fact, include the non-autonomous one through the standard reformulation: $\dot{x} = f(x,t)$ becomes $\dot{x} = f(x,y),\ \dot{y} = 1$. Under our assumption (\ref{eq:H0}), all relevant quantities remain bounded, so no compactness issues "at infinity" arise. For example, we do not require any notion of an attractor as $t \to \infty$, since $t$ is always finite and $({\cal P}_0)$ itself stays bounded.

\subsection{Which inviscid systems can exhibit spontaneous stochasticity?}\label{which}
This section aims to provide deeper insight into systems that may exhibit (\ref{killeress}). The goal is to establish a sufficiently sharp necessary condition. Rather than examining which regularisations $({\cal P}_\kappa)$ are $\SP$ -- a question addressed in the previous section -- we focus on identifying necessary properties of the inviscid system $({\cal P}_0)$ that allow $({\cal P}_\kappa)$ to be $\SP$. In light of Section \ref{secmea}, this question solely amounts to investigate the lack of uniqueness for $({\cal P}_0)$, namely systems for which $S_0$ (and therefore ${\cal M}_0$) is not a singleton.

To clarify the goal, \cite{Drivas24} established a sharp condition for a class of systems to be $\SP$. These systems correspond to dynamical systems with an isolated H\"older singularity. As we will see, they fall within the conditions under which $\SP$ can occur. Examples of this kind are provided in Subsection \ref{Ex3}, as they are expected to play an important role in physics.

We therefore aim at identifying a necessary condition for nonuniqueness in the inviscid system $({\cal P}_0)$. The anticipated scenario is that the inviscid flow hits a singular set in finite time. However, the precise definition of such sets must be clarified. A typical example is a non-Lipschitz isolated singularity $x^\star \in \mathbb{R}^n$, where $f(x^\star) = 0$ lacks Lipschitz continuity, though this is not a sufficient condition. There are also cases where $f(x^\star) \neq 0$, yet $x^\star$ still leads to a loss of uniqueness for flows going through it (see Subsection \ref{Ex3}). 

In all the following, we assume $f$ at least continuous to ensure Peano’s classical solution existence.
Numerous theorems addressing uniqueness exist for general nonautonomous systems of the form $\dot x = f(x,t),~x(t_0) = x_0$ (see, in chronological order \cite{Walter1970,Agarwal_Lak,Hartman2002,Bahouri2011}). These results 
typically take the form:
{\it if condition $(A)$ holds then the solution to the Cauchy problem is unique}.
Consequently by contraposition, there are many possible necessary conditions of the form
$\neg{(A)}$ for nonuniqueness. In order to obtain some sharper condition, one must look at 
a "disjunction" of different uniqueness theorems. The 1-D autonomous case is well understood.
First the singular set, denoted $\Gamma^\star$, is necessarily reduced
to an isolated critical point $x^\star$ i.e. such that $f(x^\star) = 0$ (by a straightforward use of the implicit theorem, see e.g. Theorem 1.2.7 in 
\cite{Agarwal_Lak}). Second, $f$ must increases sufficiently rapidly in the neighborhood of
$x^\star$ either to the left or right, in particular it cannot be decreasing or being Lipschitz at $x^\star$. The optimal growth rate condition is convergence of at least one of the integral $\int_{{x^\star}^\pm} \frac{dx}{f(x)}$. With respect to the initial condition, it is highly constrained in 1-D, either $x(0)=x^\star$ or if $x_0 < x^\star$, then $f$ must be positive so that $x^\star$ is reached in finite time (and conversely $x_0 > x^\star$, $f$ must be negative). Typical examples giving nonunique solutions are $\dot x = {\rm sign}(x) |x|^\alpha$, $\dot x = \pm |x|^\alpha$, $\alpha \in (0,1)$. But systems like $\dot x = |x|^\alpha + cst$
or $\dot x = -{\rm sign}(x) |x|^\alpha$ have unique solutions no matter the initial condition. 

\subsubsection{A necessary condition for nonuniqueness in the inviscid system}\label{nd}
The situation in dimension above one is much less favorable, in particular 
$f(x^\star) = 0$ is no more necessary which is the main reason this problem becomes nontrivial. 
The aim here is thus to modestly characterise regions where 1) $f$ has some expanding behaviour
2) $f$ is not Lipschitz. 

We consider here a similar idea to that of Dini derivatives (see Appendix \ref{DiniApp}). 
They correspond to a generalised notion of derivatives in contexts where $f$ is not even continuous. Dini derivatives share many desirable properties with classical derivatives.
For instance, being Lipschitz continuous is equivalent to having
finite Dini derivatives. They 
therefore provide a natural and practical way to express non-Lipschitz behaviour.
We slightly extend the notion to the Osgood property.
Let us introduce the following definitions:
\begin{definition}\label{DiniME}
Let $f: \mathbb{R}^n \to \mathbb{R}^n$, 
Let $v \in \mathbb{S}^{n-1}$, then
$$
\Lambda_\Omega^+(x,v) := \limsup_{t \to 0^+} 
\left \langle \frac{f(x+t v)-f(x)}{\Omega(t)},v
\right\rangle .
$$
where $\Omega$ is a modulus of continuity: $\Omega: \mathbb{R}^+ \to \mathbb{R}^+, \Omega(0) = 0$ and $\Int_{0^+} \frac{dz}{\Omega(z)} = \infty$.
\end{definition}
 \noindent
We also introduce a notion of stable finite-time set:
\begin{definition}
Let $\Phi: (t,x_0) \mapsto \Phi_t x_0$ be the flows associated to 
$\dot x = f(x), x(0)=x_0$, and consider some arbitrary set $\Gamma \subset \mathbb{R}^n$
\begin{align}\label{Wm_G}
{\cal W}^-(\Gamma) \equiv  \bigcup_{\Phi} \left\{ x \in \mathbb{R}^n
 \left| \right. \exists t^\star < \infty,~t^\star \geq  0, \right. \nonumber \\ \left. \lim_{t \to +t^\star} {\rm dist}(\Phi_t x,\Gamma) = 0 \right\} 
\end{align}
\end{definition}
From this definition, one can derive a useful necessary condition for breaking uniqueness
and thus having (\ref{killeress}):
\begin{theorem}\label{CN}
Assume that $({\cal P}_0)$ has nonunique solutions, then one 
can find some $\Omega$ and a nonempty set $\Gamma^\star \subset \mathbb{R}^n$ such that
$x_0 \in {\cal W}^-(\Gamma^\star)$ 
where
$$
{\displaystyle \Gamma^\star = \big\{ x \in \mathbb{R}^n ~|~ \exists v, ||v||=1, 
 ~\Lambda^+_\Omega(x,v)= +\infty \big\}.}
$$
\end{theorem}
\noindent {\bf Proof}: see Appendix \ref{DiniApp}.\\
Obviously, this is also a necessary condition for all regularisations $({\cal P}_\kappa)$ having $\SP$. The idea is to replace the classical notion of Jacobian of $f$
by some weaker notion involving Dini (directional) derivatives 
(say with $\Omega(z)=z$). The unit vectors $v$ indeed plays the role of an eigenvector
and $\Gamma^\star$ simply detects regions in phase space where the eigenvalues blow up to $+\infty$. 
 
In addition, one can also characterise for free the set of initial conditions giving
spontaneous stochasticity by a straightforward generalisation of a stable set.
Note that such result is not sharp and cannot be sufficient. However, there is still room to improve this condition. We let it for future work.

The 1-D case is rather particular and is of less interest, in particular the critical 
condition $f(x^\star) = 0$ is not distinguished. This is expected
since Theorem \ref{CN} is aimed at situations where nonuniqueness arises from 
more general situations than those due solely to the presence of non-Lipschitz 
critical points.
We next illustrate these results through various examples (Subsections \ref{Ex1}--
\ref{Ex4}). 

\subsubsection{Example 1}\label{Ex1}
We consider the well-known example $\dot x = f(x), x(0) = x_0$ with $f(x) = x^\frac13$. The function $f$ has a Lipschitz singularity at 0.
Following \cite{Drivas21}, we consider the regularised system 
$\dot x = f(x,\kappa)$, and $f(\cdot,\kappa)$ is defined as
\begin{equation}\label{toy1_3}
({\cal P}_\kappa):~\left\{
\begin{array}{lllcccc}
\dot x & = & f(x) & |x| & > & \kappa \\
\dot x & = & \kappa^\frac13 G(\frac{x}{\kappa},\kappa) & |x| & \leq & \kappa
\end{array}\right.,
\end{equation}
where
$G(x,\kappa) =\xi(x) f(x) + (1-\xi(x)) \frac{\omega(\kappa)+x}{2}$ 
with $\xi(x) = 3x^2 - 2 |x|^3$.
Since hypothesis (\ref{eq:H0}) is obviously satisfied, 
it guarantees that ${\cal A}$ is well-defined as a function. The regularised system is
well-posed. We will use two different regularisations denoted ${\cal A}_1$ and
${\cal A}_2$ with resp.
$\omega_1(\kappa) = \sin \frac{1}{\kappa}$ and $\omega_2 = 0.5 + \sin (\frac{1}{\kappa}+{\rm cst})$.
Note that these deterministic functions rapidly change their sign when  $\kappa$ becomes small. They introduce a source of uncertainty that vanishes in the limit as $\kappa \to 0$,
playing a role similar to that of a random variable.

\begin{figure}[htbp]
\centerline{\includegraphics[width=0.9\columnwidth]{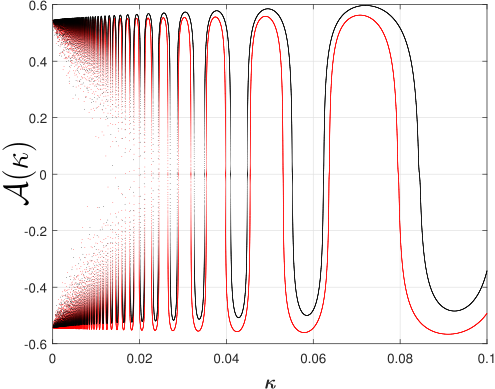}}
\caption{${\cal A}(\kappa) = x^\kappa(1)$ solution of (\ref{toy1_3}) for $\kappa$ uniformally distributed in $[10^{-4},10^{-1}]$: in red ${\cal A}_1$ and in black ${\cal A}_2$. Both measures concentrate on $\pm a$ with $a \approx 0.544$.}
\label{Toyess}
\end{figure}

Considering Definition \ref{SPST}, one then defines
${\cal A}_{1,2}(\kappa) = {\cal O}(x^\kappa) = x^\kappa(1)$ for the two functions $\omega_{1,2}$. 
This is a continuous function due to well-posedness. The result is shown in Fig.~\ref{Toyess}.
For the two regularisations  ${\cal A}_1$ and ${\cal A}_2$, the pushforward measures are not equal but their support is the same: the first perturbation yields 
$\frac{\delta_{+a} + \delta_{-a}}{2}$
whereas the second perturbation is $\theta \delta_{+a} + (1-\theta) \delta_{-a}$ with $\theta \approx 0.7$.
To summarise:\hspace*{-0,2cm}
\begin{itemize}
\item[$\bullet$] $({\cal P}_\kappa)$ is $\SP$ for 
${\cal A}_1,{\cal A}_2$ and regularisation by additive noise. One must necessarily have
$x_0=0$ for any choice $t > 0$.
\item[$\bullet$] $({\cal P}_\kappa)$ is not $\SP$ when taking $\omega(\kappa) = {\rm cst}$.
\item[$\bullet$] $({\cal P}_\kappa): dX_t = {\rm sign}(\sin X_t) |\sin X_t|^\alpha~dt + 
\sqrt{2\kappa} dW_t, x(0) = x_0 ~\alpha \in (0,1)$ is $\SP$ for $x_0 = 2k\pi, k \in \mathbb{Z}$. Theorem \ref{CN} is satisfied with Dini derivatives equal to $+\infty$ only at $x_0 = 2k\pi, k \in \mathbb{Z}$ so that $\Gamma^\star = (2k\pi)_{k \in \mathbb{Z}}$. Moreover ${\cal W}^-(\Gamma^\star) = \Gamma^\star$. 
\end{itemize}
\subsubsection{Example 2}\label{exAmbrosio}
\begin{figure}[htbp]
\centerline{\includegraphics[width=1.1\columnwidth]{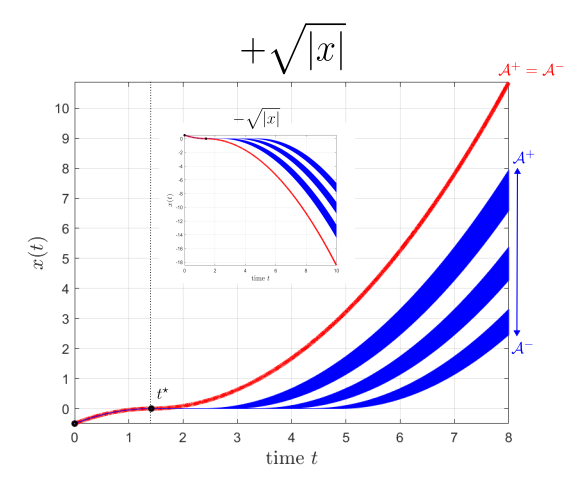}}
\caption{Solutions selected in the limit $\kappa \to 0$ for the system 
$\dot x = \sqrt{|x|}, x(0)=-c^2$ for two different regularisations, in red by additive noise: the system is not $\SP$ (${\cal A}^+={\cal A}^-$), in blue using the regularisation (\ref{regamb}) with $T \in [1,\frac32] \cup [2,\frac52]
\cup [3,\frac72]$ uniformally distributed, the system becomes $\SP$ for all $t > 2|c|+1$ (${\cal A}^+ > {\cal A}^-$). The black dots are for $c=1/\sqrt{2}$, $(t_0,x_0)=(0,-0.5)$ and $(t^\star,x^\star)=(\sqrt{2},0)$. The inset shows the case with $-\sqrt{|x|}$ and
$x_0 = +c^2$ (minus the regularisation (\ref{regamb})).}
\label{Ambrosio_ex}
\end{figure}

We discuss again a 1-D example which deserves some important comments. It is inspired
by \cite{ambrosio2004transport,Flandoli2009}. The inviscid system is the simple $\dot x = \sqrt{|x|}$ with
initial condition $x(0)=-c^2 < 0$. It has an infinity of solutions: the pre-blowup
life is until time $t^\star = 2|c|$ where it reaches the H\"older singularity 0. Then the solution can
wait an arbitrary long time $T\geq 0$ at 0 until it takes off: $x(t) = \frac14 (t-T-2|c|)^2$.
We consider two different regularisations, one is adding some additive noise 
$dx = \sqrt{|x|} dt + \sqrt{\kappa} dW_t, x(0)=-c^2$, the
other takes the form
\begin{align} \label{regamb}
f(x,\kappa)= \mathbbm{1}_{(-\infty,-\kappa^2]}(x) \sqrt{|x|} + 
\mathbbm{1}_{[-\kappa^2,\kappa T-\kappa^2]}(x) \kappa \nonumber \\ + \mathbbm{1}_{]\kappa T - \kappa^2,+\infty)}(x)
\sqrt{x - \kappa T + 2 \kappa^2}, 
\end{align}
where $\mathbbm{1}_I(x)$ is the characteristic function equals to 1 if $x \in I$ and 0 else.
As discussed in \cite{ambrosio2004transport}, this regularisation selects a solution that remains at 0 for a duration of time $T$. Consequently, the regularised system can be fully controlled to match any target distribution. In Fig.~\ref{Ambrosio_ex}, we deliberately randomise $T$ so that the system becomes $\SP$ and selects three equally likely, disjoint intervals! Alternatively, one could carefully design some rapidly oscillating function $T = T(\kappa)$ inspired by the technique (\ref{controlswitch}), as in the previous example which would give the same result (not shown).
There is in fact no need to randomise $f(\cdot,\kappa)$.
The system is not $\SP$ when regularised by classical additive noise.
This example once again demonstrates that one cannot expect (\ref{killeress}) to give simple universality properties, not only at the measure level but also regarding the set of selected solutions. This system (or any other) exemplifies Theorem~\ref{M0M}: it is possible to design a regularisation of $\sqrt{|x|}$ that converges to any target probability distribution in ${\cal M}_0$ with support $S_0 = [0,\frac14 (t-2|c|)^2], t \geq 2|c|$ (see Eq. (\ref{calM0})) as $\kappa \to 0$.

\subsubsection{Example 3}\label{Ex3}
We consider a system with an isolated H\"older singularity at the origin. 
Isolated H\"older singularities are believed to play an important role in shell models of turbulence \cite{Dombre1998,Maily2012}. A detailed measure-theoretic analysis can be found in \cite{Drivas21,Drivas24}. These systems take the form
$$
\dot x = |x|^\alpha F(y),~y = \frac{x}{|x|} \in \mathbb{S}^{n-1},
F:\mathbb{S}^{n-1} \to \mathbb{R}^n.
$$ 
Using the convention of \cite{Drivas21}, one can write $F$ as a decomposition into
radial and tangential components: $F(y) = F_r(y) y + F_s(y)$, with
$F_r: \mathbb{S}^{n-1} \to \mathbb{R}$ and $F_s: \mathbb{S}^{n-1} \to T\mathbb{S}^{n-1}$.
We note that the Dini directional derivatives along some vector $v \in \mathbb{S}^{n-1}$ at the singularity $x=0$, reduce in this case to 
$$
\Lambda^+_z(0,v) = \limsup_{t \to 0^+} t^{\alpha-1} F_r(v) \in \{-\infty,0,+\infty\}.
$$
In order for Theorem \ref{CN} to apply, one must necessarily have directions $v$ for which
$F_r(v) > 0$. This property is indeed the same than the {\it defocusing}
property considered
in \cite{Drivas24} (see (b), p.1859). In general, such singularities can be seen as generalised 'saddles', including purely unstable critical points, and are distinguished by having both nontrivial stable and unstable sets.

Note that we also allow the initial condition to be on the singularity
like in 1-D but the stable set ${\cal W}^-(0)$ in fact provides a detailed characterisation of all "pre-blowup" initial conditions. 
The example presented here is a slight modification of the 2 d.o.f. example studied in \cite{Drivas21}, but, unlike the original, it does exhibit (\ref{killeress}).
It takes the form
\be \label{Ex2eq}
F(y) = y(y_1+y_2) - y^\perp y_1y_2^2,~y^\perp = (-y_2,y_1).
\de 
This example is not trivial and has in particular homoclinic orbits with overlapping stable and unstable sets. Indeed, using the conventions of \cite{Drivas21,Drivas24}, the
renormalised system $\dot y = (F(y) \cdot y^\perp) y^\perp$ on $\mathbb{S}^1$ has a well-defined focusing attractor at $\varphi= \frac{3\pi}{2}$ (it guarantees finite-time blowup), but the defocusing attractor at $\varphi=0$ in 
$\mathbb{S}^1$ is non-hyperbolic and one-sided (and a physical measure $\delta_{\varphi=0}$).

\begin{figure}[htbp]
\centerline{\includegraphics[width=0.95\columnwidth,keepaspectratio]{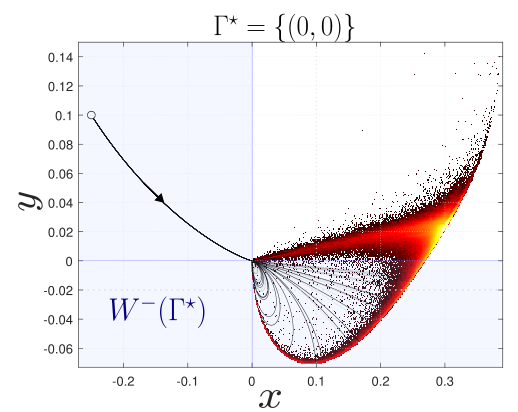}}
\caption{Distribution of the solutions of (\ref{Ex2eq}) at $t=1.5$ starting from the initial condition
$x_0=(-0.25,0.1)$ (log-scaled colormap). The system has been regularised by additive noise of amplitude
$\sqrt{\epsilon}$ with $\epsilon \to 0$. The solution for this initial condition reach the singularity at finite blowup time $t^\star 
\approx 0.875$. The region in light blue color corresponds to the stable set
associated with initial conditions giving (\ref{killeress}).}
\label{figex2}
\end{figure} 
\subsubsection{Example 4}\label{Ex4}
\begin{figure}[htbp]
\centerline{\includegraphics[width=0.95\columnwidth]{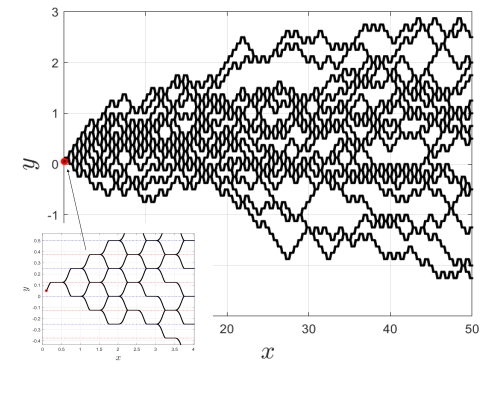}}
\caption{System (\ref{ex3}) for  $t \in [0,50]$, $\mu=1$ and $\sigma=4$ starting from the initial condition 
$x_0 = (0.1,0.05)$ (the system has been regularised by additive noise of amplitude
$\sqrt{\epsilon}$, $\epsilon \to 0$ integrated for 20 different realisations).
The blowup time is $t^\star \approx 0.5$ (see zoom: the levels $x_2 = Y_{p,1},Y_{p,2}$ of $\Gamma^\star$ are also shown in red and blue dash-dotted lines).}
\label{figex3}
\end{figure}

We now discuss a simple example which has (\ref{killeress}) although there is no critical points. It is in fact a 1-D nonautonomous system translated into an autonomous one. Let us define $s_{\alpha,\sigma}(x_2) = 
{\rm sign}(\sin 2 \pi \sigma x_2) 
|\sin 2 \pi \sigma x_2|^\alpha$ and for $\alpha \in (0,1)$, $\mu \in \mathbb{R}$, 
\be \label{ex3}
\begin{array}{lll}
\dot x_1 & = & \mu,  \\
\dot x_2 & = & s_{\alpha,\sigma}(x_2) \sin(2\pi x_1)
\end{array}
\de
then by an abuse of notation, we can write 
\begin{align}
\Lambda^+_z(x,v) = 
(s_{\alpha,\sigma}(x_2) \cos 2\pi x_1) 2 \pi v_1 + \nonumber \\ (\sigma \alpha s_{\alpha-1,\sigma}(x_2) \cos(2\pi \sigma x_2) \sin 2\pi x_1)) 2 \pi v_2,
|v|=1.
\end{align}
Therefore, one can easily identify a nonempty singular set $\Gamma^\star$ corresponding
to all $x=(x_1,x_2)$ such that $\sin 2\pi \sigma x_2=0$, i.e. $x_2 = \frac{j}{2\sigma},
j \in \mathbb{Z}$ such that $(-1)^j \sin (2 \pi x_1) > 0$ or equivalently
introducing the two open intervals $I_{k,1} = (2k \pi,(2k +1)\pi)$, $I_{k,2} = 
I_{k,1} - \pi$ and $Y_{p,1} = \frac{p}{\sigma}$, $Y_{p,2} = Y_{p,1} + \frac{1}{2\sigma}$:
$$
\Gamma^\star = \bigcup_{j,p \in \mathbb{Z}^2} I_{j,1} \times \{Y_{p,1} \}
\cup I_{j,2} \times \{Y_{p,2} \}.
$$
For all $x \in \Gamma^\star$, one has $\Lambda^+_z(x,v) = +\infty$ provided $v_2 \neq 0$.
The stable set ${\cal W}^-(\Gamma^\star)$ is more difficult to characterise and depends crucially on $\mu$, but it includes
$\Gamma^\star$ by definition. We claim that, at least for the parameters chosen in Fig.~\ref{figex3}, 
${\cal W}^-(\Gamma^\star) = \mathbb{R}^2$,
so that all initial conditions meet the singular set in finite time. 
This system is obviously $\SP$, leading to the formation of space-filling beehive-like patterns. The hexagonal cells are constrained by the set $\Gamma^\star$ in the $x_2$-direction and by $\mu$ in the 
$x_1$
direction.
These patterns result from a periodic alternation: pairs of trajectories merge into one or split into two, dictated by the 
$x_1$  
position and driven by alternating phases of contraction and expansion.
The number of trajectories reaching the axis $x_1 = t$ with initial condition at $t_0$ scales like
 $2^{\lfloor \lambda(\sigma,\mu) (t-t_0) \rfloor}$. We do not show the distribution at fixed time, it can be guessed from Fig.~\ref{figex3} as simply corresponding to a sum of Dirac terms with complicated weights.

\subsection{PDE case: Lagrangian versus Eulerian Spontaneous Stochasticity}\label{subEuler_Lag}
As explained elsewhere, Lagrangian Spontaneous Stochasticity (LSS) is: 1) originally framed for passive scalar transport with vanishing diffusivity; 2) tantamount to anomalous diffusion (\cite{Drivas17}); 
3) tied to dissipative weak solutions of $({\cal P}_0)$—solutions to the inviscid problem that break energy conservation.
This raises the question: how does our Definition \ref{SPST} relate to LSS? Here, $\SP$ should be read as an Eulerian version of spontaneous stochasticity in the PDE setting. Since this connection depends on the chosen observable, we denote
$\SP\left|_{{\cal O}}\right.$
to emphasise that a continuous observable ${\cal O}$ exists such that (\ref{killeress}) is satisfied.

The PDE context requires special care, as assumption (\ref{eq:H0}) must be reformulated for infinite-dimensional settings. The problem $({\cal P}_\kappa)$ becomes $\dot u = F(u,\kappa),\ u(0)=u_0$, with $F : H \times \mathbb{R}^+ \to H$ is now an operator. Since many results are better expressed in the language of dynamical systems, it makes sense to consider the semigroup associated with the PDE. Compactness, frequently used in earlier sections, should involve weaker topologies. Extending the results from Section \ref{secmea} may be even more difficult. We will not pursue this further, as the details depend heavily on the specific PDE.
Still, to get a feel for the new phenomena in the PDE setting, recall that in finite dimensions, linear systems can never be $\SP$ under (\ref{eq:H0}). As shown below, this is no longer necessarily true for the linear transport PDE.

With that said, the following is a formal discussion. Anomalous diffusion in (\ref{ad_dif0}) is defined by
$\limsup_{\kappa \to 0} \kappa
||\nabla \theta||_{L^2_{t,x}}^2 > 0$ as explained in Section \ref{sec_AV}.
Why not define anomalous diffusion using the liminf instead? A sharp answer is provided in recent work by \cite{colombo} (see also \cite{Titi2023}), where they construct a transport system such that one subsequence converges to an energy-conserving weak solution, and another to a dissipative one (Theorem B).
This implies that the liminf is zero, while the limsup remains positive. But this is exactly the setup for Eulerian Spontaneous Stochasticity. In fact, Definition \ref{SPST} asks for an observable such that liminf $<$ limsup. This is simply rephrased as
$$
\SP\left|_{{\cal L}_2}\right. \Rightarrow \mbox{LSS},
$$
where ${\cal L}_2:
\theta(t,\cdot) \mapsto ||\theta(t,\cdot)||_{L^2(\mathbb{T}^d)}$.
Still, it is plausible that there are observables ${\cal O}$ for which $\SP\left|_{\cal O}\right. \not\Rightarrow {\rm LSS}$. Of course, it must involve a solution which conserves energy.

A notable counterexample to the converse, i.e.  
${\rm LSS} \not\Rightarrow \SP\left|_{{\cal L}_2}\right.$,
is the Kraichnan model.
Let us briefly explain why. The Lagrangian formulation of the advection-diffusion PDE for a passive scalar, say $\theta^\kappa(x,t)$ with
$\kappa$ the diffusion, leads to the SDE 
$dX_s = {\bf b}^\omega(X_s,s) ~ds + \sqrt{2\kappa} dW_s, X_t = x, s \leq t$. Here, ${\bf b}^\omega$ is the divergence-free H\"older Kraichnan velocity field for some fixed realisation $\omega$, $X_s=X(s;t,x)$ corresponds to 
a particle's position at time $s$ going through the position $x$ at time $t$. By the Feymann-Kac formula,
we have $\theta^\kappa(x,t) = \mathbb{E}[\theta_0(X_0)] = \Int p^\kappa(y,0|x,t)\theta_0(y)~~dy$
where $p^\kappa$ is the transition density.
It turns out that in the Kraichnan model the (weak) limit $p^\kappa \rightharpoonup p^0$ exists.
The field $\theta$ is therefore fully deterministic and there is no Eulerian Spontaneous Stochasticity.
However, the limiting density $p^0$ is nontrivial (i.e. not a Dirac distribution), implying Lagrangian Spontaneous Stochasticity \cite{Chaves2003}.

\section{Lack of selection principle in the Armstrong-Vicol passive scalar \cite{AV24}}
\label{Theirproof}
Like in the first part I, for the sake of consistency, we restate the lack of selection principle
in the vanishing diffusivity limit $\kappa \to 0$ proven in \cite{AV24}.
 \begin{propositionav}
 Fix $\beta \in [68/67,4/3)$. There exists a constant $C_\star = C_\star(\beta) \geq 1$ such that the following holds. For every parameters $A \in (0,1]$ and $B>1$, assume that $\Lambda$ is taken sufficiently large with respect to $\beta,A,B$ to ensure that
$$
C_\star \Lambda^{-\delta} \leq \min \left\{ 
A,B^{-\frac{2}{2+\gamma}(2-\beta+\frac{2\beta}{q+1})}
\right\}.
$$
 Fix the sequence $\{\epsilon_m \}_{m \geq 0}$ according to (\ref{2_8av}). Choose an initial datum $\theta_0 \in \dot H^2(\mathbb{T}^2)$. Define the length scale 
$L_{\theta_0}:= \frac{||\theta_0||_{L^2(\mathbb{T}^2)}}{||\nabla \theta_0||_{L^2(\mathbb{T}^2)}}$ 
and let $m_\star \geq 1$ be the unique integer such that 
$\epsilon_{m^\star}
\leq C_\star L_{\theta_0}^{\frac{2}{2+\gamma-2q\delta}} < \epsilon_{m^\star}-1$.
Assuming that $\theta_0$ satisfies 
$$
\frac{||\nabla \theta_0||^4_{L^2(\mathbb{T}^2)}}{||\theta_0||^2_{L^2(\mathbb{T}^2)}
||\Delta \theta_0||^2_{L^2(\mathbb{T}^2)}} \geq A,~~{\rm and}~ C_\star L_{\Theta_0}^{\frac{2}{2+\gamma-2q \delta}} \leq B \epsilon_{m^\star},
$$
there exist two sequences of diffusivities $\{\kappa_m^{(1)} \}_{m \in \mathbb{N}}$ and
$\{\kappa_m^{(2)} \}_{m \in \mathbb{N}}$, both converging to 0 as $m \to \infty$, such that the corresponding solutions $\{\theta^{\kappa_m^{(1)}} \}_m$ and $\{\theta^{\kappa_m^{(2)}} \}_m$ of the advection-diffusion equation with initial data $\theta_0$ and drift ${\bf b}$, converge as $m \to \infty$ in $C^{0,\mu}((0,1);L^2(\mathbb{T}^2))$ (for some $\mu > 0$)
to two distinct weak solutions of the associated transport equation.
\end{propositionav}
As mentioned several time in this work, we believe that this result is even more striking than the main Theorem 1.1  on anomalous diffusion. Indeed, this result automatically implies
our Definition \ref{SPST} (see Section \ref{subsec1}). 
 
\section{Eulerian Spontaneous Stochasticity in the Armstrong-Vicol passive scalar}
\label{secMAIN}
We present here our own mathematical proof that the passive scalar $\theta^\kappa$
in (\ref{ad_dif0}) becomes Eulerian spontaneously stochastic in the sense of Definition \ref{SPST} as $\kappa \to 0$.  This proof is based on the \cite{AV23} version and
relies on key estimates in \cite{AV23}, in particular their
section 5. It differs in some aspects from Proposition 5.5 in \cite{AV24} above, we believe
that it is related to our choice of taking $m^\star=0$ which simplifies the demonstration.

While the result below is a corollary of their work and relatively simple in concept, it cannot be directly inferred from Theorem 1.1 of \cite{AV23} (or \cite{AV24}), as it stems from the proof strategy itself. The main objective is to demonstrate that 
$({\cal P}_0)$ in (\ref{ad_dif0}) satisfies Definition \ref{SPST}  
with respect to the regularisation function ${\cal A}(\kappa) = \kappa || \nabla \theta^\kappa||^2_{L^2((0,1) \times 
\mathbb{T}^2)}$ (see Definition (\ref{generic_F})). Note that, 
from the energy equality, it is also equivalent to choose ${\cal A}(\kappa)
= ||\theta^\kappa(1,\cdot)||_{L^2(\mathbb{T}^2)}$. Let ${\cal A}^+,{\cal A}^- = \limsup_{\kappa \to 0},\liminf_{\kappa \to 0} {\cal A}(\kappa)$, $\kappa_m$ the renormalized diffusivity at scale $m$ (see Eq. (\ref{avseq})) and ${\cal A}_m^+,{\cal A}_m^- = \limsup_{\kappa \to 0},\liminf_{\kappa \to 0} 
\kappa_m ||\nabla \theta_m||^2_{L^2((0,1)\times \mathbb{T}^2)}$ (see details in (\ref{Anotations})).
Our result is the following: 
\begin{theorem} \label{LAMODUS}
Let $\epsilon > 0$, $m^\star = 0$, and
$\rho_0 := 
{\displaystyle \frac{||\nabla \theta_0||_{L^2_x}^2}{||\theta_0||^2_{L^2_x}}}
\geq (2\pi)^2$. For $\Lambda$ large enough, larger than 
\be \label{LBHyp}
 \Lambda \geq \max \left\{ \left( \frac{C}{\epsilon} \right)^\frac{1}{\delta},
 \left( \pi \sqrt{\rho_0} \right)^\frac{q-1}{\beta}
 \right\},
\de 
then ${\cal A}^-_{m_\star} > 0$, ${\cal A}^- > 0$, and
\be \label{paslemodus}
\frac{{\cal A}^+}{{\cal A}^-} \geq e^{-4\epsilon} 
\frac{{\cal A}_{m_\star}^+}{{\cal A}_{m_\star}^-} > 1.
\de 
In particular one has (\ref{killeress}) for the problem (\ref{ad_dif0}).
\end{theorem}
\noindent {\bf Proof}: 
during the preparation of the manuscript, an official published version of \cite{AV23} has appeared as \cite{AV24}. An explicit proof demonstrating the existence of two distinct subsequences in the inviscid limit $\kappa \to 0$, and thereby confirming that (\ref{killeress}) holds, can now be found in Proposition 5.5 of \cite{AV24} (see above). 
For the sake of consistency, we have still decided to retain our proof, as it differs in some aspects, but we have postponed it to Appendix \ref{APP_LAMODUS}.
\\\\
Consider two subsequences $(\kappa_{1,j})_{j \in \mathbb{N}}$, and
$(\kappa_{2,j})_{j \in \mathbb{N}}$, $\kappa_{1,j},\kappa_{2,j} \in {\cal I}_j$ (see definition
\ref{Kset}) such that 
$$
\varrho^2_p:= \lim_{j \to \infty} \kappa_{p,j} \frac{ 
||\nabla \theta^{\kappa_{p,j}}||_{L^2}^2}{||\theta_0||_{L^2}^2}  > 0, p \in \{1,2\}
$$
As a consequence of Theorem \ref{LAMODUS}, one can find two subsequences such that $\varrho_1 \neq \varrho_2$. One thus obtains 
the lower bound
\begin{property} \label{diffk}
One has 
$\varrho_1^2,\varrho_2^2 \in [\varrho^2,\frac12]$, where $\varrho^2$ is such that
$\inf_{\kappa \in {\cal K}_A} ||\nabla \theta^{\kappa}||_{L^2}^2 \geq \varrho^2 
||\theta_0||^2_{L^2}$ and $\varrho^2 \in (0,\frac12]$ (see Theorem 1.1 of \cite{AV23}),
\be 
\lim_{j \to \infty}  ||(\theta^{\kappa_{1,j}}- \theta^{\kappa_{2,j}})(1,\cdot) ||_{L^2}
\geq 2 ||\theta_0|| ~\varrho |\varrho_1-\varrho_2| > 0.
\de 
\end{property}
\noindent{\bf Proof}: see Appendix \ref{proofdiffk}.
\\\\
As mentioned early in \cite{AV23}, this is a demonstration that there is no selection principle
in the zero dissipation limit and that (\ref{ad_dif0}) has non-unique bounded weak solutions. In our view, the selection principle is now in term of measures in the space of weak solutions with a well-defined distribution. This distribution can be inferred from Lemma \ref{lemmasharp} by looking at the $L^2$ norm of $\theta$.
Let us consider a subsequence $(\kappa_{\omega,j})_{j \in \mathbb{N}}$ with 
$\kappa_{\omega,1} = \omega$ and $\kappa_{\omega,j} \in {\cal I}_j$ and
such that 
$\lim_{j \to \infty} \kappa_{m^\star}(\kappa_{\omega,j}) = K_\star^\omega$.
One has
\be \label{Lom}
||\theta_{m^\star}(1,\cdot)||_{L^2} = \left( \sum_{\bf k \in \mathbb{Z}^2}
|\widehat{\theta}_{0,{\bf k}}|^2 e^{-|{\bf k}|^2 K_\star^\omega}
\right)^\frac12 := {\cal L}(K_\star^\omega) := {\cal L}^\omega.
\de 
From Lemma \ref{lemmasharp}, we infer that $\forall \epsilon > 0$, one can choose
some $\Lambda$ large enough like in Theorem \ref{LAMODUS} such that
\be 
\left| ~||\theta(1,\cdot)||_{L^2} - {\cal L}^\omega ~\right| \leq (e^{2\varepsilon}-1) \varsigma(\theta_0)||\theta_0||^2_{L^2},
\de 
where $\varsigma(\rho_0)$ is a constant which depends on $\theta_0$ but not $\varepsilon$.
Of course the random variable ${\cal L}^\omega$ has a nontrivial density probability which is not a Dirac measure. We will show in the numerical Section \ref{num} how those densities look like.

\section{Numerical behaviour of the renormalisation sequence}\label{num}
Let $c_0 =  \frac{9}{80}, \Lambda > 1$, $\beta \in (1,\frac43)$
and $q = \frac{\beta}{4(\beta-1)}$, we consider the sequence
(\ref{avseq}) (see notations Appendix \ref{AppdefI}), that we rewrite 
$$
\kappa_{m-1} = F_m(\kappa_m),~~\kappa_M = \kappa,~~{\rm with}~~
F_m(\kappa) := \kappa + 
c_0 \frac{\epsilon_m^{2\beta}}{\kappa}.
$$
The diffusivities are described by a two-parameter semigroup
$\kappa_m = \varphi(m,M) \kappa$ and we are interested in the
pullback limit, for $m$ fixed, 
\begin{equation}\label{seq}
    {\cal K}(m;\kappa) = \lim_{M \to \infty} \varphi(m,M) \kappa.
\end{equation}
We then need to understand the behaviour of
${\cal K}(m;\kappa)$ when
$\kappa \to 0$. Of practical importance is the fact that
the double asymptotics, $M \to \infty$ first, then $\kappa \to 0$
, can be replaced by a single asymptotics $\kappa \to 0$ using
\begin{equation}
    M(\kappa) ~~\mbox{such that}~~\sqrt{c_0} \epsilon_{M(\kappa)}^\beta \propto  \kappa.
\end{equation}
Indeed, $F_m(\kappa) \sim \kappa$ unless
$c_0 \epsilon_m^{2 \beta}/\kappa$ has the same order of magnitude
than $\kappa$.\\\\
Let
$z_m := \frac{\kappa_m}{\sqrt{c_0} \epsilon_m^\beta}$,
$z_M = \frac{\kappa}{\sqrt{c_0} \epsilon_M^\beta}$,
then 
\begin{equation}\label{seqz}
    z_{m-1} = G_m(z_m),~~{\rm with}~~ G_m(z) := g_m
    \left(z + \frac1z \right), g_m = \left(\frac{\epsilon_m}{\epsilon_{m-1}} \right)^\beta.
\end{equation}
In the geometric case, the scale separation is given by $\epsilon_m = \lceil \Lambda^m \rceil^{-1}$. Assuming, for simplicity, that $\Lambda \in \mathbb{N}$, it follows that $g_m = \Lambda^{-\beta}$. Therefore, Eq. (\ref{seqz}) becomes autonomous.
It is important to note that, in this case, the estimates presented in \cite{AV23} are no longer valid.
Still, it is very instructive to look at the behaviour of the diffusivities sequence. A close inspection shows that
the diffusivities converge in the small $\kappa$ limit:
\begin{equation}
    {\cal K}_m := \lim_{\kappa \to 0} {\cal K}(m;\kappa)
    = \Lambda^{-\beta m} \sqrt{\frac{c_0}{\Lambda^\beta-1}}, \forall m \geq 0.
\end{equation}

\begin{figure}[H]
\centerline{\includegraphics[width=0.95\columnwidth,keepaspectratio]{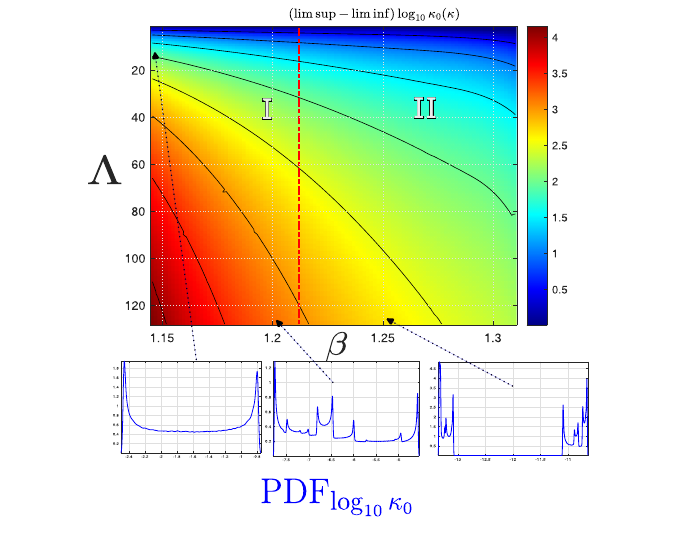}}
\centerline{\includegraphics[width=0.95\columnwidth,keepaspectratio]{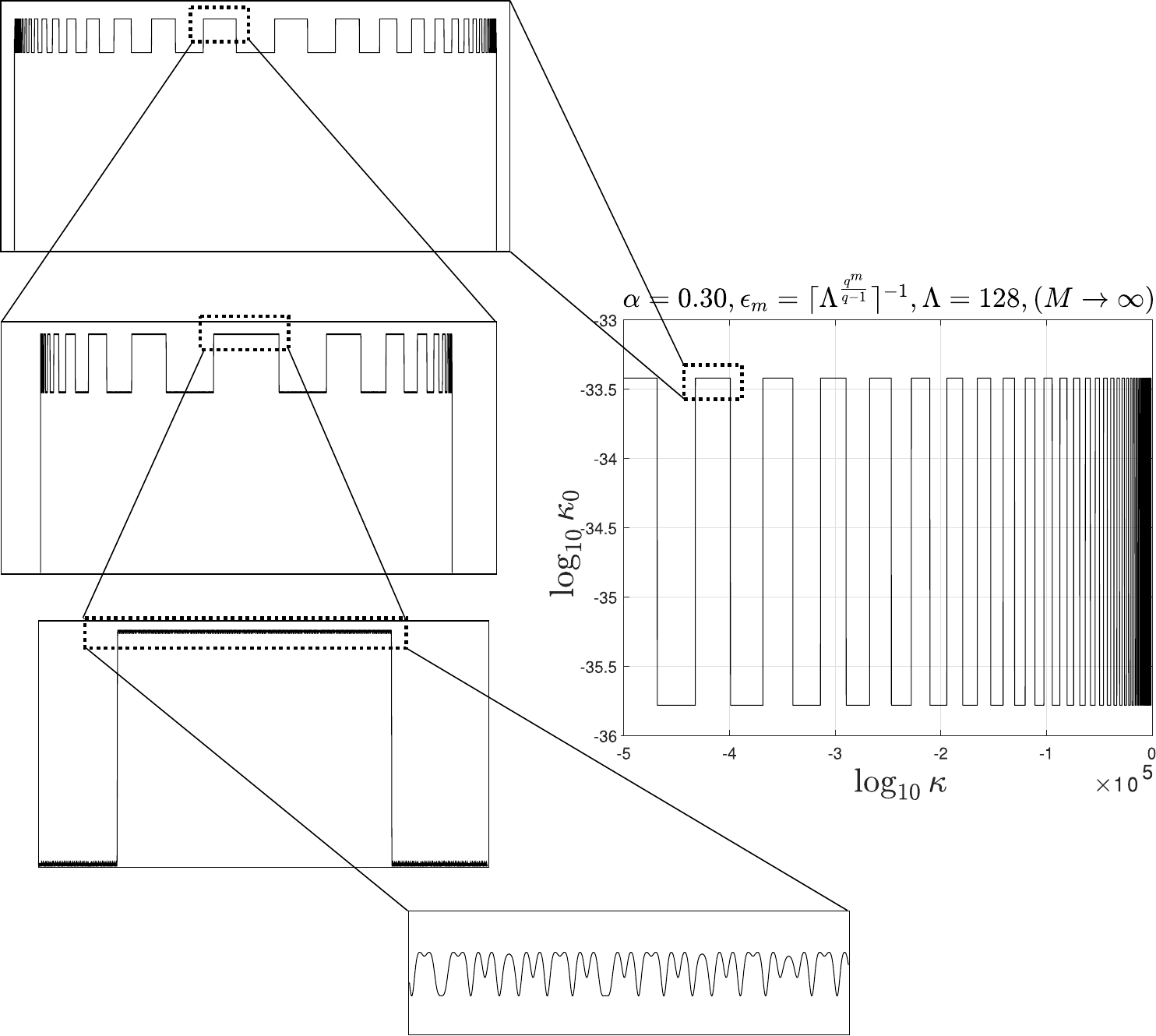}}
\caption{{Upper} panel: standard deviation of $\log_{10} \kappa_0$ as a function of $\beta$ and
$\Lambda$. Regime I corresponds to $\kappa_0$ distributions having a
simply-connected compact support,
for $1.14 \lessapprox \beta \lessapprox 1.21$, and regime II to a bimodal qualitative behaviour of the $\kappa_0$ distribution. Three probability densities for the limit $\kappa \to 0$ are shown for $\beta=1.20$, $\beta=1.25$ ($\Lambda=128$) and $\beta = 1.143$ ($\Lambda=10$). {Lower} panel: behaviour of $\log_{10} \kappa_0$ in $\log \kappa$ scale for $\beta=1.30$, $\Lambda=128$ with successive zooms revealing the fractal structure. Note that the frequency slows down suggesting a $\log \log$ scaling. Numerics are achieved using BigFloat precision.}
\label{fig:betaLBall}
\end{figure}
Therefore, if one were able to establish a link between the renormalised sequence and the inviscid system in the geometric case as well, it would follow that the system is well-posed, yielding an atomic (Dirac) distribution, i.e. the absence of Eulerian spontaneous stochasticity.

A radically different situation occurs when one considers an hypergeometric scale separation. The sequence (\ref{seqz}) becomes nonautonomous with
\begin{equation}
  G_m(z) = \Lambda^{-\beta q^{m-1}} \left( z + \frac{1}{z} \right).
\end{equation}
Figure \ref{fig:betaLBall} illustrates the behaviour of $\limsup - \liminf > 0$ of the function $\kappa \mapsto \kappa_0(\kappa)$ as a function of the parameters $\beta$ and $\Lambda$ in the limit $\kappa \to 0$. The corresponding distributions are shown on a logarithmic scale. The lower panel presents the function itself for selected parameter values, highlighting its intricate fractal structure. 
In general, $\kappa_0$ is not of order unity and depends significantly on $\beta$ (and $\Lambda$). It is $O(1)$ only in the limit $\beta \to 1^+$. As $\beta \to \frac43$, $\kappa_0$ becomes increasingly small -- dropping by several order of magnitude. Moreover, we numerically identify a critical value $\beta^\star \approx 1.1428$ above which the constraint permissible set ${\cal K}$ is not required to control the sequence. Figure
\ref{fig:betaLBall} uses $\beta$ values above this critical threshold $\beta^\star$. Additional numerical details on these aspects are provided in Appendix \ref{Odeun}.

These results can be reinterpreted using the concepts of Section \ref{secmea}. 
The function $\kappa \mapsto {\cal K}(0;\kappa)$ in (\ref{seq}),
which emerges from the proof in \cite{AV23,AV24}, plays the role of a scalar regularisation function ${\cal A}$. In light of the discussion in Section \ref{secmea}, in order to have a well-defined probability measure, the ambient measure must be in $\log \log$ scale as suggested by Fig.~\ref{fig:betaLBall}. Then, the $\kappa_0$ distribution corresponding to ${\cal A}(0)$, is the weak limit of $\kappa_0$ in the inviscid limit w.r.t. to the ad-hoc ambient measure. Moreover, the available observable takes the form of a
discrete nonautonomous dynamical system. Due to that, we can only conjecture, based on
numerical simulations, the existence of a 
nontrivial (pullback) attractor with fractal dimension. A rigorous treatment of the nonautonomous case would likely require at least a skew-product interpretation of the dynamics
(see e.g. \cite{RDS1998}). We have not explored these aspects further and leave them for future work.

\section{Conclusion and perspectives}\label{Conclu}
This work has two parts. First, we provide a detailed historical review of spontaneous stochasticity.
We then describe the Armstrong-Vicol model, emphasising the Lagrangian interpretation of their construction as well as the RG phrasing. We then show clear numerical evidences for anomalous diffusion.

In the second part, we introduce a very general definition of spontaneous stochasticity ($\SP$: see Definition \ref{SPST}). Our analysis is restricted to deterministic and autonomous regularisations of the inviscid system. This framework is, in effect, a reformulation of the absence of a selection principle for the inviscid limit -- that is, a selection principle that may only hold along subsequences, for fixed time and fixed initial data (see Section \ref{subsec1}). The definition makes explicit the requirement that regularisations approach the inviscid system in a rapidly oscillatory manner.

We then examine the singular limit from a statistical perspective in Section \ref{secmea}, focusing on the conditions under which a probability measure can be defined on the space of solutions at the limit. For a given regularisation, the resulting probabilistic limit depends crucially on the choice of an ambient measure -- or, equivalently, on the way the regularisation is parameterised. As a consequence, no selection principle exists in general at the level of probability measures either. It is only under particular circumstances that a {\it measure selection principle} holds, i.e., a situation in which the statistics converge to a unique probability measure.

This motivates a broader viewpoint developed in Section \ref{secmea}. We consider the space ${\cal M}$ of probability measures obtained along subsequences across all admissible regularisations of a fixed inviscid system (see Eq. (\ref{Mg})). One of the main results, stated in Theorem \ref{M0M}, shows that spontaneous stochasticity can generate any statistical behaviour consistent with the fixed-time and fixed-position statistics of the inviscid system, represented by the space ${\cal M}_0$ (see Eq. (\ref{calM0})). In other words, we establish that ${\cal M} = {\cal M}_0$. Remarkably, the proof is constructive and provides a near-explicit procedure for building regularisations targeting any probability measure in ${\cal M}_0$ (see Appendix \ref{M0Mproof}).

The space ${\cal M}$ of measures generated via spontaneous stochasticity is shown to be the convex hull of the Dirac measures associated with regularisations that are not spontaneously stochastic. This characterisation enables a rigorous definition of universality classes of regularisations, identified by the statistical behaviour they produce and distinguished solely through their ergodic properties (see Eq. (\ref{B_mu})). This framework further permits the introduction of a continuous (semi)group of renormalisation acting on the space of regularisations associated to a given inviscid system. This dynamical system corresponds to the Bebutov flow (see Definition \ref{RegBebu}). In this reformulation, it becomes apparent that chaotic features in the inviscid system should give rise to broader and statistically more robust universality classes. This important aspect will be explored in future work.

We emphasise that in our deterministic framework, stochasticity can emerge in its purest form in the singular limit, entirely independent of whether the regularisations themselves are stochastic. See Subsection \ref{stochno} for details.

A direct consequence of these results is that, as soon as ${\cal M}_0$ fails to be a singleton, there exists at least one regularisation (and, in fact, as an indirect consequence of Kneser's theorem, there are infinitely many) for which spontaneous stochasticity occurs. Since this situation arises if and only if the inviscid system admits non-unique trajectories, we seek a necessary condition for such non-uniqueness in Section \ref{which}. To this end, we employ a dynamical characterisation based on Dini derivatives, which allows one to detect where (in phase space) non-Lipschitz singular sets emerge in the inviscid system and for which initial data (Theorem \ref{CN}).
These theoretical results are illustrated through several nontrivial numerical examples (Subsections \ref{Ex1}--\ref{Ex4}).

Finally, we show that the advection-diffusion transport equation for a passive scalar proposed by \cite{AV23,AV24} is also spontaneous stochastic in the sense of Definition \ref{SPST}. Thus, it exhibits not only anomalous diffusion, or equivalently Lagrangian spontaneous stochasticity, but also Eulerian spontaneous stochasticity (Sections \ref{sec_AV} and \ref{secMAIN}).
We add that the peer-reviewed version \cite{AV24} confirms this finding (their Section 5.5).
It is described as a lack of selection principle rather than the passive scalar being Eulerian Spontaneous Stochastic. These two notions are rigorously equivalent (see Section \ref{subsec1}).
We also provide numerical evidences that the model exhibits Eulerian spontaneous stochasticity but are unable to observe it within direct numerical simulations, limited somewhat by the lack of resolution of the very small scales due to the hypergeometric constraint.
On the other hand, the renormalised diffusivity sequence is well illustrated and exhibit nontrivial behaviour.

Several comments are in order. At least six important papers have recently appeared
\cite{DeLellis2021,Drivas_Elgindi,colombo,AV23,Titi2023,burczak2023anomalous} (in chronological order).
All of them address the inviscid limit $\kappa \to 0$ of the linear transport equation $\partial_t \theta
+ {\bf b} \cdot \nabla \theta = \kappa\Delta \theta, \theta(0,\cdot)=\theta_0(\cdot)$ where ${\bf b}$
is a divergence-free velocity field. Each paper considers different choices for  ${\bf b}$ and  focuses on a rigorous mathematical proof of anomalous diffusion: $\limsup_{\kappa \to 0} \kappa ||\nabla \theta||^2_{L^2_{x,t}} > 0$ (except for \cite{DeLellis2021}, which addresses the selection mechanism only).

The works of C. De Lellis, V. Giri \footnote{
who, together with R.-O. Radu, recently proved the Onsager conjecture for the 2D Euler equations \cite{Giri}}, T.D. Drivas et al. \cite{DeLellis2021,Drivas_Elgindi} (see also \cite{colombo}) are, to our knowledge, the first rigorous proofs that the transport PDE can exhibit Eulerian spontaneous stochasticity in the limit $\kappa \to 0$ according to our interpretation (see Section \ref{subEuler_Lag}). These studies demonstrate that, in addition to having anomalous diffusion, the inviscid limit admits multiple distinct solutions.

However, these results are often framed negatively as evidence of the failure of a selection principle or the nonuniqueness of solutions in the inviscid limit. In contrast, we emphasise that Eulerian spontaneous stochasticity is precisely equivalent to the absence of a selection principle, that is, whenever one can distinguish distinct
subsequences. We recognise that this definition -- expressed simply as the absence of a selection principle -- is fundamentally equivalent to ours. While we do not claim anything beyond this, we wish to state it explicitly and in greater detail.

In \cite{Titi2023}, the authors once again identified the phenomenon of Eulerian spontaneous stochasticity (i.e., non-unique convergent subsequences in the inviscid limit) in the transport PDE, as previously observed in \cite{Drivas_Elgindi,colombo} but for different velocity fields. Their finding, however, raises an important question. They constructed a particular vector field ${\bf b}$ such that the vanishing limit $\kappa \to 0$ selects a non-physical solution (a "wild Euler weak solution") that unexpectedly unmixes back to its initial state. 

These earlier works thus highlight a fundamental question: the potential lack of realism in the inviscid limit and its implications for physics. 
However, since the velocity field lacks realism and a lack of a selection mechanism appears in the inviscid limit, what should be questioned is not the absence of a selection mechanism itself, but rather the way it manifests nonphysically through the emergence of non-entropic solutions.

The study in \cite{AV23} takes this a step further by constructing a significantly more realistic vector field than those in \cite{Drivas_Elgindi, colombo, Titi2023}. In particular, it remains arbitrarily close to a $C^\alpha$ weak Euler solution for $\alpha \in (0,1/3)$, making it, in a loose sense, almost an Euler PDE solution. As announced in \cite{AV23} and now proven in \cite{AV24}, our work provides additional mathematical evidence that the inviscid limit leads to nonunique solutions -- once again confirming the presence of (Eulerian) spontaneous stochasticity.

The most advanced contribution is likely the recent work in \cite{burczak2023anomalous}. It shares many features with the fractal homogenisation approach in \cite{AV23,AV24} but employs in addition convex integration scheme for constructing the velocity field using Mikado-like flows. In doing so, they
are able to construct a large class of velocity fields ${\bf b}$ which are now
exact weak solution of the 3-D Euler equations with optimal Onsager regularity $C^{1/3^-}$ and such that they induce anomalous diffusion for the passive scalar. It is unclear at the present time, whether there is Eulerian Spontaneous Stochasticity or not in this model. We strongly believe it is so. {Indeed, they obtain a sequence of renormalised diffusivities similar to the one of \cite{AV23,AV24} with an additional space and time dependence. This sequence retains the same ingredients (e.g hyper-geometric scale separation and non-autonomous behaviour) as the one of  \cite{AV23,AV24}, leading us to think that the passive scalar will be spontaneously stochastic.} More advanced investigations are required for confirming this. 

From a physicist's perspective, the lack of selection principle for 3-D Euler equations, was intuited long ago by Lorenz \cite{Lorenz69}.
This insight marks a paradigm shift in how we perceive turbulence: as a multi-world realm with possibly universal statistical aspects. 

We also stress once again that chaos and spontaneous stochasticity are not the same phenomenon, though they often appear together. To distinguish between them, especially in turbulence studies, it is often more convenient to examine the behaviour of Lyapunov exponents (or their ansatz) in the inviscid limit as Re
$\to \infty$: chaos yields finite positive values, whereas spontaneous stochasticity leads
to (positive) infinite values. This is of course, a direct rephrasing of our Definition \ref{SPST}. 

A key aspect, particularly in the second part of our work, is the crucial role of singularities. We have emphasised that a necessary condition for spontaneous stochasticity is that the inviscid system must hit a singular set (in phase space) in finite time. Therefore, we believe that investigating the existence of such sets, though still challenging, is a more achievable goal. Once these sets are identified, the next question concerns renormalisation. This is precisely the focus of the RG program explored in \cite{Maily2012, Maily2016, AM_24, AM_24end, Bandaketal_PRL24}, albeit in the context of simpler (shell) models of turbulence such as Sabra.

Even in the most favorable scenario -- where the system encounters an isolated H\"older singularity (a non-Lipschitz critical point) -- severe "hell-after-death" challenges immediately arise. 
Indeed, there are infinitely-many possible dynamical scenarios which can hide
within an isolated singularity.  
We believe that the nature of singular sets in the Euler equations, and even in shell models, must be highly complex to capture multifractality and the emergence of possibly non-unique self-similar blowups \cite{CiromeSimon}.
These difficulties can be managed using renormalisation techniques, as derivatives diverge, indicating that, at the very least, understanding non-smooth singular vector fields under renormalisation is crucial. 

\section*{Acknowlegment} We would like to thank S.Thalabard and J.Bec for stimulating discussions on the subject and the IDEX summer school "100 years of cascades", from which this project has started. We also gratefully acknowledge the Calisto team at INRIA for their warm hospitality and continuous support throughout the project.

\bibliography{REFESS.bib}
\bibliographystyle{plainnat}

\addtocontents{toc}{\protect\setcounter{tocdepth}{0}}  

\section{Appendix Part I}\noindent
\subsection{Notations and definitions}\label{AppdefI}
This section introduces various definitions used in \cite{AV23,AV24}.
\begin{center}
\begin{tabular}{|l||l|l|l|l|}
\hline
 & definition & $\inf \alpha$ & $\sup \alpha$ & meaning \\
\hline 
$\alpha$ & $\alpha$ & $0$ & $\frac13$ & velocity H\"older exponent \\
$\beta$ & $\alpha+1$ & $1$ & $\frac43$ & streamfunction regularity \\
$q$     & $\frac{\beta}{4(\beta-1)}$ & $+\infty$ & $1$ & rate of scale separation \\
$\delta$ & $\frac{(q-1)^2}{4(q+1)(4 q-1)}$ & $\frac{1}{16}$ & $0$&  \\
$\gamma$ & $\frac{q-1}{q+1} \beta$ & $1$ & $0$&  \\
\hline
\end{tabular}
\end{center}
\begin{itemize}
\item[$\bullet$] the minimal scale separation $\Lambda \in [2^7,\infty)$,
\item[$\bullet$] \emph{the (hypergeometric) scale separation} $\epsilon_m$
\be \label{2_8av}
\epsilon_m^{-1} = \lceil \Lambda^{\frac{q^m}{q-1}} \rceil,~~m \in \mathbb{N}.
\de
\item[$\bullet$] Finally, we define a slightly different permissible set than in \cite{AV23,AV24}, for some fixed $A > 1$:
\be \label{Kset}
{\cal K}_A = \bigcup_{j=1}^\infty {\cal I}_j,~~{\rm with}~{\cal I}_j = \left[ \frac{1}{A},A  \right] \sqrt{c_0} \epsilon_j^\frac{2\beta}{q+1},~~c_0 = \frac{9}{80}.
\de
Appendix \ref{proofs_ESS_AV} provides detailed proofs using this set, and in particular how the sequence (\ref{avseq}) stays under control.
\item[$\bullet$] Let $R_{\theta_0} > 0$, 
we consider an initial scalar $\theta_0$ which satisfies the analyticity condition:
\be
\max_{|\AL| = n} || \partial^{\AL} \theta_0||_{L^2} \leq ||\theta_0||_{L^2} \frac{n!}{R_{\theta_0}^n}, \forall n \in \mathbb{R}.
\de
\item[$\bullet$] One assumes in the following $m \in \{ m^\star,\cdots,M\}$ with
\be \label{mstar}
m^\star = \min \left\{ m \in \mathbb{N}, m \geq 2, \epsilon_{m-1}^{1+\frac{\gamma}{2}} 
\leq R_{\theta_0} \right\} - 1.
\de
We will discuss afterwards the possibility to consider $m^\star=0$.
\end{itemize}

\subsection{How to numerically implement the Armstrong-Vicol model}\label{ALGO} 
In the following section, we present the numerical methods used to study the Armstrong-Vicol model, with particular emphasis on the numerical computation of the vector field, which is the most challenging aspect. Before proceeding, we briefly recall the construction of the streamfunction as detailed in \cite{AV24}. The streamfunction $\phi$ of the two dimensional vector field $\vect= \nabla 
^\perp\phi$ is defined as the limit of a sequence
of streamfunctions $( \phi_m)_{m\geq 0}$ which are generated recursively according to the induction relation for $m \geq1$:
\begin{equation}
\begin{cases}\label{eq:DefAVField}
& \hspace*{-0.25cm} \phi_{m}(t, \pos)=\displaystyle\phi_{m-1}(t,\pos)+ \sum \limits_{l\in \mathbb{Z}} \sum \limits_{k\in \mathbb{Z}} \zeta_{m,k}(t) \hat{\zeta}_{m,l}(t)\psi_{m,k}(\flow^{-1}_{m-1,l}(t,\pos,l \tau_m'')), \\
& \hspace*{-0.25cm} \phi_0(t,\pos)=0,\quad (t,\pos) \in [0,1]\times \mathbb{T}^2.
\end{cases}
\end{equation}
The functions $\psi_{m,k}$ are the building blocks of the flow and are chosen in~\cite{AV24} to be,
$$\psi_{m,k}(\pos)= a_m \epsilon_m^2 \left[\mathbbm{1}_{k\in 4 \mathbb{Z}+1 } \sin\left( \dfrac{2\pi x_1}{\epsilon_m} \right) +  \mathbbm{1}_{k\in 4 \mathbb{Z}+3 } \sin\left(  \dfrac{2\pi x_2}{\epsilon_m} \right) \right]. $$
This streamfunction corresponds to a cosine shear flow with Lipschitz norm $a_m$
and period $\epsilon_m$. By definition, $\psi_{m,k}$ itself is time-independent.
The smooth alternation between shear flows in the $x_1$ and $x_2$-directions is controlled by the time cutoff function $\zeta_{m,k}$, which is smooth and compactly supported:
$$ \mathrm{Supp}\, \zeta_{m,k}  \subset \left[ \left(k -\frac{2}{3}\right) \tau_m, \left(k +\frac{2}{3}\right) \tau_m \right].$$
This implies that for any given value of $t$, there is at most one $k\in \mathbb{Z}$ such that $\zeta_{m,k}(t) \psi_{m,k}(\pos) \neq 0$. As mentioned in the Lagrangian description~\ref{sechom}, simply adding shear flows is not sufficient to generate anomalous diffusion. Therefore, the perturbation added to $\phi_{m-1}$ in \eqref{eq:DefAVField} 
is introduced within the reference frame of the vector field $\vect_{m-1}$. This is the role of $\flow^{-1}_{m-1,l} $ entering \eqref{eq:DefAVField}, which is the inverse flow map of the field $ \vect_{m-1}$ (on a given time interval).
The inverse flow map is obtained as solution of:  \be \label{eq:InverseFlow}
\displaystyle \partial_t \flow^{-1}_{m-1,l}+ \vect_{m-1}\cdot \nabla \flow^{-1}_{m-1,l}=0,~~
\flow^{-1}_{m-1,l_k}( l \tau_m'',\pos)=\pos,
\de 
with $(t,\pos) \in \left[\left(l -\frac{1}{2} \right) \tau_m'', \left(l +\frac{1}{2} \right) \tau_m'' \right] \times \mathbb{T}^2$.

The inverse flow map of the field $\vect_{m-1}$ is not computed on the whole $[0,1]$ time interval but rather on sub-intervals of length $\tau_m''\ll a_{m-1}^{-1}$ ensuring that the inverse flow maps are not deformed too much. These inverse flow maps are refreshed every $\tau_m''$ unit of time, the time cutoff $\hat{\zeta}_{m,l}$ ensures that this refreshing is done smoothly in time. The new timescale $\tau_m''$ must be homogenised as well, necessitating the introduction of a third timescale $\tau_m'\ll \tau_m''$. This time cutoff $\hat{\zeta}_{m,l}$ is therefore smooth and compactly supported, such that:
\begin{equation*}
      \mathbbm{1}_{t\in [(l-\frac{1}{2})\tau_m'' +2\tau_m',(l+\frac{1}{2})\tau_m'' -2\tau_m']} \le \hat{\zeta}_{m,l}\le \mathbbm{1}_{ t \in [(l-\frac{1}{2})\tau_m'' +\tau_m',(l+\frac{1}{2})\tau_m'' -\tau_m']}. \label{eq:supportzetahat}
\end{equation*}
In practice, for a given $t$ and $k\in \mathbb{Z}$, there is only one value of $l\in \mathbb{Z}$ such that $\zeta_{m,k}(t) \hat{\zeta}_{m,l}(t) \neq 0$. This implies that the sums over $k,l$ appearing in~\eqref{eq:DefAVField} can effectively be written as a sum over $k$ only. 

With this construction, the authors of~\cite{AV24} are able to prove that, roughly speaking, the $L^2$ norm drop of the passive scalar advected by $\vect_m$ and with diffusivity $\kappa_m$ is of the same order as that of the passive scalar advected by $\vect_{m-1}$ and with diffusivity $ \kappa_{m-1}= \kappa_m + c (a_m^2 \epsilon_m^4)/\kappa_m$. We emphasised in section~\ref{sechom} that this homogenisation occurs only if the scales $(\epsilon_m,a_m,\tau_m,\tau_m',\tau_m'')$ are carefully chosen. In particular, the following constraints must be satisfied:
\be
\begin{array}{llll}
\tau_m\ll \tau_m'\ll \tau_m''\ll a_{m-1}^{-1}, &
\dfrac{a_m\epsilon_m^3 }{\kappa_m} \ll \epsilon_{m-1}, & 
\dfrac{\epsilon_m^2}{\kappa_m} \ll \tau_m.
\end{array}
\de 
The goal of this procedure is to obtain a large-scale effective diffusivity $\kappa_0$ of order unity. We refer the reader to~\cite{AV24} for a more detailed discussion on these constraints. In particular, the H\"older regularity of the velocity field $\vect$ cannot be greater than $1/3$ for these constraints to be satisfied. 

In~\cite{AV24}, the sequence of length scales $\epsilon_m$ is chosen to decrease hyper-geometrically, and all other scales are built from $\epsilon_m$ such that the set of constraints is met. However, with the sequence used in~\cite{AV24}, the spatial resolution required to numerically compute $\phi_2$ would already be prohibitive, while $\phi_1$ is explicit and thus not numerically interesting. 
Therefore, we adopt a different choice of sequences, which retains the hyper-geometric decay of $(\epsilon_m)_{m\geq1}$, namely:
\be 
\begin{array}{llllllll}
\epsilon_m^{-1}= \left\lceil \Lambda^{  \frac{q^m}{q-1}} \right\rceil, &
a_m=\epsilon_m^{\beta-2}, & 
\tau_m= \tau_m', \\
\tau_m'= \frac{\tau_m''}{3(4\left\lceil \epsilon_m^{-\delta} \right\rceil +1) }, & 
\tau_m''=2^{-2} \left\lfloor\frac{a_{m-1}}{ \epsilon_m^{2\delta}}\right\rfloor^{-1},
\end{array}
\de 
where $q$ and $\delta$ are defined in Appendix~\ref{AppdefI} and the choice of parameters $\Lambda$, $\beta$ and $M$ are detailed in Table~\ref{tab:ParamsB}. 

With this choice of sequences, the hypotheses of Theorem~(1.1) from~\cite{AV24} are not strictly satisfied. However, our numerical simulations show that the passive scalar still exhibits anomalous diffusion. As predicted in~\cite{Drivas17}, this anomalous diffusion of the passive scalar is tantamount to the spontaneous stochasticity of Lagrangian trajectories. We also give the full algorithm below. The purpose of the Algorithm~\ref{alg:cap} is to compute the streamfunction $\phi_m$. For this we need to integrate the backward flow map equation~\eqref{eq:InverseFlow}. We do this using pseudo spectral method. However, we have that $\flow^{-1}_{m-1,l}(l\tau_m'',\pos)=\pos$ which obviously is not a periodic function. In order to tackle this issue we introduce $\mathbf{Y}:=\flow^{-1}_{m-1,l}-\mathbbm{1}$ which is periodic and satisfies 
\be\label{eq:PeriodicBackFlow}
\partial_t\mathbf{Y}=-\vect_{m-1}\cdot\nabla \mathbf{Y}-\vect_{m-1},~~\mathbf{Y}(l\tau_m'',\pos)=0,
\de 
where $(t,\pos)\in [(l-\frac12 )\tau_m'',(l+\frac12 )\tau_m'']\times \mathbb{T}^2$.
The variable $\mathbf{Y}$ in Algorithm~\ref{alg:cap} is therefore the numerical approximation of the solution of~\eqref{eq:PeriodicBackFlow} and the function $\mathrm{RHS}$ is the numerical approximation of the right hand side of~\eqref{eq:PeriodicBackFlow} computed using a pseudo-spectral dealiased differentiation.

\begin{algorithm}[hptb]
\caption{Compute $\phi_m$ given $\phi_{m-1}$}\label{alg:cap}
\begin{algorithmic}
\Require $\phi_{m-1}$, $m$, $\epsilon_m$, $\tau_m$, $\tau^{'}_m$, $\tau^{''}_m$
\State find the $n_{0}^l$ such that $t^{n_{0}^l}=l \tau_m^{''}, \, l \in \left \lbrace 0, \dots , 1/\tau_m^{''} \right \rbrace.$
\State find the $n_{\pm}^l$ such that $t^{n_{\pm}^l}=(l\pm1/2) \tau_m^{''}, \, l \in \left \lbrace 0, \dots , 1/\tau_m^{''} \right \rbrace.$

\For{$l$ ranging from $0$ to $1/\tau_m^{''}$}
\State $n_m \gets n_0^l$, $n_i \gets n_{-}^l$, $n_s \gets n_{+}^l$
\If{ $l\neq0$ and $l\neq1/\tau_m^{''}$ }
\State $ \mathbf{Y}_+, \mathbf{Y}_-  \gets \mathrm{zeros}(N_x,N_x) $  \Comment{Inverse flow maps going forward  $(t> l_k \tau_m^{''})$ and backward $ (t< l_k \tau_m^{''})$  }
\State $\phi_{m-1}^{n_0^l} \gets  \phi_{m-1}^{n_0^l} + \sum \limits_{k \in \mathbb{Z}} \hat{\zeta}_{m,l}^{n_0^l} \zeta_{m,k}^{n_0^l} \psi_{m,k} \left( \mathbf{Id} \right)$ \Comment{In practice the sum contains at most one non-zero element } 
\State $f^{\pm}_0 \gets $ RHS$(\mathbf{Y}_\pm,\vect^{n_0^l})$
\State $\mathbf{Y}_\pm \gets  \mathbf{Y}_\pm  \pm \delta t (c_{0,0} f_0^\pm) $ \Comment{The $c_{i,r}$ are the coefficient of A-B order $r$ scheme}
\State $f^{\pm}_1 \gets $ RHS$(\mathbf{Y}_\pm,\vect^{n_0^l\pm 1})$
\State $\phi_{m-1}^{n_0^l\pm 1} \gets  \phi_{m-1}^{n_0^l \pm 1} + \sum \limits_{k \in \mathbb{Z}} \hat{\zeta}_{m,l}^{n_0^l\pm 1} \zeta_{m,k}^{n_0^l \pm 1} \psi_{m,k} \left(\mathbf{Y}_\pm + \mathbf{Id} \right)$ 
\State $\mathbf{Y}_\pm \gets  \mathbf{Y}_\pm  \pm \delta t (c_{0,1} f_1^\pm +c_{1,1} f_0^\pm) $

\State $f^{\pm}_2 \gets $ RHS$(\mathbf{Y}_\pm,\vect^{n_0^l\pm 2 })$
\State $\phi_{m-1}^{n_0^l\pm 2} \gets  \phi_{m-1}^{n_0 ^l\pm 2} + \sum \limits_{k \in \mathbb{Z}} \hat{\zeta}_{m,l}^{n_0^l\pm2} \zeta_{m,k}^{n_0^l \pm 2} \psi_{m,k} \left(\mathbf{Y}_\pm + \mathbf{Id} \right)$ 
\State $\mathbf{Y}_\pm \gets  \mathbf{Y}_\pm  \pm \delta t (c_{0,2} f_2^\pm +c_{1,2} f_1^\pm +c_{2,2} f_0^\pm) $

\State $f^{\pm}_3 \gets $ RHS$(\mathbf{Y}_\pm,\vect^{n_0^l\pm 3 })$

\State $\phi_{m-1}^{n_0^l\pm 3} \gets  \phi_{m-1}^{n_0^l \pm 3} + \sum \limits_{k \in \mathbb{Z}} \hat{\zeta}_{m,l}^{n_0^l\pm3} \zeta_{m,k}^{n_0^l \pm3} \psi_{m,k} \left(\mathbf{Y}_\pm + \mathbf{Id} \right)$ 
\For{$n_t$ ranging from $4$ to $n_\pm^l-n_0^l$}
\State $n_\pm \gets n_0^l \pm n_t$
\State $\mathbf{Y}_\pm \gets  \mathbf{Y}_\pm  \pm \delta t (c_{0,3} f_3^\pm +c_{1,3} f_2^\pm +c_{2,3} f_1^\pm  +c_{3,3} f_0^\pm) $
\State $ f_0^\pm \gets f_1^\pm$, $ f_1^\pm \gets f_2^\pm$, $ f_2^\pm \gets f_3^\pm$
\State $f^{\pm}_3 \gets $ RHS$(\mathbf{Y}_\pm,\vect^{n_\pm })$
\State $\phi_{m-1}^{n_\pm} \gets  \phi_{m-1}^{n_\pm} + \sum \limits_{k \in \mathbb{Z}} \hat{\zeta}_{m,l}^{n_\pm} \zeta_{m,k}^{n\pm } \psi_{m,k} \left(\mathbf{Y}_\pm + \mathbf{Id} \right) $
\EndFor
\ElsIf{$l=0$}
\State Integrate only forward in time
\Else
\State Integrate only backward in time
\EndIf
\EndFor
\end{algorithmic}
\end{algorithm}

\subsubsection{Numerics: Vector field, passive scalar \& Lagrangian trajectories }
From a numerical perspective, the computation of the inverse flow map is the main bottleneck. This is due to its role in equation \eqref{eq:DefAVField}, which causes the streamfunction $\phi_m$ at a given time $t$ to depend not only on $\phi_{m-1}$ at the same time but also on its values at earlier and later times. In other words, the inverse flow map renders the recurrence relation \eqref{eq:DefAVField} non-local in time. Figure~\ref{fig:RecurPhi} provides a schematic illustration of this temporal non-locality.

This complex dependence of $\phi_m$ on $(\phi_{p})_{p<m}$, together with the requirement to integrate equation~\eqref{eq:InverseFlow} both forwards and backwards in time, represents the primary limitation on the numerical resolution we have been able to attain. Among the various methods explored, the simplest approach proved to be the most effective in achieving the highest spatial resolution.
 
Our procedure is as follows. First, we choose $M$ and construct a space-time grid fine enough to resolve $\tau_M$ and $\epsilon_M$. We then compute $\phi_1$ explicitly on this fine grid and store it. Since $\phi_1$ is now available over the entire space-time grid, we can compute $\phi_2$, store it, and continue this process iteratively until $\phi_M$ is obtained. The main constraint is the size of the file containing $\phi_M$. In our numerical experiment, $\phi_M$ is computed over the time interval $[0,1/2]$ on a $2048^2$ grid, resulting in a file several hundred gigabytes in size. This presents a significant limitation, preventing us from reaching higher resolutions. Nevertheless, this approach has the advantage of making the full field immediately accessible, which is particularly useful for studying passive scalar dynamics and Lagrangian trajectories.

The backward flow map equation~\eqref{eq:InverseFlow} is integrated in time using a third-order Adams-Bashforth scheme, with dealiased pseudo-spectral differentiation. 
For further details on the implementation of this scheme, we provide a pseudo-code in Algorithm~\ref{alg:cap}. All the numerical results presented here are obtained using a vector field $\vect_M$ computed on a Cartesian spatial grid with resolution $\delta x$ in each direction and a temporal resolution $\delta t$. The parameters used to compute 
$\vect_M$ in all our numerical simulations are listed in Table~\ref{tab:ParamsB}.
\begin{table}[h!]
\centering
\begin{tabular}{||c c c c c||} 
 \hline
 $M$ & $\Lambda$ & $\beta$ & $\delta x$ & $\delta t$ \\ [1ex] 
 \hline\hline
 $ 3$& $2.5$ & $1.2$ & $ \frac{1}{4}2^{\left\lfloor  \log_2\left( \epsilon_M\right) \right\rfloor }  $ & $\tau_M/8$ \\  [1ex] 
 \hline
\end{tabular}
\caption{Parameters used in numerical simulations.}
\label{tab:ParamsB}
\end{table}
Theses parameters correspond to a space grid of $2048^2$ points and a time step $\delta t \simeq 4 \times 10^{-5} $. 

To determine whether anomalous diffusion can be observed numerically, we need to integrate the passive scalar equation. This equation is solved using a third-order exponential Adams-Bashforth scheme. This exponential integration method enable us to better handle the numerical stiffness of the Laplace operator: 
\be 
\partial_t \theta^\kappa + e^{-t \kappa  \Delta}\left[\vect_M \cdot\nabla (e^{t\kappa\Delta}\theta^\kappa)\right]= \kappa \Delta \theta^\kappa, \quad \theta^\kappa(0,\pos)=0, \label{eq:PassivNum}
\de
The solution $\theta^\kappa$ to this advection-diffusion equation is computed on the same space-time grid as $\phi_M$. From the numerical integration of \eqref{eq:PassivNum}, we extract several observables, such as $\|\theta^\kappa\|_{L^2}$ and $ \|\nabla\theta^\kappa\|_{L^2}$ to assess whether the system exhibits anomalous diffusion. The passive scalar field is computed on the same space-time grid as $\vect_M$, and in the results we presented, the initial condition is a smooth random gaussian field given by
\be
\theta_0(\pos)= \sum_{ \underset{ 0 <|\mathbf{n}| \leq \frac{2}{\epsilon_{M-1}}}{\mathbf{n} \in \mathbb{Z}^2}}  \dfrac{g_\omega(\mathbf{n})}{ \left< \mathbf{n} \right>^{2-\frac{\beta}{2}} }e^{2i\pi \mathbf{n} \cdot \pos}.
\de
Here, $g_\omega(\mathbf{n})$ is a complex random Gaussian variable with zero mean and unit variance. These variables are independent and satisfy the Hermitian symmetry condition $\overline{g_\omega(\mathbf{n})}= g_\omega(-\mathbf{n})$. We use the Japanese bracket notation $\left< \mathbf{n} \right>= \sqrt{1 + \mathbf{n}^2}$, and the exponent $2-\frac{\beta}{2}$ is chosen such that the initial condition remains smooth for finite $M$ while exhibiting large-scale regularity $\frac{1-\alpha}{2}$, which corresponds to the Onsager critical regularity for anomalous diffusion of the passive scalar to occur. 

For the integration of the backward Lagrangian trajectories governed by Equation~\eqref{eq:BackLag}, we follow the approach of~\cite{Drivas17}, transforming the backward SDE into a forward SDE by introducing the time variable $s$ for $\hat{s}:= t_{f}-s$ where $t_f=0.5$ is the final simulation time. The forward SDE is integrated using an Euler-Maruyama scheme with the same time step as 
that used for computing $\vect_M$. The variance is then estimated using the unbiased estimator
$$ \sigma^2_\kappa(s) :=  \dfrac{2}{N-1} \sum_{n=1}^N  | \mathbf{X}_n(s) -\overline{\mathbf{X}}_N(s) |^2 $$
where $\mathbf{X}_n$ are independent realisations of the process \eqref{eq:BackLag} and 
$$ \overline{\mathbf{X}}_N(s)= \dfrac{1}{N} \sum_{n=1}^N   \mathbf{X}_n(s).$$

\section{Appendix Part II}
\subsection{Proof of Proposition \ref{equivdef} Subsection \ref{subsec1}\label{proof2flows}}
We first prove that if (\ref{killeress}) holds then  necessarily there are least two
distinct trajectory solutions of the problem (${\cal P}_0$) starting from the same initial condition $x_0$: one associated with the limsup and one
with the liminf. We proceed by contradiction by assuming that (\ref{killeress}) holds
and there is a unique solution
of (${\cal P}_0$). We denote $\Phi_\kappa(s) := \Phi_s[f(\cdot,\kappa)] x_0$ and $\Phi_0(s) :=
\Phi_s[f_0(\cdot)] x_0$. We want to show that, $\forall s \in [0,t]$, $||\Phi_\kappa(s)-\Phi_0(s)|| \to 0$ uniformally. Under hypothesis (\ref{eq:H0}), this is straightforward. However, for self-consistency we provide the details of the proof.

We have $\Phi_\kappa(s) = x_0 + \int_0^s f(\Phi_\kappa(s'),\kappa) ds'$ so that $||\Phi_\kappa(s)|| \leq ||x_0|| + t \sup_{x} ||f(x,\kappa)|| \leq C$ uniformally (in $\kappa$ and $s$). Similarly, since $\dot \Phi_\kappa = f(\Phi_\kappa,\kappa)$, 
$||\dot \Phi_\kappa|| \leq C$ uniformally, the time derivatives being uniformally bounded, one obtains uniform equicontinuity (by the mean value theorem). One can therefore use the
Arzel\`a-Ascoli theorem which gives uniformally convergent subsequences 
$\{ \Phi_{\kappa_j} \}$
converging to some $\Phi^\star$ (depending on the subsequence). We can write
$\Phi_{\kappa_j}(s) = x_0 + \int_0^s f(\Phi_{\kappa_j}(s'),\kappa_j) ds'$ and by uniform convergence of the subsequence together with the uniform convergence $ ||f(\cdot,\kappa_j)-f_0(\cdot)||_\infty \to 0$ (and dominated convergence),
one obtains $\Phi^\star(s)  = x_0 + \int_0^s f_0(\Phi^\star(s')) ds'$.
Since by hypothesis, one has assumed the solution of $({\cal P}_0)$ to be unique, then $\Phi^\star = \Phi_0$ (for all subsequences). We therefore obtain the uniform convergence
$\Phi_\kappa \to \Phi_0$. Then using the continuity of the observable, one 
also has ${\cal O}(\Phi_t[f(\cdot,\kappa)] x_0) 
\to {\cal O}(\Phi_t[f_0(\cdot)] x_0)$, or using the notations of Definition \ref{SPST}
$\limsup_{\kappa \to 0} {\cal A}(\kappa) = 
\liminf_{\kappa \to 0} {\cal A}(\kappa)$ giving a contradiction. 

One has therefore two well-defined trajectories, say $\Phi^+_s x_0$ and $\Phi^-_s x_0$, $s \in [0,t]$
corresponding to limsup and liminf resp. starting from the same $x_0$ such that
$\Phi^+_t x_0 \neq \Phi^-_t x_0$. This is just finite-time separation of trajectories in
phase space since $t < \infty$ by hypothesis.

The proof that $(B) \Rightarrow (\ref{killeress})$ is easy: if two trajectories with same initial condition $x_0$ separate in the inviscid limit $\kappa \to 0$ before time $t$, then one can exhibit two subsequences corresponding to these two trajectories. One then 
uses an observable measuring for instance the distance between the two positions at time $t$. One has obviously $(\ref{killeress}) \Longleftrightarrow (C)$. We write the proof for self-consistency. Proposition (C) is a lack of selection principle which means that one can find two subsequences $\kappa_1,\kappa_2 \to 0$ such that $x_{\kappa_j}(t;x_0) \to x_j,j = 1,2$ where $x_{\kappa_j}$ are solutions of the well-posed problems $({\cal P}_{\kappa_j})$,  and $x_1 \neq x_2$ are two distinct solutions of the inviscid problem $({\cal P}_0)$. One can then distinguish a continuous observable such that ${\cal O}(x_1) \neq {\cal O}(x_2)$ (one can invoke a general result in topology for normal spaces, holding trivially for $\mathbb{R}^n$) and by continuity (\ref{killeress}) follows. The converse is trivial, also using the continuity of the observable.
\qed

\subsection{Proof of Property \ref{ObsSP}}\label{ProofObsSP}
Assume that ${\cal O}|_{S_0} = {\rm cst}$. Since $S_0$ is compact, for all sequences $\kappa_j \to 0$, there exists
a subsequence such that $\Phi_t[f(\cdot,\kappa_{j_k})]x_0 \to \phi^\star$ and $\phi^\star \in S_0$. Since ${\cal O}$ is continuous by hypothesis, one has therefore ${\cal A}(\kappa_{j_k}) \to {\cal O}(\phi^\star) = {\rm cst}$ and thus ${\cal A}^+= {\cal A}^-$ giving a contradiction.
Note that the connectedness of $S_0$, from Kneser's theorem, can be used together with the continuity of ${\cal O}$ to infer that ${\cal O}(S_0)$ is a closed interval. If ${\cal O}$ is constant, then the contradiction becomes immediate.
\qed

\subsection{Pedestrian proof of Lemma \ref{Birkhoff}}\label{proofBirkhoff}
We denote $J(T) =
\int_1^\infty F \circ \gamma(\tau T) \frac{d\tau}{\tau^2}
=\int_T^\infty F \circ\gamma (\tau) \omega_T(\tau) d\tau$ with $\omega_T(\tau) = \frac{T}{\tau^2}$ and $R(T)= \frac{1}{T} \int_1^T F \circ \gamma(t) dt$. We split the integral like
\begin{align}
J(T) = \sum_{n \geq 0} \frac{1}{2^{n+1}} \int_{2^nT}^{2^{n+1}T}
F \circ \gamma (\tau ) g_T(\tau),~g_T(\tau) = \frac{2^{n+1}T}{\tau^2}, \nonumber \\ \int_{2^nT}^{2^{n+1}T} g_T(\tau) d\tau = 1. \nonumber
\end{align}
We have
$$
\left| \int_{2^nT}^{2^{n+1}T} F \circ \gamma (\tau) g_T(\tau) d\tau - \bar F \right|
\leq 
$$
$$ \left| \underbrace{\int_{2^nT}^{2^{n+1}T} F \circ \gamma (\tau) \left( g_T(\tau) -
\frac{1}{2^nT} \right) d\tau }_{I} \right| + \left| \underbrace{ \frac{1}{2^nT} 
\int_{2^n T}^{2^{n+1}T}  F \circ \gamma (\tau)d\tau - \bar F}_{II} \right|.
$$
The second term is just $II= 2 R(2^{n+1} T)-R(2^n T)$ which converges to zero uniformally in $n$, and we write $|II| \leq \frac{\epsilon}{2}$.
The term I is more difficult. First notice that $\int_{2^n T}^{2^{n+1}T} \Omega(\tau) d\tau = 0$ where $\Omega(\tau) = g_T(\tau)-\frac{1}{2^nT} = \frac{2^{n+1}T}{\tau^2} - \frac{1}{2^nT}$. 
Then $I = \int F \circ \gamma  \Omega d\tau = \int (F \circ \gamma -\bar F) \Omega d\tau + \int \bar F \Omega d\tau = 
\int (F \circ\gamma -\bar F) \Omega d\tau + \bar F \int \Omega d\tau  = \int (F \circ \gamma -\bar F) \Omega d\tau$ and
therefore 
$$
|I| \leq \sup_{\tau \in [2^nT,2^{n+1}T]} |\Omega(\tau)| \left|\int_{2^n T}^{2^{n+1}T} (F \circ \gamma (\tau)-\bar F)d\tau \right|
$$
Writing $\Omega(\tau) = 2a/\tau^2-1/a$ for $\tau \in [a,2a]$ with $a=2^nT$, an easy calculation shows that the sup is $1/a$. We thus obtain
$$
|I| \leq \left| \frac{1}{2^n T} \int_{2^n T}^{2^{n+1}T} (F\circ\gamma(\tau)-\bar F)d\tau \right|.
$$
But the right-hand side is just the term II, i.e. $|I| \leq |II|$.
We are thus ready to conclude, noticing that $\sum_{n \geq 0} \frac{1}{2^{n+1}} = 1$, one has
$$
|J(T) - \bar F| \leq \sum_{n \geq 0} \frac{1}{2^{n+1}}
\left|
\int_{2^nT}^{2^{n+1}T} F\circ \gamma (\tau) g_T(\tau) d\tau - \bar F
\right| \leq 2|II| \leq \epsilon.
$$ \qed
\subsection{Proof of Prop. \ref{propMg}} \label{alternative}
\begin{itemize}
\item ${\cal M}_0$ is compact and convex. It is convex since taking $\mu_1,\mu_2 \in {\cal P}(S_0)$ then for $\theta \in [0,1]$, $\mu(S_0)=(1-\theta) \mu_1(S_0) + \theta \mu_2(S_0) = 1$. Moreover, for all Borel sets $B \subset S_0$, $\mu(B)\geq 0$ since
$\mu_1(B),\mu_2(B) \geq 0$ and thus $\mu \in {\cal M}_0$.
Due to the compacity of $S_0$, one has immediate tightness: for all $\mu \in {\cal P}(S_0)$, $\mu(S_0)=1$ so that the probability measures in ${\cal P}(S_0)$ are tight. Prokhorov theorem then states that tightness of probability measures in a Polish space is equivalent to be relatively compact in the weak topology. It thus remains to show that the set is closed. Let $\mu_k \rightharpoonup \mu$ then since $\limsup_{k \to \infty} \mu_k(S_0) \leq \mu(S_0)$, then $\mu(S_0) \geq 1$. By weak convergence $\mu(\mathbb{R}^d) = 1$ and $\mu(B) \geq 0$ for all Borel sets, therefore $\mu(S_0) \leq 1$ i.e. $\mu(S_0)=1$ and $\mu \in {\cal M}_0$.
\item ${\cal M}(\gamma) \subseteq {\cal M}_0$.
Let $\mu \in {\cal M}(\gamma)$, from (\ref{eq:H0}) any convergent subsequences $\tau_k \to \infty$, $\gamma(\tau_k) \to \gamma_k$ is such that $\gamma_k$ is solution of $({\cal P}_0)$ (since $f(\cdot,1/\tau_k)$ converges uniformally to $f_0$), i.e. $\gamma_k \in S_0$.
Denote the closure of the  set of all accumulation points by $\bar \gamma$.
From the previous remark, $\bar \gamma \subset S_0$. Let $N$ be some  open set with $N \cap S_0 = \emptyset$, and thus $N \cap \bar \gamma = \emptyset$, then $\mu(N) = \lim_{T_n \to \infty} 1/T_n \int_0^{T_n} \chi_{\tau \in N}(\gamma(s)) ds = 0$. Since the probability measure $\mu$ has support contained in $S_0$ then $\mu \in {\cal M}_0$.
\item ${\cal M}(\gamma)$ is compact.
The space ${\cal M}(\gamma)$ is non-empty due to tighness. We first show that it is 
compact in the weak topology. It suffices to show that it is closed and since it is in the compact set ${\cal P}(S_0)$ it is compact as well. The proof is a classical diagonal argument. Let $\mu_k
\in {\cal M}(\gamma) \to \mu$.
By hypothesis, there is a sequence $T_{n,k} \to \infty$ such that $
\lim_{n \to \infty} 1/T_{n,k} \int_0^{T_{n,k}} F(\gamma(s)) ds = \langle \mu_k,F \rangle$. For each $n$, one can choose
$k=k(n)$ such that $| 1/T_{n,k} \int_0^{T_{n,k}} F({\gamma(s)}) ds - \langle \mu_k,F\rangle | \leq 1/n$.
The sequence $T'_n = T_{n,k(n)}$ is such that $\lim_{n \to \infty} 1/T_n' \int_0^{T_n'} F({\gamma(s)}) ds = {\left\langle \mu, F \right \rangle }$, i.e. $\mu \in {\cal M}(\gamma)$.
\item Nonconvexity. It is useful to give a counter-example. Let $\gamma(s)=e^{i \log s}$
and write the test functions like $F(z) = \sum_{k \geq 0} F_k z^k$. Let
$C(T) = 1/T \int_0^T F(\gamma(s)) ds = \sum_{k \geq 0} G_k e^{i k \log T}$ with $G_k = F_k/(1+ik)$. Then we can exhibit subsequences of the form $T_n^{(a)} = e^{2\pi n + a_n}$
with $a_n \to a$. It gives the family of measures $\langle \mu_a,F \rangle
= \sum_{k \geq 0} G_k e^{i k a}$. Consider two distinct measures $\mu_{a_1},\mu_{a_2}$ then one must find a sequence $T_n$ such that $C(T_n) \to \sum_{k \geq 0} G_k (\theta e^{i k a_1} + (1-\theta) e^{i k a_2})$. It gives the constraint: $\forall k \geq 0,
e^{i \log T_n} \to \left( \theta e^{i k a_1} + (1-\theta) e^{i k a_2} \right)^\frac1k$.
This is not possible (unless $T_n$ depends on $k$).
\item Genericity. This is a tautology: suppose that ${\cal M}(\gamma)$ does not reduce to a singleton, and there is a measure $\mu$ in ${\cal M}(\gamma)$ such that $\gamma$ is generic for $\mu$, then 
${\cal M}(\gamma)$ must reduce to a singleton (since all subsequences would have the same limit), giving a contradiction.
\end{itemize}
\qed
\subsection{Convex example}\label{exconv}
We consider $\gamma(s) =  \tanh (s \sin \log s)$ and want to compute
$I(T) = 1/T \int_0^T F(\gamma(s)) ds$ for large $T$. We write $F(z) = \sum_{ k \geq 0} F_k z^k$.
For $k$ even, one has $I(T) \sim 1$. One must discuss the case $k$ odd.
We write $I(R) = e^{-R} \int_{-\infty}^R \tanh^k(e^s \sin s) e^s ds$ with $R=\log T$.
Let $R^{-1} \ll  \delta \ll 1$ such that 
$|\tanh^k (e^s \sin s) - {\rm sign}(\sin s)| \leq \delta$
then $I(R) = e^{-R} \int_{-\infty}^{\delta^{-1}} \tanh^k(e^s \sin s) e^s ds
+ e^{-R} \int_{\delta^{-1}}^R \tanh^k(e^s \sin s) e^s ds$ since 
$|\tanh^k(e^s \sin s)| \leq 1$, the first term is negligible (it is bounded by $e^{-R(1-\delta^{-1}/R)}$). We thus focus on the second term only, i.e.
$C_R = e^{-R} \int_{\delta^{-1}}^R {\rm sign}(\sin s) e^s ds$ up to correction terms
uniformally bounded  by $\delta - e^{-R(1-\delta^{-1}/R)}$.
\\ We obtain
$C_R = e^{-R} \sum_{j = j_0}^{j_1} \int_{ j \pi}^{(j+1) \pi} (-1)^j e^s ds + 
e^{-R} \int_{\delta^{-1}}^{j_0 \pi} (-1)^{j_0-1} e^s ds + 
e^{-R} \int_{(j_1+1) \pi}^R (-1)^{j_1+1} e^s ds$ where $j_0 = \lceil \delta^{-1}/\pi \rceil$ and $j_1 = \lfloor R/\pi \rfloor -1 $. It is
$C_R = e^{-R} \sum_{j=j_0}^{j_1} (-1)^j (e^{(j+1)\pi} - e^{j\pi}) -
(-1)^{j_0} e^{-R} (e^{j_0 \pi} - e^{\delta^{-1}}) - (-1)^{j_1} (1 - e^{\lfloor R/\pi \rfloor \pi-R})$.
$$
C_R \sim \Gamma(R):= (-1)^{\lfloor \frac{R}{\pi} \rfloor} \left( 1 - \frac{2 e^\pi}{1+ e^{\pi}} {\rm exp}
\left( -(R - \pi \left\lfloor \frac{R}{\pi} \right\rfloor ) \right) \right).
$$
Coming back to ${\cal M}(\gamma)$, one must look at subsequential limits of ad-hoc subsequences $T_n \to \infty$, taking $R_n = \log T_n =  2n \pi + \alpha \pi$,
and $R_n = (2n+1)\pi + \beta$, where $\alpha,\beta \in [0,1]$ one
obtains two families for $0 \leq x \leq 1$:
$$
\mu_x = \frac12 (1\pm \Theta(x)) \delta_{+1} + \frac12 (1\mp \Theta(x)) \delta_{-1},~~
\Theta(x)  = 1-\frac{2e^{-\pi x}}{1+e^{-\pi}}.
$$
One can equivalently write: $\mu_\theta = \theta \delta_{+1} + (1-\theta) \delta_{-1}$ where $\theta \in [\sigma,1-\sigma], \sigma = \frac{1}{1+e^\pi}$. For the sake of rigour, let $F \in C_b$, there is a sequence of analytic functions $F_n \to F$ uniformally such that using continuity of the functional $F \mapsto \langle \mu,F \rangle$, the result holds for all $F \in C_b$.
This example gives a situation where ${\cal M}(\gamma)$ is indeed convex.

\subsection{Proof of Theorem \ref{M0M}} \label{M0Mproof}
The fact that ${\cal M}_0 = \overline{\rm co}({\cal E})$ is a standard result which is a consequence of ${\cal M}_0$ being compact and convex and Krein-Milman theorem (see proof Appendix \ref{alternative}).
\begin{itemize}
\item Proof that for all $x \in S_0$, there exists $\gamma_x \in {\cal R}_{\rm eg}$ such that ${\cal M}(\gamma_x) = \{\delta_{x} \}$.\\
Consider the set of trajectories of the inviscid system $\{ \Phi_s x_0, s \in [0,t] \}$. One can select a trajectory call $g$ in this set such that $g(t)=x$ (otherwise it would contradict $x \in S_0$). This trajectory is $C^1$ since $\dot g = f_0(g)$ and $f_0$ is continuous. We first use Stone-Weierstrass to approximate $\dot g$ by $C^\infty$ functions and then one obtain $C^\infty$ $g_\epsilon$ such that $g_\epsilon(0) = x_0$ and $||g_\epsilon-g||_{C^1} \to 0$. We then construct a vector field $F_\epsilon = 
\dot g_\epsilon$ in a small tubular neighborhood of $g$ of radius size $\epsilon$. We use a partition of unity so that 
$f_\epsilon = \theta_\epsilon F_\epsilon + (1-\theta_\epsilon) f_0$, namely $f_\epsilon$ is equal to $f_0$ outside the tube but connects smoothly to $F_\epsilon$ inside. We show that this vector field satisfies (\ref{eq:H0}). Inside the tube, one has $||f_\epsilon(x)-f_0(x)|| = ||F_\epsilon(x)-f_0(x)|| \leq ||F_\epsilon(x) - f_0(g(x))|| + ||f_0(x)-f_0(g(x))||$. The first term goes to zero by the definition of $F_\epsilon$ ($\dot g_\epsilon \to \dot g = f_0(g)$ uniformally). The second term goes to zero uniformally by continuity of $f_0$ and the definition of the tubular neighborhood. Outside the tube $f_\epsilon = f_0$. In order to conclude, one has by construction $||g_\epsilon -g||_{C^0} \to 0$ and is the unique solution of $\dot x = f_\epsilon(x), x(0)=x_0$.
\qed

\item Proof that any finite convex combination of Dirac measure is in ${\cal M}$.
\\
Let $x,y \in S_0$, $x \neq y$. We denote $f_x(\cdot,\epsilon)$ the regularisation constructed above such that the inviscid limit attain $x$, and similarly for $y$. 
Let $\theta \in (0,1)$,
we will construct a regularisation of the form $f_\theta(\cdot,\epsilon) = a_\theta(\epsilon^{-1})
f_x(\cdot,\epsilon) + (1-a_\theta(\epsilon^{-1})) f_y(\cdot,\epsilon)$. We consider
\begin{align} \label{controlswitch}
a_\theta(\tau) = \frac12 \left( 1+ \tanh (\tau p_\theta(\tau)) \right)~\nonumber \\ {\rm with}~p_\theta(\tau) = \sin 2\pi \tau + \sin \pi(\theta - \frac12).
\end{align} 
It is designed such that $a(\tau)$ is asymptotically behaving like a periodic step function with values 1 during an interval of length converging to $\theta$ and 0 during an interval of length converging to $1-\theta$. The function $f_\theta$ satisfies (\ref{eq:H0}) since 
$||f_\theta-f_0||_\infty = ||a(f_x-f_0) + (1-a)(f_y-f_0)||_\infty \leq ||f_x-f_0||_\infty + ||f_y-f_0||_\infty$ and $f_x,f_y$ satisfy (\ref{eq:H0}) by hypothesis. One then write
$1/n \int_0^n F(\gamma(s+s_0)) ds =  1/n \sum_{k = 0}^{n-1} \int_0^1 F(\gamma(s+s_0+k)) ds \approx 1/n \sum_{k = 0}^{n-1} (\theta F(x) + (1-\theta) F(y))$
where $s_0 \in (0,1)$ is such that $\gamma(s+s_0+k) \approx x$ for $s \in [0,\theta]$ and $\approx y$ for $s \in [\theta,1]$. As usual, the boundary terms does not contribute to the weak limit which is thus $\theta \delta_x + (1-\theta) \delta_y$. We do not write all the details but the rapid switches between 0 and 1 yield a small remainder term in the Ces\`aro mean of size $O(1/n)$.\\
In order to conclude, the proposed construction can be extended to switches between any finite number of arbitrary points $x \in S_0$. One has thus ${\rm co}({\cal E}) \subseteq {\cal M} \subseteq {\cal M}_0 = \overline{\rm co}({\cal E})$, taking the closure one concludes that ${\cal M} = {\cal M}_0$.\qed
\end{itemize}
\subsection{${\cal M}(\gamma_{\rm un}) ={\cal M}_0$: a universal non-renormalised regularization.}\label{proofgun}
Let $b_n := (x_n,y_n,\theta_n) \in B:=S_0 \times S_0 \times [0,1]$ a sequence which is sequentially dense, meaning that for all $b \in B$ one has a convergent subsequence $b_{n_k} \to b$. Let $T_n \to \infty$ a given sequence to choose. Call $\chi_n:\mathbb{R}^+ \to [0,1]$ a partition of unity such that $\sum_n \chi_n(s) = 1$ and
$\chi_n(s)=1, s \in [T_n+\delta,T_{n+1}-\delta]$ and zero if not. We define $\gamma_{\rm un}(\tau) = \sum_n \chi_n(\tau) \gamma_{x_n,y_n,\theta_n}(\tau)$ where $\gamma_{x,y,\theta}$ is the regularization curve constructed in the proof above (\ref{M0Mproof}).
We must choose $T_n$ so that $T_{n+1}-T_n$ is large enough to distinguish at least one period of $a_\theta$ in (\ref{controlswitch}).
In the interval $(T_n,T_{n+1})$, the curve is approximating $\theta_n x_n + (1-\theta_n) y_n$. Due to the sequential density, one can exhibit subsequences $n_k \to \infty$ such that $x_{n_k},y_{n_k},\theta_{n_k} \to x,y,\theta$, namely $\mu_{T_{n_k}} \to  \theta \delta_{x} + (1-\theta) \delta_{y}$. This construction shows that ${\cal M}(\gamma_{\rm un}) = {\rm co}({\cal E})$. Taking the closure, one obtains ${\cal M}(\gamma_{\rm un}) = {\cal M}_0$.\qed
\\
\subsection{Dini derivatives and proof of Theorem \ref{CN}}\label{DiniApp}
Although we do not use Dini derivatives explicitly, it is closely related to
Definition \ref{DiniME}. We just recall very few well-known properties.
\begin{definition}
Let $f(x)$ be a  real valued function defined in $(a,b)$, the four Dini derivatives at $x_0$ are defined as
\begin{align}
D^\pm f(x_0) = \limsup_{x \to x_0^\pm} \frac{f(x)-f(x_0)}{x-x_0}, \nonumber \\ 
D_\pm f(x_0) = \liminf_{x \to x_0^\pm} \frac{f(x)-f(x_0)}{x-x_0}. \nonumber
\end{align}
$\pm \infty$ is allowed.
\end{definition}
We state some Dini properties for general functions (not necessarily continuous)
\begin{itemize}
\item if $f$ is continuous in $(a,b)$, and $D(x) \in \{D_\pm^\pm f(x)\} > 0 (<0),~\forall x \in (a,b)$ then $f$ is
strictly increasing (decreasing). In other words, one needs only
one of the Dini derivative. 
\item Let $f$ be continuous on $(a,b)$ with at least one of the 
Dini derivatives bounded  (e.g. $|D^+ f(x)| \leq C, \forall x \in (a,b)$)
then f is  Lipschitz on $(a,b)$.
\end{itemize}
Deeper results can be found in \cite{giorgi1992dini,bruckner1978}.
\\\\
The proof of Theorem \ref{CN} is simple, since it is mostly a rephrasing of non-Lipschitz
property using Dini-like quantities but translated in the "Osgood framework".
We show the contrapositive of Theorem  \ref{CN}. We restrict $x$ to a compact set
$M \subset \mathbb{R}^n$. Therefore, $\exists C, \forall (x,v) \in M \times \mathbb{S}^{n-1}, 
\Lambda^+_\Omega(x,v) \leq C < + \infty$. We need to show that in this case
$\dot x =f(x), x(0)=x_0 \in M$ has a unique solution. Note that $f$ might not be necessarily Lipschitz. Consider $v = \frac{y-x}{||y-x||}$. Using the uniform bound above, for small enough $t>0$, one has
$\langle f(x+tv)-f(x),v \rangle \leq C \Omega(t)$, and taking $t=||y-x||$  gives the one-sided inequality
$$
\langle f(y)-f(x),y-x \rangle \leq C ||y-x|| \Omega(||y-x||).
$$
Denote $g(||y-x||^2) = ||y-x|| \Omega(||y-x||)$. Then by the one-sided 
Osgood Lemma, uniqueness
occurs if $\Int_{0^+} \frac{dz}{g(z)} = \Int_{0^+} \frac{dz}{\Omega(z)} = +\infty$ which is just the hypothesis of Theorem \ref{CN}. Note that the nonautonomous case holds
as well: this is Giuliano's uniqueness Theorem with a time-dependent measurable constant (see Theorem 3.5.1 in 
\cite{Agarwal_Lak}).\qed

\subsection{Notations used in the rest of Part II}
It will be useful to introduce the compact notations
\be \label{Anotations}
\renewcommand*{\arraystretch}{0.9}
\begin{array}{llllll}
{\cal A}(\kappa) & \hspace*{-0.15cm} :=  
{\displaystyle \kappa ||\nabla \theta^\kappa||^2_{L^2((0,1) \times \mathbb{T}^2)}}
& 
{\cal A}_m(\kappa_m) & \hspace*{-0.15cm}:= 
{\displaystyle \kappa_m ||\nabla \theta_m||^2_{L^2((0,1) \times \mathbb{T}^2)}}
\\\\
{\cal A}^+ & \hspace*{-0.15cm}:=  {\displaystyle \limsup_{\kappa \to 0} {\cal A}(\kappa)} &
{\cal A}^- & \hspace*{-0.15cm}:=  {\displaystyle \liminf_{\kappa \to 0} {\cal A}(\kappa)} \\
{\cal A}_{m_\star}^+ & \hspace*{-0.15cm}:= {\displaystyle \limsup_{\kappa \to 0} {\cal A}_{m^\star}(\kappa_{m^\star}(\kappa))} &
{\cal A}_{m_\star}^- & \hspace*{-0.15cm}:=   {\displaystyle \liminf_{\kappa \to 0} {\cal A}_{m^\star}(\kappa_{m^\star}(\kappa))} \\
\kappa^+ & \hspace*{-0.15cm}:=  {\displaystyle \limsup_{\kappa \to 0} \kappa_{m^\star}(\kappa)} & \kappa^- & \hspace*{-0.15cm}:=  {\displaystyle \liminf_{\kappa \to 0} \kappa_{m^\star}(\kappa)}
\end{array},
\de 
with $\kappa \in {\cal K}_A$.
Note that $\kappa_m$ is always understood as depending on the initial condition $\kappa_M = \kappa$
(see (\ref{avseq})). In fact, it is important to notice that the mapping
$\kappa_{m} \mapsto \kappa_{m-1}$ is non-autonomous 
and any limit $M \to \infty$ must be understood as a pullback limit
for fixed $m$. In principle, one should always
write the term $\kappa_m$ as depending on both $\kappa$ and $M$. In order
to avoid heavy notations, we will write $\kappa_m$, unless needed.
\subsection{Proofs of Theorem \ref{LAMODUS} and Property \ref{diffk} in section \ref{secMAIN}} \label{proofs_ESS_AV}
As announced in the main text, the following proofs are based on the ArXiv version
\cite{AV23}.
In the following, it will be useful to have  lower and upper bounds for the
products
\be \label{defSharprod}
\Pi_{\pm a,b,m} := \prod_{k = m}^{M-1} (1 \pm a \epsilon_k^b),~\epsilon_k = \epsilon_{k-1}^q,~a > 0, b > 0.
\de 
The upper bound is found by using $\log(1+x) \leq x$ giving
$S_{a,b,m} \leq a \sum_{k=m}^{M-1} \epsilon_k^b = a \epsilon_m^b \sum_{k=m}^{M-1}
(\epsilon_k/\epsilon_m)^b$ and since $\epsilon_{k+1}/\epsilon_k = \epsilon_k^{q-1}
\leq \epsilon_m^{q-1} = \Lambda^{-q^m}$ then
$S_{a,b,m} \leq a \epsilon_m^b \sum_{k=m}^{M-1} g^k = a \epsilon_m^b \frac{1-g^{M-m}}{1-g}\leq a \epsilon_m^b/(1-g)$ where $g = \Lambda^{-b q^m}$.
For the lower bound, we use $\log(1+x) \geq \lambda x$ for $0 <\lambda < 1$ valid for
$x \leq C_\lambda^-$ where $\log(1+C_\lambda^-) = \lambda C_\lambda^-$
(for $\lambda \sim 1$, $C_\lambda^- \sim 2(1-\lambda))$. It gives
$S_{a,b,m} \geq \lambda a \epsilon_m^b \sum_{k=m}^{M-1}
(\epsilon_k/\epsilon_m)^b \geq \lambda a \epsilon_m^b $ provided $ \epsilon_{M-1}^b < \cdots <
\epsilon_m^b \leq C_\lambda^-$. 
We are also interested in a lower bound for $\Pi_{-a,b,m}$, we use
$\log(1-x) \geq -\lambda x$ for $x$ small enough and $\lambda > 1$,
namely $x \leq C_\lambda^+$ with $\log(1-C_\lambda^+) = -\lambda C_\lambda^+$.
It gives
$\Pi_{-a,b,m} \geq -\lambda a \epsilon_m^b \sum_{k=m}^{M-1}
(\epsilon_k/\epsilon_m)^b \geq -\lambda a \epsilon_m^b \sum_{k=m}^{M-1}
g^k = -\lambda a \epsilon_m^b (1-g^{M-m})/(1-g) \geq 
-\lambda a \epsilon_m^b/(1-g)$. There are now two constraints, first
$1-a\epsilon_{M-1}^b > \cdots > 1-a \epsilon_m^b > 0$ and $\epsilon_m^b \leq C_\lambda^+$.
To summarise, we have
\be\label{Sharprod}
\left\{
\begin{array}{lllll}
\Pi_{a,b,m} &  \hspace*{-1.5cm} \geq  &
{\displaystyle {\rm exp} \left( \lambda a \epsilon_m^b \right)} \geq 1 \\
 {\rm for}~\epsilon_m^b \leq C_\lambda^-,0 < \lambda < 1 \\\\
 \Pi_{a,b,m} & \hspace*{-1.5cm}\leq & 
 {\displaystyle {\rm exp} \left( a \epsilon_m^b (1-\Lambda^{-b q^m})^{-1}
 \right)}    \\\\
 \Pi_{-a,b,m} & \hspace*{-1.5cm}\geq & 
{\displaystyle {\rm exp} \left( -\lambda a \epsilon_m^b (1-\Lambda^{-b q^m})^{-1}\right)}  \\
{\rm for}~\epsilon_m^b \leq \min \{ C_\lambda^+,a^{-1} \},
\lambda > 1.
 \end{array} \right.
 \de 
The key aspect to remember is that $\Pi_{\pm a,b,m}$ can be made arbitrarily close to 1 by increasing $\Lambda$.
\subsubsection{Lemma \ref{lemmasharp}}\label{proof_1}
\begin{lemma}
\label{lemmasharp}
Let $\varepsilon > 0$. Provided $\Lambda \geq (\frac{C}{\epsilon})^\frac{1}{\delta}$ and $M$ is  large enough, one has
\be \label{essbounds}
\sigma^- {\cal A}_{m^\star}(\kappa_m^\star) \leq {\cal A}(\kappa)
\leq \sigma^+ {\cal A}_{m^\star}(\kappa_m^\star).
\de
where
\be	
\sigma^\pm := \left( e^{\pm \varepsilon} \pm \gamma_M  \right)^2,~~
\gamma_M := C \epsilon_M^{ \beta (q - \frac{2}{q+1})} \ll 1.
\de
\end{lemma}
{\bf Proof}: it relies on several "hard" estimates which are 
deeply rooted in \cite{AV23}. The main one is (5.72) from their Proposition 5.2, that we reproduce 
$$
(5.72)~~~\left| \frac{{\cal A}_m(\kappa_m)}{{\cal A}_{m-1}(\kappa_{m-1})} -1 \right| \leq
C \epsilon_{m-1}^\delta.
$$
The aim is to iterate this estimate several times from $M$ to $m^\star$
using the quantities $\Pi_{\pm C,\delta,m^\star}$ introduced in (\ref{defSharprod}).
From (5.72)\cite{AV23} (or see (5.100)\cite{AV23}), one easily obtains
\be \label{5_100}
{\cal A}_{m^\star}(\kappa_m^\star) \Pi_{-C,\delta,m^\star} \leq {\cal A}_M(\kappa_M) \leq {\cal A}_{m^\star}(\kappa_m^\star) \Pi_{+C,\delta,m^\star} ,~~\kappa_M = \kappa.
\de
From (\ref{LBHyp}), one has
$C \epsilon_{m^\star}^\delta \leq C \epsilon_1^\delta \leq C \Lambda^{-\delta} \leq \varepsilon < 1$
and (\ref{Sharprod}) can be used. It reads
\begin{align}
e^{-2\varepsilon} \leq {\rm exp} \left(-\lambda C \epsilon_{m^\star}^\delta(1-\Lambda^{-\delta q^{m^\star}})^{-1} \right) 
\leq \Pi_{-C,\delta,m^\star},~~\Pi_{+C,\delta,m^\star} \leq \nonumber \\ {\rm exp} \left(C \epsilon_{m^\star}^\delta(1-\Lambda^{-\delta q^{m^\star}})^{-1} \right) \leq e^{2\varepsilon} \nonumber
\end{align}

One can now prove Lemma \ref{lemmasharp}.
We then write Eq.(5.103) in \cite{AV23} using our notations, it is
\be \label{ttm}
||\theta-\theta_M||_{L^\infty((0,1);L^2(\mathbb{T}^2))}^2
+ \kappa ||\nabla \theta - \nabla \theta_M||_{L^2((0,1)\times \mathbb{T}^2)}^2
\leq \gamma_M^2 {\cal A}_{m^\star}(\kappa_m^\star).
\de 
For the upper bound, one starts from
 $\sqrt{\kappa} ||\nabla \theta|| \leq \sqrt{\kappa}||\nabla \theta - \nabla \theta_M || + \sqrt{\kappa}||\nabla \theta_M||$. It is using the above notations and $\kappa_M = \kappa$,
$\sqrt{\kappa} ||\nabla \theta|| \leq \sqrt{\kappa} ||\nabla \theta - \nabla \theta_M || + {\cal A}_M^\frac12(\kappa_M)$. Then one uses (\ref{5_100}), (\ref{ttm}) so that $\sqrt{\kappa} ||\nabla \theta|| \leq \gamma_M
{\cal A}_{m^\star}^\frac12(\kappa_m^\star) + {\cal A}_M^\frac12(\kappa_M)
 \leq (\gamma_M + e^\varepsilon)
{\cal A}_{m^\star}^\frac12(\kappa_m^\star).
$ 
For the lower bound, one starts from the reverse inequality and the definition
of ${\cal A}_M$:
$ \sqrt{\kappa} ||\nabla \theta|| \geq  {\cal A}_M^\frac12(\kappa_M) - \sqrt{\kappa} ||\nabla \theta - \nabla \theta_M || \geq {\cal A}^\frac12_M(\kappa_M) - \gamma_M {\cal A}_{m^\star}^\frac12(\kappa_m^\star)$, then using 
the lower bound of (\ref{5_100}) gives
$\sqrt{\kappa} ||\nabla \theta|| \geq \left( e^{-\varepsilon} - \gamma_M \right)
{\cal A}_m^\frac12(\kappa_m^\star)
$. Now to square this inequality, one must choose $M$ large enough to insure that
$\gamma_M < e^{-\varepsilon}$. Note that $\gamma_M = C \Lambda^{-q^M \beta \frac{q+2}{q+1}}
\sim C \Lambda^{-2 q^M}$ when $\beta \to \frac43$ and in view of (\ref{LBHyp}), $\gamma_1$ is already a very small
number, much smaller than $\varepsilon$.
\qed
\subsubsection{Lemma \ref{AHAHAH}}\label{proof_4}
\begin{lemma}
\label{AHAHAH}
Let $1 < A \leq \frac12 \epsilon_{m^\star}^{-\gamma}$. The sequence (\ref{avseq}) with admissible set ${\cal K}_A$ 
is such that 
\be \label{ratiokap} 
\frac{\kappa^+}{\kappa^-} \geq c_1 A^2 ,~~c_1 = e^{-2 \epsilon_{m^\star}^{2 \gamma(q-1)}} \approx 1.
\de 
Moreover, there exists universal constants $0 < c < C < \infty$ such that
for all $m \in \{m^\star,\cdots,M-1 \}$
\be \label{kappam}
c \epsilon_m^{\beta+\gamma} \leq \kappa_m \leq C \epsilon_m^{\beta+\gamma}.
\de 
In particular $\kappa^- > 0$ and $\kappa^+-\kappa^- \leq \sqrt{c_0} \epsilon_{m^\star}^{\beta}$.
\end{lemma}
 We notice that Lemma 3.4 in \cite{AV23} provides lower and upper bounds for the renormalised
sequence (\ref{avseq}). Unfortunately, they 
cannot be used to obtain a lower bound on the term $\kappa^+/\kappa^-$. We therefore need to investigate the sequence more closely, that is, to establish a
new sharper version of Lemma 3.4 in \cite{AV23}.
\\\\
We set
\be \label{smdef}
\kappa_m = (\epsilon_m^{\beta+\gamma} c_0^\frac12) s_m, ~m \in \{ m^\star,\cdots,M \}.
\de
It gives $(\epsilon_{m-1}^{\beta+\gamma} c_0^\frac12) s_{m-1} = 
(\epsilon_m^{\beta+\gamma} c_0^\frac12)s_m + c_0 \epsilon_m^{2\beta} 
c_0^{-\frac12} \epsilon_m^{-\beta-\gamma} \frac{1}{s_m}$, 
$s_{m-1} = \frac{\epsilon_m^{\beta+\gamma}}{\epsilon_{m-1}^{\beta+\gamma}} s_m +
\frac{\epsilon_m^{\beta-\gamma}}{\epsilon_{m-1}^{\beta+\gamma}} \frac{1}{s_m}$:
$
s_{m-1} = \left( \frac{\epsilon_m}{\epsilon_{m-1}} \right)^\beta
\epsilon_m^{-\gamma} \epsilon_{m-1}^{-\gamma} \left( \epsilon_m^{2\gamma} 
s_m + \frac{1}{s_m}
\right).
$
We recall that $\gamma = \frac{q-1}{q+1} \beta$ and $\epsilon_m = \epsilon_{m-1}^q$.
It gives $(\epsilon_m/\epsilon_{m-1})^\beta \epsilon_m^{-\gamma} \epsilon_{m-1}^{-\gamma} = 
\epsilon_{m-1}^{(q-1)\beta} \epsilon_{m-1}^{-\gamma(q+1)} = 1$. We thus consider
the prefactor map
\be 
s_{m-1} = g_m(s_m),~g_m(s) := \epsilon_m^{2\gamma} s + \frac{1}{s},~~m \in \{m^\star+1,\cdots,M \}.
\de 
We assume that there exists some $A > 1$ which does not depend on $M$ such that
\be \label{conv}
\kappa_M \in c_0^{\frac12} \epsilon_M^{\frac{2\beta}{q+1}} [\frac{1}{A},A], 
A > 1.
\de
It gives using the fact that $\epsilon_M^{2\beta/(q+1) - (\beta+\gamma)} = \epsilon_M^{-2\gamma}$:
$$
s_M \in \epsilon_M^{-2\gamma} [\frac{1}{A},A],~~{\rm and}~~s_{M-1} \in
[\frac{1}{A}+A \epsilon_M^{2\gamma},A + \frac{1}{A} \epsilon_M^{2\gamma}] \approx [\frac{1}{A},A].
$$
We thus expect that the intervals are almost invariant at leading order. The goal is to find some constraint on $A$ such 
that the mapping $[s_m^-,s_m^+] \to [s_{m-1}^-,s_m^+]$ stays nearly constant and
$\approx [\frac{1}{A},A]$. Following \cite{AV23}, one has
\be  \label{ubmn}
\max \left\{ s_{m-1},\frac{1}{s_{m-1}} \right\} \leq (1+\epsilon_m^{2\gamma}) \max 
\left\{ s_{m},\frac{1}{s_{m}} \right\}.
\de 
{\bf Proof}: denote $M_m =\max \{ s_{m},1/s_m\}$, assume that
$s_{m-1} \geq 1/s_{m-1}$ then $s_{m-1} = M_{m-1} \leq (1+\epsilon_m^{2\gamma}) M_m$.
Assume now that $s_{m-1} \leq 1/s_{m-1}$, then $s_{m-1} \geq (1+\epsilon_m^{2\gamma})
\min \{ s_m, 1/s_m \} =(1+\epsilon_m^{2\gamma}) / M_{m} $ and thus $1/s_{m-1} =
M_{m-1} 
\leq (1+\epsilon_m^{2\gamma})^{-1} M_m \leq (1+\epsilon_m^{2\gamma}) M_m$. \qed
\\
Using the fact that $M_{M-1} = A + \frac{1}{A} \epsilon_M^{2\gamma}$, one has, iterating from $m$ to $M-1$:
\be \label{Mn}
M_m \leq (A + \frac{1}{A} \epsilon_M^{2\gamma}) \Pi_{1,2\gamma,m+1},~~
\de 
Since $\Pi_{1,2,\gamma,m+1}$ can be made arbitrarily close to one by
increasing $\Lambda$, $s_0$  stays in the interval $\approx [\frac{1}{A},A]$ up to an error which can be made arbitrarily small by increasing $\Lambda$. As mentioned before, this is not enough to infer a lower bound for $s_m^+/s_m^-$. 
However, based on (\ref{Mn}) and the definition of $M_m$, one can write:
\be \label{30}
s_m^+ \leq A \left( 1+ \frac{1}{A^2} \epsilon_M^{2\gamma} \right)
\Pi_{1,2\gamma,m+1} \leq A (1+\epsilon_M^{2\gamma}) e^{2 \epsilon_{m^\star+1}^{2\gamma}}
\leq 2 A.
\de
One has also $s_m^- \geq \frac{1}{2 A}$.
By hypothesis $1 < A \leq \frac12 \epsilon_{m^\star}^{-\gamma}$, i.e. 
$s_m^+ \leq \epsilon_m^{-\gamma}$
so that  $g_m'(s) \leq 0$ for $s \leq s_m^+$.
One can therefore assume that the sequence 
is decreasing in the interval $[s_m^-,s_m^+]$. As a consequence, 
$$
\frac{s_{m-1}^+}{s_{m-1}^-} = \frac{s_m^+}{s_m^-} 
~\frac{1 + (s_m^-)^2 \epsilon_m^{2\gamma}}{1+(s_m^+)^2 \epsilon_m^{2\gamma}}
\leq \frac{s_m^+}{s_m^-}, ~~\forall m \in \{m^\star+1,\cdots,M-1 \}.
$$
It gives, using (\ref{Sharprod}) twice,
\begin{align}
\frac{s_{m^\star}^+}{s_{m^\star}^-} = \frac{s_{M-1}^+}{s_{M-1}^-}
\prod_{k=m^\star+1}^{M-1} \frac{1+(s_k^-)^2 \epsilon_k^{2\gamma}}{
1+(s_k^+)^2 \epsilon_k^{2\gamma}} \geq \frac{s_{M-1}^+}{s_{M-1}^-} 
\frac{1}
{\Pi_{4A^2,2\gamma,m^\star+1}} \nonumber \\ 
\geq \frac{s_{M-1}^+}{s_{M-1}^-} {\rm exp} \left( -
\frac{4A^2 \epsilon_{m^\star+1}^{2\gamma}}{1-\Lambda^{-2\gamma q^{m^\star+1}}}
\right)
\end{align}
Therefore, with $A \leq \frac12 \epsilon_{m^\star}^{-\gamma}$, one obtains the lower
bound for $M$ large 
\be  \label{lbsh}
\frac{s_{m^\star}^+}{s_{m^\star}^-}  \geq   ~\frac{A^2+ \epsilon_M^{2\gamma}}{1+A^2 \epsilon_M^{2\gamma}} 
~{\rm exp} \left( - 2 \epsilon_{m^\star}^{-2\gamma} \epsilon_{m^\star+1}^{2\gamma}
\right) \gtrsim A^2 ~~ {\rm exp} \left( - 2 \epsilon_{m^\star}^{2\gamma(q-1)} 
\right).
\de 
From (\ref{smdef}) and (\ref{30}), one has $\kappa^+-\kappa^- \leq
\kappa^+ \leq 2 A \sqrt{c_0} \epsilon_{m^\star}^{\beta+\gamma} \leq \epsilon_{m^\star}^\beta$.
The inequality (\ref{kappam}) also follows 
with the constants $c  
\approx (2A)^{-1} \sqrt{c_0}$
and $C 
\approx (2A) \sqrt{c_0}$.
\qed
\subsubsection{Proof that ${\cal A}^-_{m_\star} > 0$}\label{proof_2}
This is a simple consequence of the Poincar\'e inequality and energy estimate:
$\frac{d}{dt} ||\theta_{m^\star}(t,\cdot)||^2_{L^2} = 
-2 \kappa_{m^\star} ||\nabla \theta_{m^\star}(t,\cdot)||^2 \leq 
-8 \pi^2 \kappa_{m^\star} ||\theta_{m^\star}(t,\cdot)||^2_{L^2}$ and thus
$2 {\cal A}_{m^\star}(\kappa_{m^\star}) = ||\theta_0||^2_{L^2} - 
||\theta_{m^\star}(1,\cdot)||^2_{L^2} \geq \left(1 - e^{-8 \kappa_{m^\star} \pi^2}
\right) ||\theta_0||^2$. Therefore, one can write 
${\cal A}_{m_\star}^- \geq \frac12 \left(1 - e^{-8 \kappa^- \pi^2} \right) > 0$
using the fact that $\kappa^- > 0$ by Lemma \ref{AHAHAH}.
\subsection{Proof of the main Theorem \ref{LAMODUS}}\label{APP_LAMODUS}
The proof can be achieved with only few steps.
First, one has ${\cal A}^-_{m_\star} > 0$
(see previous subsection) and in light of (\ref{essbounds})
${\cal A}^- > 0$.
The consequence of the sharp estimates in Lemma \ref{lemmasharp} is
that one can investigate the limit $\kappa \to 0$.
In view of
Lemma 3.4 in \cite{AV23}, condition (3.45), one takes $M$ such that
\be\label{Mkappa}
\frac1A \sqrt{c_0} \epsilon_M^\frac{2\beta}{q+1} \leq \kappa \leq A \sqrt{c_0} \epsilon_M^\frac{2\beta}{q+1}.
\de
One therefore considers the limit $\kappa \to 0$ with $M = M(\kappa)$ satisfying
(\ref{Mkappa}). Due to the hypergeometric scale separation,
(\ref{Mkappa}) implies that $M$ scales like a double logarithm of $\kappa$.
We then want to obtain a sufficient condition that guarantees the 
lack of convergence of ${\cal A}(\kappa)$ to a well-defined value.
As a consequence of Lemma \ref{lemmasharp}, one has $ \limsup_{\kappa \to 0} \sigma^- {\cal A}_{m^\star}(\kappa_{m^\star}(\kappa))
\leq \limsup_{\kappa \to 0} {\cal A}(\kappa)$ and $\liminf_{\kappa \to 0} {\cal A}(\kappa)
\leq  \liminf_{\kappa \to 0} \sigma^+ {\cal A}_{m^\star}(\kappa_{m^\star}(\kappa))$. Since $\sigma^\pm
= \sigma^\pm(M(\kappa))$ converge to $e^{\pm 2 \varepsilon}$, one has
${\cal A}^+ \geq e^{-2\varepsilon} {\cal A}^+_{m_\star}$ and 
${\cal A}^- \leq e^{2\varepsilon} {\cal A}^-_{m_\star}$. We can thus state
\be  \label{cor1}
\frac{{\cal A}^+}{{\cal A}^-}
\geq e^{-4 \varepsilon} \frac{{\cal A}^+_{m_\star}}{{\cal A}^-_{m_\star}}.
\de
The sufficient condition for having spontaneous stochasticity is therefore to have
$\frac{{\cal A}^+_{m_\star}}{{\cal A}^-_{m_\star}} > e^{4 \varepsilon}$.
The rest of the proof is to insure this is indeed possible.
We first prove in Appendix \ref{proof_3} that for $m^\star = 0$, one has
\be \label{Akappa}
\frac{{\cal A}_{m_\star}^+}{{\cal A}_{m_\star}^-} \geq \frac{\kappa^+}{\kappa^-}
\frac{4 \pi^2}{\rho_0} e^{-4 \pi^2 (\kappa^+-\kappa^-)},~~
\rho_0 := 
{\displaystyle \frac{||\nabla \theta_0||_{L^2}^2}{||\theta_0||^2_{L^2}}}
\geq (2\pi)^2.
\de 
Lastly, from Lemma \ref{AHAHAH}:
$\frac{{\cal A}^+}{{\cal A}^-} \geq e^{-4\varepsilon} 
\frac{{\cal A}^+_{m_\star}}{{\cal A}^-_{m_\star}} \geq e^{-4 \varepsilon}~
\frac{4\pi^2}{\rho_0} e^{-4 \pi^2 (\kappa^+-\kappa^-)} c_1 A^2$
and 
$ e^{-4 \pi^2 (\kappa^+-\kappa^-)} 
\geq 
e^{-4\pi^2 \sqrt{c_0} \epsilon_0^{\beta}}$. 
Let us call $f_\lambda(A) = cA^2 e^{-\lambda A}$
with $c = c_1 e^{-4\epsilon} \frac{4 \pi^2}{\rho_0}$ and 
$\lambda = 4\pi^2 \sqrt{c_0} \epsilon_0^{\beta}$. It gives the constraint
$\sup_A f_\lambda(A) = f_\lambda(2/\lambda) > 1$, i.e. $\lambda < 2 \sqrt{c} e^{-1}$, namely
$\Lambda >
\left( (e\pi \sqrt{c_0}) \frac{e^{2\epsilon} \sqrt{\rho_0}}{\sqrt{c_1}} \right)^{\frac{q-1}{\beta}}> \left(\pi \sqrt{\rho_0} \right)^{\frac{q-1}{\beta}}$.
Under this constraint, one has two roots
$f^{-1}(1) = {A_1, A_2}$
with $A_1 \leq A_2$. Since $\gamma < \beta$, it follows that
$\lambda A_1 \leq 2\pi^2 \sqrt{c_0} , \epsilon_0^{\beta - \gamma} \ll 1$.
Therefore, when $\lambda A_1 \ll 1$, one has
$A_1 \approx \frac{1}{\sqrt{c}} = \frac{\sqrt{\rho_0}}{2\pi} \cdot \frac{e^{2\epsilon}}{\sqrt{c_1}}$.
One finally obtains (\ref{LBHyp}). \qed

\subsubsection{Proof of (\ref{Akappa})}\label{proof_3}
We first establish a lower bound for ${\cal A}^+_{m_\star}$. One has by Poincar\'e inequality and energy estimate (see Appendix \ref{proof_2}): ${\cal A}_{m^\star}(\kappa_{m^\star}) \geq \frac12 (1-e^{-8 \pi^2 \kappa_{m^\star}}) ||\theta_0||_{L^2}^2$.
Applying limsup on both sides (and also the fact that $\kappa \mapsto 1-e^{-8 \pi^2 \kappa}$ is monotonically increasing)
gives ${\cal A}^+_{m_\star} \geq 
\frac12 \left(1-e^{-8 \pi^2 \kappa^+}\right) ||\theta_0||_{L^2}^2$. A classical result (see \cite{Poon},\cite{Drivas_Elgindi}) provides a lower bound for the passive scalar:
$$
||\theta_{m^\star}||_{L^2}^2 \geq ||\theta_0||_{L^2}^2 {\rm exp} \left( - \kappa_{m^\star} \rho_0 
I_{{\bf b}_{m^\star}}
 \right),
$$
with
\be \label{rho0}
\rho_0 = \frac{||\nabla \theta_0||^2_{L^2}}{||\theta_0||^2_{L^2}},~~{\rm and}~~
I_{{\bf b}_{m^\star}} = 2 \Int_0^1 {\rm exp} \left( C \Int_0^s ||\nabla {\bf b}_{m^\star}||_{L^\infty} ds' \right) ds.
\de
Therefore ${\cal A}_{m^\star}(\kappa_{m^\star}) \leq \frac12 \left( 1-e^{-\kappa_{m^\star} \rho_0 
I_{{\bf b}_{m^\star}}} \right)
||\theta_0||^2_{L^2}$. We apply liminf on both sides giving
${\cal A}_{m_\star}^- \leq 
\frac12 ( 1-e^{-\kappa^- \rho_0 I_{{\bf b}_{m^\star}}} )
||\theta_0||^2_{L^2}$ and thus ${\cal A}_{m_\star}^+/{\cal A}_{m_\star}^-
\geq (1-e^{-\kappa^+ 8\pi^2})/(1-e^{-\kappa^- \rho_0 I_{{\bf b}_{m^\star}}})$.
For $m^\star = 0$, one can write ${\cal A}_{m_\star}^+/{\cal A}_{m_\star}^-
\geq (1-e^{-2 {\cal E} \chi})/(1-e^{-2\chi}) \geq {\cal E} e^{-\chi ({\cal E}-1)}$
with ${\cal E} = 4\pi^2 \rho_0^{-1} \kappa^+/\kappa^-$ and $\chi = \kappa^- \rho_0$.
Since $\rho_0 \geq 4 \pi^2$, one can conclude.
\qed 
\subsubsection{Proof of Property \ref{diffk}}\label{proofdiffk}
Let us use the short-hand notations 
$\theta_p := \theta^{\kappa_{p,j}}$, $\kappa_p:= \kappa_{p,j}$ and
$\varrho_p^2 := 2 \kappa_p \frac{||\nabla \theta_p||^2}{||\theta_0||^2}$
for $p \in \{1,2\}$. One has $\min\{ \varrho_1^2,\varrho_2^2 \} \geq \varrho^2$.
The energy equations are
$||\theta_p||^2 = 
||\theta_0||^2 (1 - 2 \kappa_p \frac{||\nabla \theta_p||^2}{||\theta_0||^2}) 
= ||\theta_0||^2 (1-2 \varrho_p^2)$. In particular $\varrho_p^2 \in [\varrho^2,\frac12)$.
 By Cauchy-Schwarz
$||(\theta_1-\theta_2)(1,\cdot)||^2 \geq (||\theta_1(1,\cdot)||-||\theta_2(1,\cdot)||)^2$, giving 
\begin{align}
||(\theta_1-\theta_2)(1,\cdot)|| \geq ||\theta_0|| \left| \sqrt{1-2\varrho_1^2}
-\sqrt{1-2\varrho_2^2} \right|  \nonumber \\ 
\geq ||\theta_0|| |\varrho_1-\varrho_2| \frac{2\varrho}{\sqrt{1-2\varrho^2}}
\geq 2 ||\theta_0||  \varrho |\varrho_1-\varrho_2|. \nonumber
\end{align}
Going to the limit $j \to \infty$ gives the the result. \qed
\subsection{Discussion on ${\cal K}_A$ and Lemma 3.3, Lemma 3.4 \cite{AV23}.}
Lemma 3.4 in \cite{AV23} is of crucial importance for establishing the many estimates
giving Theorem 1.1.
First, for $m \in \{1,\cdots,M\}$, one defines
\begin{equation}\label{trueavseq}
\kappa_{m-1} = \overline{{\bf K}}_m^{\kappa_m},~\kappa_M = \kappa,
\end{equation}
where $\overline{\bf K}_m^\kappa$ is introduced in (3.16)\cite{AV23} 
and corresponds to a complicate expression involving the space-time average of flux correctors. This is the "true" sequence one wishes to study rather than
(\ref{avseq}).
The sequence defined in (\ref{avseq}) is close to (\ref{trueavseq})
 in the sense of Lemma 3.3 \cite{AV23}: there exists a constant $C=C(\beta) \geq 1$
such that for $\kappa > 0$ and all $m \in \mathbb{N}$
\begin{equation} \label{trueavesti}
\left| \overline{\bf K}_m^\kappa -
 f_m(\kappa) {\bf I}_2 \right|
\leq \frac{\epsilon_m^{2\beta}}{\kappa} \left( C \frac{\epsilon_m^2}{\kappa \tau_m}
+ C \epsilon_{m-1}^\delta \right),~~
f_m(\kappa) := \kappa + c_0 \frac{\epsilon_m^{2\beta}}{\kappa}.
\end{equation}
In particular, to ensure that the two map iterates $\kappa \mapsto \overline{\bf K}_m^\kappa$
and $\kappa \mapsto f_m(\kappa)$ stay close to each other, one must have $\frac{\epsilon_m^2}{\kappa \tau_m} \ll 1$. This is the condition of Lemma 3.4
 (3.47): $c \epsilon_{m-1}^{2\delta} 
\leq \frac{\epsilon_m^2}{\kappa_m \tau_m} \leq C \epsilon_{m-1}^{2\delta}$ with universal
constants $0 < c < C < \infty$.
\\\\
We show now that, indeed one has the same control on $\frac{\epsilon_m^2}{\kappa_m' \tau_m}$, when considering ${\cal K}_A$. Here, and in the following, the prime means that $\kappa_m'$ follows (\ref{avseq}). We need to specify what is $\tau_m$, it is
given in (2.16)\cite{AV23}: $\tau_m \simeq \epsilon_{m-1}^{2-\beta+4\delta}$.
We have using (\ref{kappam}) for $\kappa_m'$, $\frac{\epsilon_m^2}{\kappa_m' \tau_m} \simeq \epsilon_m^2 \epsilon_m^{-\beta-\gamma}
\epsilon_{m-1}^{-(2-\beta+4\delta} = \epsilon_{m-1}^{(2-\beta)(q-1)-\gamma q - 4\delta} $.
 Then, one has
$(2-\beta)(q-1) -q \gamma = 2 (q-1)(1 - \frac{2q+1}{2q+2}\beta) = 8 \delta$ 
(instead of $4 \delta$ in \cite{AV23}). It then gives instead of (3.53) with $2\delta$:
\be \label{controltrue}
c \epsilon_{m-1}^{4\delta} \leq \frac{\epsilon_m^2}{\kappa_m'\tau_m} 
\leq C \epsilon_{m-1}^{4\delta},
\de  
where $c,C$ are universal (now depending also on the constant $A$).
The typo in \cite{AV23} has of course no consequences.
\\\\
The following is an induction proof that (\ref{trueavseq}) and (\ref{avseq}) can be
made arbitrarily close to each other for $\Lambda$ large enough. 
It is again possible mainly due to the hypergeometric scale separation. Sequence (\ref{avseq})
will be denoted with a prime with the mapping $\kappa' \mapsto f_m(\kappa') = \kappa' g_m(\kappa')$
with $g_m(\kappa') = 1 + \frac{c_0 \epsilon_m^{2\beta}}{\kappa'^2}$.
We assume that there is some $1 < B \leq 2$ such that
\be \label{H}
\frac{1}{B} \leq \frac{\kappa'_m}{\kappa_m} \leq B,~~
\forall m \in \{m_0,\cdots,M\} \tag{H}
\de 
First, it is valid for $m_0 = M$ since both sequences start from the same initial condition, i.e. $\kappa_M'/\kappa_M = 1$.
We have $c \epsilon_m^{\beta+\gamma} \leq \kappa_m' \leq C \epsilon_m^{\beta+\gamma}$
where $C =2 A \sqrt{c_0}$, $c = \sqrt{c_0}/(2A)$,
and using (H), $c B^{-1} \epsilon_m^{\beta+\gamma} \leq \kappa_m \leq C B 
\epsilon_m^{\beta+\gamma}$.
It gives
$$
\frac{1+ (2A)^{-2} \epsilon_m^{2\gamma}}{1 + (2AB)^2 \epsilon_m^{2\gamma}}
\left(\frac{\kappa_m}{\kappa_m'} \right)^2
\leq \frac{g_m(\kappa'_m)}{g_m(\kappa_m)} \leq \frac{1+ 4 A^2  \epsilon_m^{2\gamma}}
{1+ (2 A B)^{-2}  \epsilon_m^{2\gamma}} 
\left(\frac{\kappa_m}{\kappa_m'} \right)^2,
$$
and 
\be \label{ggpr}
 (1 - 4 A^2B^2 \epsilon_m^{2\gamma}) \left(\frac{\kappa_m}{\kappa_m'} \right)^2
\leq \frac{g_m(\kappa'_m)}{g_m(\kappa_m)} \leq (1+ 4 A^2 \epsilon_m^{2\gamma})
\left(\frac{\kappa_m}{\kappa_m'} \right)^2.
\de 
 We now rewrite (\ref{trueavesti}) using our notations and replacing $\kappa$ 
 by $\kappa_m$. It is
 \begin{align} \label{trueavesti2}
 \left| \frac{\kappa_{m-1}}{\kappa_m} - g_m(\kappa_m) 
 \right| \leq \frac{\epsilon_m^{2\beta}}{\kappa_m^2} \left(
 C \frac{\epsilon_m^2}{\kappa_m \tau_m} + C \epsilon_{m-1}^\delta
 \right) \leq C_m g_m(\kappa_m) \nonumber \\ {\rm with}~C_m = 
  B C \epsilon_{m-1}^{4 \delta} + C \epsilon_{m-1}^\delta,
 \end{align}
 where one has used (\ref{controltrue}). We then estimate the quantities
 $\kappa_{m-1}'/\kappa_{m-1}$ and $\kappa_{m-1}/\kappa_{m-1}'$.
 One has $\frac{\kappa_{m-1}}{\kappa_{m-1}'} = 
 \frac{\kappa_{m-1}}{\kappa_m} \frac{\kappa'_m}{\kappa_{m-1}'} \frac{\kappa_m}{\kappa_m'}
 = \frac{\kappa_{m-1}}{\kappa_m} \frac{1}{g_m(\kappa_m')} \frac{\kappa_m}{\kappa_m'}$
 using the relation $\kappa_{m-1}' = \kappa_m' g_m(\kappa_m')$. One then uses (\ref{trueavesti2}) and (\ref{ggpr}) to obtain
 $$
 \frac{\kappa_{m-1}}{\kappa_{m-1}'} \leq \left(1 + C_m \right) 
 \frac{g_m(\kappa_m)}{g_m(\kappa_m')} \frac{\kappa_m}{\kappa_m'} 
 \leq (1+C_m)(1+4 A^2 B^2 \epsilon_m^{2\gamma}) \frac{\kappa_m'}{\kappa_m}.
 $$
 We can do the same reasoning for the inverse $\frac{\kappa_{m-1}'}{\kappa_{m-1}}$. 
 It gives
 $$
 \frac{\kappa_{m-1}'}{\kappa_{m-1}} \leq  \frac{1+4 A^2 \epsilon_m^{2\gamma}}{1-C_m}
 \frac{\kappa_m}{\kappa_m'}.
 $$
 Finally, one can write, from the definition of $C_m$, and for an ad-hoc constant $C$
 depending on $\beta$ and $A$ (and also $2\gamma q > \delta$), 
 \be
 \max \left\{ \frac{\kappa_{m-1}'}{\kappa_{m-1}},\frac{\kappa_{m-1}}{\kappa_{m-1}'}
 \right\}
 \leq   (1 + C \epsilon_{m-1}^\gamma )\max \left\{ \frac{\kappa_{m}'}{\kappa_{m}},\frac{\kappa_{m}}{\kappa_{m}'}
 \right\},
 \de 
 $\forall m \in \{ m_0,\cdots,M \}$.
 One cannot use (H), it is necessary to compute the full product term making use
 of (\ref{Sharprod}). It gives
 $$
 \max \left\{ \frac{\kappa_{m_0-1}'}{\kappa_{m_0-1}},\frac{\kappa_{m_0-1}}{\kappa_{m_0-1}'}
 \right\} \leq \Pi_{C,\gamma,m_0-1} \leq {\rm exp}(2 C \epsilon_{m_0-1}^\gamma) 
 \approx 1 + 2C \epsilon_{m_0-1}^\gamma.
 $$
 The constant $B$ in (\ref{H}) can  therefore be chosen arbitrarily close to 1, by taking
 $\Lambda$ large enough. In particular, both (\ref{kappam}) and (\ref{controltrue}) are also valid for the sequence (\ref{trueavseq}). \qed
 \subsection{Discussion on $m^\star$}
Strictly speaking,
estimates in \cite{AV23} are obtained only for $m^\star \geq 1$ defined in (\ref{mstar}). However, looking more closely, all the necessary estimates in \cite{AV23} also hold when one assumes $\epsilon_0^{-1} = 
\lceil \Lambda^{\frac{1}{q-1}} \rceil$ instead of defining $\epsilon_0 = 1$ in \cite{AV23}. In particular, 
the hypothesis of (4.61) on which all the estimates of Sections 4 and 5 in \cite{AV23} rely, are satisfied in this case.
\subsection{Renormalised sequence: numerics} \label{Odeun}
\begin{figure}[htbp]
\centerline{\includegraphics[width=0.95\columnwidth]{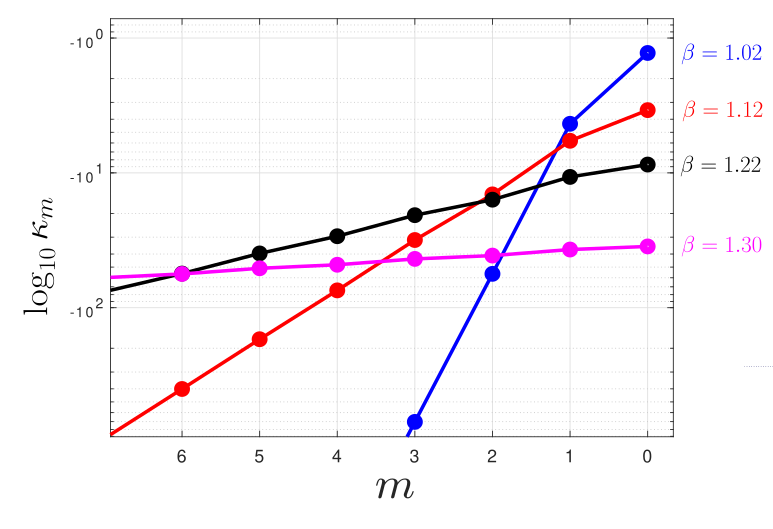}}
\vspace*{-0.3cm}
\caption{The diffusivity sequence $\kappa_m$ in loglog scale for four different values of $\beta$ using $\Lambda=128$. One can notice that as $\beta \to \frac43$, the slope tends to zero while the intercept diverges to minus infinity.}
\label{valkap}
\end{figure}

\begin{figure}[htpb]
\centerline{\includegraphics[width=0.9\columnwidth]{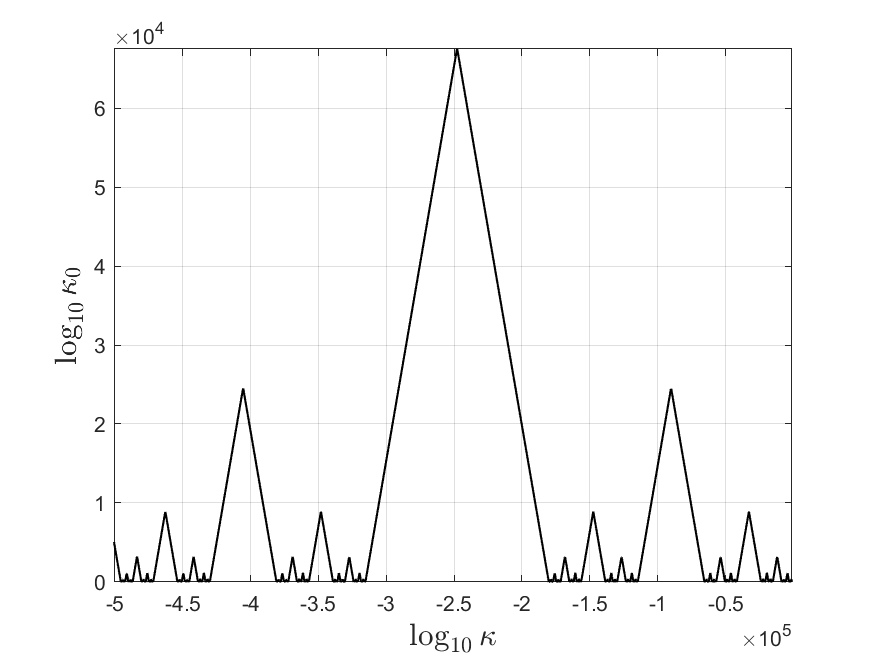}}
\vspace*{-0.3cm}
\caption{Values of $\kappa_0$ as a function of $\kappa_M = \kappa$ (in $\log_{10}$ scale) for $\beta = 1.10 < \beta^\star$ ($\Lambda = 128$). The renormalised diffusivity sequence becomes wild, with unbounded behaviour unless $\kappa_M \in {\cal K}$ is imposed.}
\label{unboundk0}
\end{figure}

We recall the definition of the diffusivity sequence: $\kappa_{m-1} = \kappa_{m} + 
c_0 \frac{\epsilon_m^{2\beta}}{\kappa_m},~~\kappa_M = \kappa \in {\cal K}$ with
$\epsilon_m^{-1} = \lceil \Lambda^{\frac{q^m}{q-1}} \rceil $, $c_0 = 9/80$ and ${\cal K} = \bigcup_{m \geq 1} \left[ \frac12,2 \right] \epsilon_m^{\frac{2\beta}{q+1}}$. To meet the requirements outlined in \cite{AV23,AV24}, we fix $M$ and set $\kappa_M$ as $\theta \epsilon_M^{\frac{2\beta}{q+1}}$ for $\theta \in  (\frac12,2)$. We fix $\theta=1$. The sequence is shown in Fig. \ref{valkap} for different values of $\beta \in (1,\frac43)$.
Only in the limit $\beta \to 1^+$, one expects $\kappa_0 = O(1)$. Note that the values of $\kappa_0$ depend on $\theta$ in all cases (not shown), indicating the absence of a selection principle at the level of $\kappa_0$. The selection occurs only at the level of the measures, which is the primary focus of this work. We also establish the existence of a critical value $\beta^\star \approx 1.1428$, above which the constraint set ${\cal K}$ is not required to control the sequence. For $\beta < \beta^\star$, however, it is necessary to enforce $\kappa_M \in {\cal K}$. This behaviour is illustrated in Fig. \ref{unboundk0} for $\beta = 1.10$.

\addtocontents{toc}{\protect\setcounter{tocdepth}{1}}

\end{document}